\documentclass[twocolumn,prd,aps,superscriptaddress,preprintnumbers,tightenlines,showpacs,nofootinbib,eqsecnum,amsfonts,amsmath,longbibliography]{revtex4-2}

\pdfoutput=1

\usepackage{epsfig}
\usepackage{graphics}
\usepackage{graphicx}
\usepackage{amsmath,amsfonts,amssymb,amsthm}
\usepackage[table,usenames,dvipsnames]{xcolor}
\usepackage{wasysym}
\usepackage{times}
\usepackage{mathptmx}
\usepackage{gensymb}
\usepackage{appendix}
\usepackage{listings}
\usepackage{url}
\usepackage[normalem]{ulem}
\usepackage{alltt}
\usepackage[colorlinks]{hyperref}
\usepackage{cleveref}
\usepackage{enumitem}
\setlist{nosep}
\usepackage{calc}
\usepackage{tensor}
\usepackage{bm}
\usepackage{times}
\usepackage[varg]{txfonts}
\usepackage{float}
\usepackage{dcolumn}
\usepackage[nolist,nohyperlinks]{acronym}
\usepackage{xspace}
\usepackage[english]{babel}
\usepackage[abs]{overpic}
\usepackage{pict2e}
\usepackage[caption=false]{subfig} 
\allowdisplaybreaks[1]
\usepackage[utf8]{inputenc}
\usepackage{gensymb}
\usepackage{stackengine}
\usepackage{boldline,multirow}
\usepackage{braket}

\usepackage{rotating}
\usepackage{booktabs}
\usepackage{longtable}

\usepackage{adjustbox}

\graphicspath{{pics/}}
\newcommand{\AEI}{Max Planck Institute for Gravitational Physics (Albert Einstein Institute), Am M\"uhlenberg 1, Potsdam 14476, Germany}
\newcommand{\Maryland}{Department of Physics, University of Maryland, College Park, MD 20742, USA}
\newcommand{\MarylandAstro}{Department of Astronomy and Joint Space-Science Institute, University of Maryland, College Park, MD 20742, USA}
\newcommand{\MPIIS}{Max Planck Institute for Intelligent Systems, Max-Planck-Ring 4, 72076 T\"ubingen, Germany}

\newcommand{\Nikhef}{Nikhef, Science Park 105, 1098 XG Amsterdam, The Netherlands}
\newcommand{\URIphysics}{Department of Physics, East Hall, University of Rhode Island, Kingston, RI 02881, USA}
\newcommand{\URICCR}{Center for Computational Research, Carothers Library, University of Rhode Island, Kingston, RI 02881, USA}
\newcommand{\UoN}{School of Mathematical Sciences, University of Nottingham, University Park, Nottingham NG7 2RD, United Kingdom}
\newcommand{\Camb}{Department of Applied Mathematics and Theoretical Physics, Cambridge CB3 0WA, United Kingdom}
\newcommand{\Kavli}{Kavli Institute for Cosmology Cambridge, Madingley Road, Cambridge CB3 0HA, United Kingdom}

\definecolor{dodgerblue}{HTML}{1E90FF}
\definecolor{viennared}{HTML}{DA0A14}

\hypersetup{citecolor=dodgerblue}

\begin{document}

\title{Evidence for eccentricity in the population of binary black holes observed by LIGO-Virgo-KAGRA}


\author{Nihar Gupte}
\affiliation{\AEI}
\affiliation{\Maryland}

\author{Antoni Ramos-Buades}
\affiliation{\Nikhef}
\affiliation{\AEI}

\author{Alessandra Buonanno}
\affiliation{\AEI}
\affiliation{\Maryland}

\author{Jonathan Gair}
\affiliation{\AEI}

\author{M. Coleman Miller}
\affiliation{\MarylandAstro}

\author{Maximilian Dax}
\affiliation{\MPIIS}

\author{Stephen R. Green}
\affiliation{\UoN}

\author{Michael P\"urrer}
\affiliation{\URIphysics}
\affiliation{\URICCR}

\author{Jonas Wildberger}
\affiliation{\MPIIS}

\author{Jakob Macke}
\affiliation{\MPIIS}

\author{Isobel M. Romero-Shaw}
\affiliation{\Camb}
\affiliation{\Kavli}

\author{Bernhard Sch\"olkopf}
\affiliation{\MPIIS}

\begin{abstract}
    Binary black holes (BBHs) in eccentric orbits produce distinct modulations
    in the emitted gravitational waves (GWs). The measurement of orbital
    eccentricity can provide robust evidence for dynamical binary formation
    channels. We analyze 57 GW events from the first, second
    and third observing runs of the LIGO-Virgo-KAGRA (LVK) Collaboration using a
    multipolar aligned-spin inspiral-merger-ringdown waveform model with two
    eccentric parameters: eccentricity and relativistic anomaly (assuming a quasi-circular merger-ringdown). 
    This is made computationally 
    feasible with the machine-learning code \texttt{DINGO}, which accelerates
    inference by 2-3 orders of magnitude compared to traditional inference
    techniques. First, when using a uniform prior on the eccentricity, we find eccentric aligned-spin against quasi-circular
    aligned-spin $\log_{10}$ Bayes factors of 1.84 to 4.75 (depending on the
    glitch mitigation) for GW200129, 3.0 for GW190701 and 1.77 for GW200208\_22. We infer 
    $e_{\text{gw, 10Hz}}$ $(e_{\text{gw, 20Hz}})$ to be $0.27_{-0.12}^{+0.10}$ ($0.16_{-0.05}^{+0.04}$)
    to $0.17_{-0.13}^{+0.14}$ ($0.1_{-0.04}^{+0.05}$) for GW200129, 
    $0.54_{-0.30}^{+0.12}$ ($0.31_{-0.13}^{+0.12}$) for GW190701 and $0.39_{-0.23}^{+0.23}$ ($0.21_{-0.08}^{+0.08}$) for GW200208\_22. 
    Second, we find $\log_{10}$ Bayes factors between the eccentric aligned-spin versus quasi-circular precessing-spin hypothesis between
    1.43 and 4.92 for GW200129, 2.61 for GW190701 and 1.23 for GW200208\_22. 
    Third, our analysis does not show evidence for eccentricity in GW190521, 
    which has an eccentric aligned-spin against quasi-circular aligned-spin $\log_{10}$ Bayes 
    factor of 0.04. Fourth, we estimate that if we neglect the spin-precession
    and use an astrophysically-motivated prior on the rate of eccentric BBHs, the
    probability of one out of the 57 events being eccentric is greater than
    99.5\% or $(100 - 8.4 \times 10^{-4})$\% (depending on the glitch mitigation).
    Fifth, we study the impact on parameter estimation
    when neglecting either eccentricity in quasi-circular models 
    or higher modes in eccentric models for GW events. These
    results underscore the importance of including eccentric parameters in the
    characterization of BBHs for the upcoming observing runs of the LVK
    Collaboration and for future detectors on the ground and in space, which will probe a more diverse BBH population. 
\end{abstract}

\date{\today}

\maketitle 

\section{\label{sec:level1}Introduction\protect} 

Gravitational waves (GWs) have been a gold mine for astrophysical characterization and discovery since the first detection in
2015  \cite{LIGOScientific:2018mvr, LIGOScientific:2020ibl, LIGOScientific:2016aoc,
LIGOScientific:2021usb, LIGOScientific:2020ibl, Nitz:2021uxj, Olsen:2022pin,
KAGRA:2018plz, KAGRA:2020tym, Islam:2023zzj, nitz20234,
TheLIGOScientific:2014jea, TheVirgo:2014hva}. They are
a clean probe for binary black holes (BBHs) due to their weak interaction
with matter. GWs have been used to infer the masses, spins 
\cite{LIGOScientific:2021psn,
galaudage2021building, tong2022population, Farr:2017uvj,
Tiwari:2018qch, belczynski2016first,
Hoy:2021rfv} eccentric parameters
\cite{Romero-Shaw:2022xko,Romero_Shaw:2020thy,Romero-Shaw:2019itr,
Iglesias:2022xfc, Ramos-Buades:2023yhy}, and test General Relativity (GR)
\cite{Maggio:2022hre, Mehta:2022pcn,
LIGOScientific:2020tif,LIGOScientific:2020tif,LIGOScientific:2016lio,Cornish:2011ys,Isi:2019asy}.
BBH properties are important as they can give insight into the binary's
formation channels~\cite{Zevin:2021rtf, Belczynski:2022wky,
Kovetz:2016kpi,Fragione:2017blf,Banerjee:2017mgr,
Kimball:2020qyd}.

While stellar-mass black holes (BHs) form from the collapse of stars with masses $\gtrsim 20
M_\odot$~\cite{Bailyn:1997xt,Fryer:1999mi,Bethe:1998bn}, and primordial
BHs may form from over-dense regions in the early Universe
\cite{Carr:2016drx}, the formation of a BBH is not trivial. One cannot
assume that all stellar binaries will become BBHs. If the initial separation of 
two $\sim 10 M_\odot$ compact objects in a quasi-circular orbit is greater than $\sim 20 R_\odot$, 
then they will not merge in a Hubble time ($\sim 14$
Gyr) from gravitational radiation alone \cite{peters1964gravitational}.
However, if the initial separation of this binary is less than $\sim 20 R_\odot$, the stars merge before forming BHs (see,
e.g., Figs.~1 and 2 in Ref.~\cite{Mandel:2018hfr}).

Thus, to explain the formation of BBHs, two classes of formation mechanisms of
BBHs have been proposed in the literature: isolated binary evolution and
dynamical formation.  Within these categories, the former has sub-channels including common-envelope evolution
\cite{Bethe:1998bn, Belczynski:2001uc, Belczynski:2016ieo,
Belczynski:2016jno, Dominik:2012kk, Mennekens:2013dja,
Barrett:2017fcw, Giacobbo:2018etu}, chemically-homogenous evolution
\cite{Mandel:2015qlu,deMink:2016vkw} and Population III stars
\cite{Belczynski:2004gu}. Similarly, the dynamical channel has sub-channels, e.g., binaries arising from triple interactions
\cite{Antonini:2017ash,Fragione:2020nib,Silsbee:2016djf,Arca-Sedda:2018qgq,Vigna-Gomez:2020fvw},
von Zeipel-Kozai-Lidov (ZKL) oscillations
\cite{1910AN....183..345V, kozai1962secular,lidov1962evolution,giacaglia1964notes,
Kimpson:2016dgk, Wen:2002km}, binary-binary
interactions \cite{Zevin:2018kzq}, and binary-single interactions
\cite{Samsing:2017rat, Samsing:2013kua}. These possibilities had been proposed in
Refs. \cite{Miller:2001ez, heggie1975binary,sigurdsson1993primordial,Kulkarni:1993fr,Sigurdsson:1994ju,Sigurdsson:1993tui,PortegiesZwart:1999nm,Gultekin:2005fd}, and further investigated in 
Refs.~\cite{Antonini:2015zsa,OLeary:2005vqo,Gondan:2017wzd,Hoang:2017fvh,Fragione:2019vgr,Takatsy:2018euo,OLeary:2008myb,Britt:2021dtg}

It is unclear what fraction of BBHs arise from each formation channel. Multiple
channels may operate simultaneously making the origin of BBHs even harder to pinpoint.
But measuring eccentricity in a BBH can serve as a reliable indicator 
of a dynamical formation channel \cite{Wen:2002km}. This
is because isolated binaries that are initially eccentric 
would circularize by the time they enter the LIGO-Virgo-KAGRA (LVK) band
\cite{peters1964gravitational}. So by studying the effect of eccentricity, we 
can help discriminate what fraction of BBHs originate from dynamical interactions 
or isolated binary (IB) evolution \cite{Zevin:2021rtf}.

There is also good reason to believe we may observe a few eccentric events in
the LVK observing runs \cite{TheLIGOScientific:2014jea, TheVirgo:2014hva, KAGRA:2018plz} or future GW observatories,  
such as the Einstein Telescope \cite{Punturo:2010zz}, Cosmic Explorer \cite{Reitze:2019iox} and LISA \cite{LISA:2017pwj}. 
For instance, there have been robust simulations
performed to determine the expected number of BBHs from the dynamical-formation
channel. Simulations which include the post-Newtonian (PN) description of the orbital dynamics in globular star clusters show that 
$\approx 5\%$ of dynamical mergers have eccentricities greater than 0.1 This estimate refers to
an astrophysical definition of eccentricity \cite{Wen:2002km,Vijaykumar:2024piy}, not the quasi-Keplerian definition. at 10 Hz (in the detector frame) 
\cite{Samsing:2017rat,Rodriguez:2018pss,Rodriguez:2017pec}. Estimates from galactic 
nuclei predict higher rates of eccentricity with up to 70\% of mergers predicted to have eccentricities 
larger than 0.1\footnotemark[1] at 10 Hz
\cite{Tagawa:2020jnc,Samsing:2020tda,Gondan:2020svr}. 
These high
eccentricity mergers could be detected by upcoming runs of the LVK collaboration. It is then important to 
have an eccentric waveform model to analyze the data. 

Taking eccentricity into account is also important as ignoring it can lead to systematic
biases in parameter estimation~\cite{Cho:2022cdy, Divyajyoti:2023rht,
Bonino:2022hkj, Wu:2020zwr,Lower:2018seu, Ramos-Buades:2023yhy,
OShea:2021faf, Favata:2021vhw, Favata:2013rwa,
Sun:2015bva, Wagner:2024ecj, LIGOScientific:2016ebw, DuttaRoy:2024aew}. So far, eccentric waveform models have not been used for analysis of the LVK GW transient
catalogues (GWTC) \cite{LIGOScientific:2018mvr,
LIGOScientific:2020ibl, LIGOScientific:2016aoc, LIGOScientific:2021usb}. This limitation may cause biases and have implications
on hierarchical BBH population studies and tests of GR~\cite{Saini:2023rto, Bhat:2022amc, Ma:2019rei, Narayan:2023vhm, Shaikh:2024wyn, Saini:2022igm}.

Several parameter-estimation studies that included one eccentric parameter in
the waveform model have been carried out in the last few
years~\cite{Sun:2015bva,Romero-Shaw:2019itr,Gayathri:2020coq, Iglesias:2022xfc,
Romero_Shaw:2020thy, Wu:2020zwr, Lenon:2020oza, Ramos-Buades:2019uvh,
OShea:2021faf, Romero-Shaw:2022xko}. There have also been parameter estimation
studies that use two parameters to explore the effect of
eccentricity~\cite{Gamba:2021gap,Ramos-Buades:2023yhy,Bonino:2022hkj}. In
particular, Ref.~\cite{Bonino:2022hkj} performed inference by fixing the binary
at apastron and sampling on the initial eccentricity and the average starting
frequency between apastron and periastron. On the other hand,
Ref.~\cite{Ramos-Buades:2023yhy} fixes
the orbit averaged starting frequency and samples directly on the relativistic
anomaly and the eccentricity. We note that sampling on starting frequency and
eccentricity at fixed relativistic anomaly may not be equivalent to sampling on
eccentricity and relativistic anomaly at a fixed starting frequency. In
principle these methods can probe the same parameter space. However, if
sampling on the starting frequency, one has to choose arbitrary prior bounds on
the starting frequency. If there are templates with a better match to the GW
outside of these bounds, this method does not cover the same parameter space as
the method which fixes the starting frequency and samples on the relativistic
anomaly and eccentricity. In addition, sampling on the starting frequency will
change the reference point where the binary properties, such as the reference
phase, are measured. Finally, Ref.~\cite{Gamba:2021gap} samples on the energy
and angular momentum of the binary at a fixed instantaneous starting frequency.
We note that whereas Ref.~\cite{Ramos-Buades:2023yhy} deals with (bound)
eccentric orbits, Ref.~\cite{Gamba:2021gap} focuses on hyperbolic orbits.
During the referee period of this paper there have been even more studies with eccentric waveform models
including Refs.~\cite{McMillin:2025hof, Morras:2025xfu, Planas:2025jny}.

Due to the computational cost, the aforementioned analyses were restricted to either sampling on
non-eccentric parameters and reweighting with eccentric models, or to a small
number of GW events when using Bayesian methods that sample directly on the
eccentric parameters. Here, we employ the Deep INference for Gravitational-wave
Observations (\texttt{DINGO}) code to sample a large number of GW events. As in
Refs.~\cite{Bonino:2022hkj, Ramos-Buades:2023yhy}, we perform inference on
the two eccentric parameters: eccentricity and relativistic anomaly. This is
important, as ignoring the relativistic anomaly can lead to biases in parameter
estimation~\cite{Clarke:2022fma,Ramos-Buades:2023yhy}. \texttt{DINGO} has
already been used to analyze a subset of the spin-precessing BBHs in the first
and third observing runs~\cite{Dax:2022pxd}.  Remarkably, \texttt{DINGO} yields
posterior distributions within $\mathcal{O}$(min - hours) without sacrificing
accuracy \cite{Dax:2022pxd, Dax:2023ozk}. However, it is currently limited to GW
signals shorter than $16\, {\rm sec}$.

In this study, we employ a set of waveform models from the effective-one-body
(EOB) family \texttt{SEOBNR}. We use the multipolar aligned-spin
quasi-circular model \texttt{SEOBNRv4HM}~\cite{Bohe:2016gbl,Cotesta:2018fcv} and the
multipolar aligned-spin eccentric model
\texttt{SEOBNRv4EHM}~\cite{Ramos-Buades:2021adz,Khalil:2021txt} to analyze 57
GW events from the first (O1), second (O2) and third (O3) observing runs of the LVK
Collaboration ~\cite{LIGOScientific:2018jsj, KAGRA:2021vkt, LIGOScientific:2021usb}. 
We find $\log_{10}$ Bayes factors, between the eccentric against quasi-circular
hypothesis, greater than 1 for three events. 
We also do a comprehensive study to understand the impact of
glitch subtraction for one of the GW events (GW200129), which  happens to be
the BBH with the largest evidence for eccentricity.
Confirming the recent results in Ref. \cite{Ramos-Buades:2023yhy} we
do not find evidence for eccentricity in GW190521, which previous analyses had
suggested to be eccentric~\cite{Romero_Shaw:2020thy, Gayathri:2020coq}.

It has been suggested that for some regions of the parameter space, and at certain stages of the coalescence, 
spin-precession can be mistaken for eccentricity in parameter-estimation studies \cite{Romero-Shaw:2022fbf,CalderonBustillo:2020xms}. Thus, 
for comparison, we also employ two multipolar spin-precessing waveform models for quasi-circular orbits. One is from 
the EOB waveform family, \texttt{SEOBNRv4PHM}~\cite{Buonanno:2005xu,Pan:2013rra,Babak:2016tgq,Ossokine:2020kjp}, 
and the other is from the numerical-relativity 
surrogate family, \texttt{NRSur7dq4}~\cite{Varma:2019csw}. Unfortunately, inspiral-merger-ringdown waveform 
models with spin-precession and eccentricity are not yet available, with the 
exception of the recent Ref.~\cite{Liu2023spinprececc}, whose code is not available to us.
We also explore the impact of ignoring eccentricity and higher harmonics in
parameter estimation.

When trying to determine if a BBH is eccentric, we must also consider the relative merger rates of eccentric and quasi-circular BBHs. This can be done by incorporating astrophysical 
knowledge from N-body simulations or semi-analytic calculations. We can also directly 
infer the rates from the GW data. In this paper, we combine both approaches to first 
generate a prior on the rate of eccentric and quasi-circular BBHs based on results from the 
literature, and then incorporate GW information. This results in a new way to quantify the probability 
that a GW comes from an eccentric BBH, similar to the $p_{\text{astro}}$ reported in 
LVK analyses \cite{Farr:2013yna}. This approach also allows us to quantify the 
probability of eccentricity in the population of BBHs. 

The paper is organized as follows. In Sec.~\ref{sec:bayesian_inference}, we give
an overview of the Bayesian inference techniques used in this study. In particular, 
in Sec.~\ref{sec:bayes_theorem} we discuss Bayes theorem and how to compute odds
ratios, in Sec.~\ref{sec:p_ecc} we compute the
probability a GW signal is eccentric based on the ensemble of 
events, and in Sec.~\ref{sec:DINGO} we review the machine-learning code \texttt{DINGO}. 
In Sec.~\ref{sec:methods_waveform_models}, we introduce the waveform models used in this 
paper, and review the method introduced
in Refs.~\cite{Shaikh:2023ypz,Ramos-Buades:2022lgf} to infer
eccentricity directly from the waveform. Furthermore, in Sec.~\ref{sec:trained_networks} we
discuss the settings of the trained networks and priors employed in the \texttt{DINGO} analysis, while 
in Sec.~\ref{sec:zero_noise} we validate the neural-networks with zero-noise synthetic signals
and comparisons to parallel \texttt{Bilby} (\texttt{pBilby}) \cite{Smith:2019ucc}. From Secs.
\ref{sec:GW200129} through \ref{sec:GW190521}, we present the analysis of 57 GWs and
discuss in depth the events that support eccentricity. In Secs.~\ref{sec:neglecting_eccentricity} and \ref{sec:hms},
we investigate the impact on parameter estimation of neglecting eccentricity and higher modes in waveform models, respectively.
In Sec.~\ref{sec:prior_odds}, we incorporate astrophysical BBH merger-rate estimates of eccentric and quasi-circular events  
into our analysis, while in Sec.~\ref{seca:p_ecc}, we estimate the probability of GW signals being eccentric 
and the probability that eccentricity exists in the population of GWs analyzed. In Sec.~\ref{sec:conclusion}, 
we summarize our main conclusions, comparing our results with the literature and suggesting future work. 
In Appendix \ref{sec:one_detector}, we perform analyses of GW200129 using either only
the LIGO-Hanford or the LIGO-Livingston detector. In Appendix
\ref{sec:lower_bound_p_ecc_pop}, we describe a method to compute the probability
there exists an eccentric event in the ensemble of GWs analyzed. 
In Appendix \ref{sec:background_distribution},
we do full parameter estimation and recover the Bayesian evidence of over 500 injections drawn from an astrophysical distribution to assess
the significance of our Bayes factors. In Appendix 
\ref{sec:gaussian_noise} we show agreement between DINGO and pBilby for an eccentric injection 
in simulated gaussian noise. In Appendix \ref{sec:kick_velocity}, we present the kick velocity posterior
distribution of GW200129 when using the multipolar aligned-spin eccentric model \texttt{SEOBNRv4EHM}, and compare it to
results using the multipolar spin-precessing quasi-circualr model \texttt{NRSur7dq4}.
In Appendix \ref{sec:occam_penalty} we explore the effect of the occam penalty 
when comparing eccentric aligned-spin and quasi-circular spin-precessing approximants.
Lastly, In Appendix \ref{sec:cross_injections}, we present the results of
synthetic-signal injections aimed at determining if eccentricity could be
mistaken for spin-precession in GW200129.

\section*{Notation}
In this paper we use natural units, and set $G = c = 1$ unless otherwise specified. 
Individual component masses are denoted $m_1, m_2$,  
while the total mass is $M \equiv m_1 + m_2$, and the chirp mass is $\mathcal{M} = (m_1 m_2)^{3/5} M^{-1/5}$.
We use the convention, $q \equiv m_2/m_1 \leq 1$, and define 
$\mu \equiv m_1 m_2 / M$ and $\nu \equiv \mu / M $. 
We quote the masses in both the source and detector frame denoting them as 
$M_{\text{src}} \equiv M$ and $M_{\text{det}} \equiv M (1 + z)$, respectively. Here $z$ is the 
redshift of the source. 

Although we mostly consider BHs with spins $\bm{S_{1,2}}$ aligned (or anti-aligned) with the direction perpendicular to the 
orbital plane, we also carry out a few studies with spin-precessing BHs. Thus, generically, we introduce the dimensionless spin 
\begin{equation}
    \label{eq:comp_spins}
    \chi_i \equiv \frac{\bm{S_i} \cdot \boldmath{\bm{\hat{L}}}}{m_i^2} \quad i = 1,2\,, 
\end{equation}
which takes values in the range $[-1,+1]$, where $\hat{\bm{L}}$ is the unit vector pointing in the direction perpendicular to the instantaneous orbital plane. In parameter estimation one
often uses $\chi_{\text{eff}}$ as it is better constrained \cite{Damour:2001tu, Racine:2008qv, Santamaria:2010yb, Ajith:2009bn}. This is defined as

\begin{equation}
    \label{eq:chi_eff}
    \chi_{\text{eff}} \equiv \frac{m_1 \chi_1 + m_2 \chi_2}{M}.
\end{equation}

Similarly, we use the effective precession parameter $\chi_p$ defined in
\cite{Schmidt:2014iyl}, 

\begin{equation}
    \chi_p \equiv\frac{\text{max}(A_1 S_{1,\perp}, A_2 S_{2, \perp})}{A_1 m_1^2}, 
\end{equation}

where $A_1 = 2 + 3/(2q)$ and $A_2 = 2 + 3/(2q)$  and $S_{i, \perp}$ is 
the magnitude of the in plane spins. That is, $S_{i, \perp} = | \bm{S_i} - \bm{S_i} \cdot \bm{L} |$, 
where $\bm{L}$ is the orbital angular momentum. We also introduce the total 
angular momentum $\bm{J} = \bm{L} + \bm{S}_1 + \bm{S}_2$.

We denote $\bm{\vartheta}$ as the set of all GW parameters including intrinsic
parameters of the source and the extrinsic parameters related to projection of
the waveform onto the detector. In the case of eccentric
aligned-spin binaries we have 13 parameters
$\bm{\vartheta} = \{\mathcal{M}, \, q, \, \chi_1, \, \chi_2, \, e, \, \zeta, \, \alpha,
\, \delta, \, d_{\text{L}}, \, t_{\text{coal}}, \, \iota, \, \psi,
\, \phi_{\text{coal}} \}$. In the case of quasi-circular precessing-spin
binaries we have 15 parameters $\bm{\vartheta} = \{ \mathcal{M},
\, q, \, | \chi_1 | , \, | \chi_2 | , \, \theta_1, \theta_2, \phi_{\text{JL}}, \, \phi_{12}, \,
\alpha, \, \delta, \, d_{\text{L}}, \, t_{\text{coal}}, \, \theta_{\text{JN}},
\, \psi, \, \phi_{\text{coal}} \}$. 

We indicate with $e$ the eccentricity and with $\zeta$ the relativistic anomaly, which are
defined more precisely in Sec.~\ref{sec:methods_waveform_models}. The
parameters $\{ | \chi_1 | , \, | \chi_2 | , \, \theta_1, \theta_2, \phi_{\text{JL}}, \,
\phi_{12} \}$ describe the (dimensionless) spins values, spin tilts, the angle between the
total and orbital angular momentum, and the angle between the spins of the
BHs, respectively \cite{Farr:2014qka}. The rest are extrinsic parameters with
$\alpha$ the right ascension, $\delta$ the declination, $d_{\text{L}}$ the
luminosity distance, $t_{\text{coal}}$ the time of coalescence, $\theta_{\text{JN}}$
the angle between the total angular momentum and the line of sight of the
observer, $\iota$ the angle between the orbital angular momentum and the
line of sight of the observer, $\psi$ the polarization angle between the GW and
detectors, and $\phi_{\text{coal}}$ the orbital reference phase at coalescence 
(e.g., see Figs.~1-3 in
Ref.~ \cite{Buonanno:2002fy} for a diagram, and Eq.~(17-20) in Ref.
\cite{Schutz:2011tw} for how the extrinsic parameters enter the detector
projections).  

In the following we assume the Planck 2015 cosmology \cite{Planck:2015fie}.

\section{\label{sec:bayesian_inference} Bayesian inference \protect}

In this section, we outline how Bayesian inference can be used to determine if a
detected GW event is more likely to be described by a BBH moving on an eccentric
or quasi-circular orbit. The approach is to first compute posterior
distributions for individual events, then to use these posteriors to compute
Bayes' factors and finally weight these Bayes' factors by astrophysical
information to calculate the probability that a BBH is eccentric.

\subsection{\label{sec:bayes_theorem} Bayes theorem, Bayes factor and odds ratios }

Here, we summarize the basics of the Bayesian inference formalism used in GW astronomy. In doing so,  
we also introduce the concepts of Bayes factors and odds ratios.

Our hypothesis, $\mathcal{H}_a$, is that in the detector data, $d$, an observed GW signal is
described by a waveform model, $a$, with parameters $\bm{\vartheta}_a$. 
The posterior probability distribution on the parameters of the model, $\bm{\vartheta}_a$, given the hypothesis, $\mathcal{H}_a$, is obtained using Bayes' theorem,
\begin{equation}\label{eq:Bayes_theorem}
p(\bm{\vartheta}_a|d, \mathcal{H}_a) = \frac{p(d|\bm{\vartheta}_a, \mathcal{H}_a)\,p(\bm{\vartheta}_a|\mathcal{H}_a) }{p(d|\mathcal{H}_a)} \,.
\end{equation}
Here, $p(\bm{\vartheta}_a|\mathcal{H}_a)$ is the prior probability distribution, $p(d|\bm{\vartheta}_a, \mathcal{H}_a)$ is the likelihood function, and $p(d|\mathcal{H}_a)$ is the evidence of the hypothesis $\mathcal{H}_a$. 
In this study, we compute most of the posterior distributions
with \texttt{DINGO} (see Sec.~\ref{sec:DINGO}). We also compute a few
posteriors with \texttt{pBilby}~\cite{Smith:2019ucc} for validation and
comparison. 

For a detector with stationary, Gaussian
noise, we use the likelihood function 

\begin{equation}
p(d|\bm{\vartheta}_a, \mathcal{H}_a) \propto \exp \left[ - \tfrac{1}{2} \left< d-h_a(\bm{\vartheta}_a)|d-h_a(\bm{\vartheta}_a) \right>\right] \,,
\end{equation}
where $h_a$ is the GW signal generated by waveform model $a$, and the brackets denote the noise-weighted inner product,
\begin{equation}
\langle A|B \rangle = 2\int_{f_{\rm low}}^{f_{\rm high}} df \, \frac{\tilde{A}^*(f) \tilde{B}(f) + \tilde{A}(f) \tilde{B}^*(f)}{S_n(f)} \, .
\end{equation}
Here, $\tilde{A}(f)$ is the Fourier transform of $A(t)$, the asterisk
denotes the complex conjugation and $S_{n}(f)$ is the one-sided power spectral density (PSD) of the detector.
The integration limits $f_{\rm low}$ and $f_{\rm high}$ are set by the bandwidth of the detector's sensitivity. 

When trying to estimate whether the GW event is eccentric or quasi-circular, we compute the odds ratio
$\mathcal{O}_{a/b}$ between the two hypotheses, $\mathcal{H}_a$ and $\mathcal{H}_b$, which is given by

\begin{equation}
    \label{eq:posterior_odds}
    \mathcal{O}_{a/b} = \frac{p(\mathcal{H}_a)}{p(\mathcal{H}_b)} \frac{p(d | \mathcal{H}_a)}{p(d | \mathcal{H}_b)} \equiv  \frac{R_a}{R_b} \: \: \mathcal{B}_{a/b} .
\end{equation}
The first ratio in the product is the \emph{priors odds}, which represent our belief about the universe prior to
observing any GW data. It can be determined through astrophysical simulations or
other non-GW observations.\footnote{If only analyzing a subset of GW events, we could 
also use the rate estimates from previous GW observing runs to determine the rates (and thus) prior odds.} The second term is the \emph{Bayes factor}, which 
is the odds of the hypotheses given the data\footnote{Note these Bayes factors are defined differently from Ref.~\cite{Romero-Shaw:2022xko} 
and cannot directly be compared. The Bayes factor in Eq. (\ref{eq:posterior_odds}) will be conservative (smaller) compared to the 
$e_{\text{10Hz}} < 0.05$ vs $e_{\text{10Hz}}>0.05$ Bayes factor considered in Ref.~\cite{Romero-Shaw:2022xko}. }. 

In this study, we consider the eccentric aligned-spin (EAS) vs quasi-circular
aligned-spin (QCAS) odds-ratio ($\mathcal{O}_{\text{EAS/QCAS}}$), for which we
need to compute the merger rate of eccentric events $R_{\text{E}}$ and
quasi-circular events $R_{\text{QC}}$. We do not consider the eccentric
aligned-spin vs quasi-circular precessing odds-ratio as to our knowledge no
rate estimates (and thus prior odds) for this fraction exist in the
literature.

\subsection{\label{sec:p_ecc} Probability  a GW event is  eccentric} 

To determine the odds ratio (\ref{eq:posterior_odds}), one needs to include the
priors odds, i.e., some astrophysical information.  But the odds ratio 
does not account for the fact that we have analyzed many GW events,
and therefore have information about the rates directly from GW
observations. Therefore, we propose an alternate way to measure the probability
that an event is eccentric. This method is analogous to the way that the
LVK Collaboration reports the probability that a GW detection is of
astrophysical origin, with the so-called $p_{\text{astro}}$
\cite{Farr:2013yna}. It relies on the formalism derived in Ref.~\cite{Farr:2013yna}.
This method has three advantages over just computing the odds ratio. 

\begin{enumerate}
    \item When computing the merger rates in Eq.~(\ref{eq:posterior_odds}), we are essentially
    drawing from a prior distribution, $p(R_{\text{E}}, R_{\text{QC}})$, which
    is motivated by astrophysics. However, the astrophysical prior has a large
    variance. With an increasing number of GW events, we can begin constraining
    the prior on the rates from observations.

    \item We can naturally incorporate selection effects which are not included when 
    considering the ratio between eccentric and quasi-circular BBH rates in Eq.~(\ref{eq:posterior_odds}).

    \item We can compute the probability that eccentric events exist in the population of GWs analyzed 
    so far without referring to a specific event.
\end{enumerate}

We now review the formalism of Ref.~\cite{Farr:2013yna} and show how it can be used to derive the probability that an 
event is eccentric. Let $\{ d_i \}$ be the collection of strain data segments of GW triggers. 
We assign each event a flag, $g_i$ which is 1 if the event is 
eccentric and 0 if the event is quasi circular. Our goal is to obtain the probability distribution over the set of flags, $\{ g_i \}$.
Let $N$ be the number of events. According to Bayes' theorem, the posterior distribution over the flags is 

\begin{equation}
    \begin{aligned}
        \label{eq:posterior_over_flags}
        & p( \{ g_i \}, R_{\text{E}}, R_{\text{QC}} \: | \: \{ d_i \}, N)   \\
        &= \frac{p(\{ d_i \} \: | \: \{g_i \}, N, R_{\text{E}}, R_{\text{QC}})\, p(\{ g_i \}, N, R_{\text{E}}, R_{\text{QC}})}{p(\{ d_i \}, N)}.
    \end{aligned}
\end{equation}
Let us recover an expression for the likelihood term, $p(\{ d_i \} | \{ g_i \}, N, R_{\text{E}}, R_{\text{QC}})$. The probability that the data segment 
$d_i$ is from an eccentric BBH is 

\begin{equation}
    \label{eq:ecc_probability}
    Z_\mathrm{E}(d_i) \equiv p(d | \mathcal{H}_\text{E}) = \int d\bm{\vartheta} \: p(d | \bm{\vartheta}, \mathcal{H}_\text{E}) p(\bm{\vartheta} | \mathcal{H}_{\text{E}}).
\end{equation}
Here the ``E'' subscript indicates that an eccentric model is being
used. Similarly, $Z_{\mathrm{QC}}(d_i)$ is the probability that $d_i$ is generated by from a quasi-circular BBH.
This is given by Eq.~(\ref{eq:ecc_probability}), but using the quasi-circular
waveform, which we denote as ``QC''. Thus, we can write the likelihood as

\begin{equation}
    \label{eq:fgmc_likelihood}
    p(\{ d_i \} \: | \: \{g_i \}, N, R_{\text{E}}, R_{\text{QC}}) = \left[ \prod_{\{ i | g_i = 1 \}}^{N_\text{E}} Z_{\text{E}}(d_i) \right] \left[ \prod_{\{ i | g_i = 0 \} }^{N_\text{QC}} Z_{\text{QC}}(d_i) \right].
\end{equation}
Here, the conditioning on the rates does not enter the expression for the rate
likelihood. Note that we are only using aligned-spin evidences in the above rate
likelihood and not considering spin-precessing evidences. We leave extending these
results to the spin-precessing case for future work. 

The prior distribution in Eq.~(\ref{eq:posterior_over_flags}) can be factorized
exactly the same way as Eqs.~(13)--(17) in Ref.~\cite{Farr:2013yna}. For
completeness, we include the final expression here: 
\begin{equation}
    \begin{aligned}
        \label{eq:fgmc_prior}
        p(\{ g_i \}, & R_{\text{E}}, R_{\text{QC}}, N) = R_{\text{E}}^{N_\text{E}} \: R_{\text{QC}}^{N_\text{QC}} \: \frac{e^{-(R_\text{E} + R_\text{QC})}}{N!} p(R_{\text{E}}, R_{\text{QC}}),
    \end{aligned}
\end{equation}
where $N_{\text{E}} = \sum_i g_i$ and $N_{\text{QC}} = \sum_i (1-g_i) = N-N_\text{E}$. 
The last term in Eq.~(\ref{eq:fgmc_prior}) is the prior over the rates. This can
either be a uniform prior or an astrophysical prior from the literature. We discuss the derivation of an astrophysical prior on the rates in Sec.~\ref{sec:prior_odds}. The 
exponential term comes from the Poisson uncertainty on the number of events \cite{KAGRA:2021vkt}.
Substituting the likelihood and 
prior term into Eq.~(\ref{eq:posterior_over_flags}) our posterior over the flags and rates becomes:
\begin{equation}
    \begin{aligned}
        \label{eq:fgmc_posterior}
        p(\{ g_i \}, & R_{\text{E}}, R_{\text{QC}} \: | \: \{ d_i \}, N) \\
        \propto & \left[  \prod_{\{ i | g_i = 1 \}}^{N_\text{E}} R_{\text{E}} Z_{\text{E}}(d_i) \right] \left[ \prod_{\{ i | g_i = 0 \} }^{N_\text{QC}} R_{\text{QC}} Z_{\text{QC}}(d_i) \right]   \\
        & \times e^{-(R_\text{E} + R_\text{QC})}
        p(R_{\text{E}}, R_{\text{QC}})\,,
    \end{aligned}
\end{equation}
where we have dropped the evidence term and the factorial, and replaced them with a proportionality. 
If we are only interested in an estimate for the rates, we can marginalize over the flags to get 
\begin{equation}
    \begin{aligned}
        \label{eq:fgmc_flag_marginalized}
        p(& R_{\text{E}}, R_{\text{QC}} \: | \: \{ d_i \}, N) \\
        \propto & \left[  \prod_{i} R_{\text{E}} Z_{\text{E}}(d_i) + R_{\text{QC}} Z_{\text{QC}}(d_i) \right]  
         e^{-(R_\text{E} + R_\text{QC})} 
        p(R_{\text{E}}, R_{\text{QC}}).
    \end{aligned}
\end{equation}
To obtain the normalization of Eq.~(\ref{eq:fgmc_flag_marginalized}) we can
numerically integrate over $R_{\text{E}}$ and $R_{\text{QC}}$.

Finally, we can compute the probability that the $m$'th event is eccentric by
marginalizing over the rates and all flags except the $m$'th flag. This results in the expression
\begin{equation}
    \begin{aligned}
        \label{eq:p_ecc}
        & p_{\text{ecc}}(m) = p(g_m = 1 \: | \: \{ d_i \}, N)  \\
        & = \int dR_\text{E} \: dR_{\text{QC}} \: R_\text{E} Z_\text{E}(d_m) \: \frac{p(R_\text{E}, R_{\text{QC}} | \{ d_i \}, N)}{R_\text{E} Z_\text{E}(d_m) + R_{\text{QC}} Z_{\text{QC}}(d_m)}.
    \end{aligned}
\end{equation}
We can gain intuition on this statistic by considering a limiting case which 
relates to the odds ratio. Suppose that the set of GW observations holds no information on
$R_{\text{E}}$ and $R_{\text{QC}}$. 
This means that $p(R_\text{E}, R_{\text{QC}} | \{ d_i \}, N) = p(R_\text{E},
R_{\text{QC}})$. We can define $(\tilde{R}_{\text{E}}, \tilde{R}_{\text{QC}}) =
\text{argmax} \{ p(R_{\text{E}}, R_{\text{QC}}) \}$. That is,
the pair $(\tilde{R}_{\text{E}}, \tilde{R}_{\text{QC}})$ are the merger rates 
for which the function $p(R_{\text{E}}, R_{\text{QC}})$ is maximized. In this context, $\tilde{R}_{\text{E}}$ and $\tilde{R}_{\text{QC}}$
are the rates we would use to estimate the odds ratio in Eq.~
(\ref{eq:posterior_odds}). If we draw one sample from $p(R_{\text{E}},
R_{\text{QC}})$ to estimate Eq.~ (\ref{eq:p_ecc}) we will find the most likely
estimate for $p_\text{ecc}$ is:

\begin{equation}
    \label{eq:p_ecc_intuition}
    \begin{aligned}
        \tilde{p}_{\text{ecc}}(m) &\equiv \frac{\tilde{R}_{\text{E}} Z_{\text{E}}(d_m)}{\tilde{R}_{\text{E}} Z_{\text{E}}(d_m) + \tilde{R}_{\text{QC}} Z_{\text{QC}}(d_m)} \\
        &=  \frac{\mathcal{O}_{\text{EAS/QCAS}}}{1 + \mathcal{O}_{\text{EAS/QCAS}}}.
    \end{aligned}
\end{equation}
Thus, if we use the prior instead of the rate posterior in Eq.~(\ref{eq:p_ecc}),
$p_\text{ecc}$ can be thought of as a mapping of the odds ratio to the interval
[0, 1]. As more events are observed, $p(R_\text{E}, R_{\text{QC}} | \{ d_i \},
N)$ becomes tighter than $p(R_\text{E}, R_{\text{QC}})$. This means we lose the
large uncertainty due to the astrophysical uncertainty on the rates and get a
better estimate on $p_{\text{ecc}}$.

We can also compute the probability that there exists at least one eccentric event in the 
population of GW events analyzed so far. This can be done by defining 
\begin{equation}
    \label{eq:p_ecc_population}
    p_{\text{ecc, pop}} \equiv 1 - p(\{ g_i = 0 \} \: | \: \{ d_i \}, N).
\end{equation}
That is, the probability there exists eccentricity in the population is one
minus the probability that all events are quasi-circular. This can be computed 
by marginalizing Eq.~(\ref{eq:fgmc_posterior}) over the rates 
and normalizing by the integral over the rates and flags. Explicitly: 

\begin{equation}
    \begin{aligned}
        \label{eq:prob_quasi-circular}
        &p(\{ g_i = 0 \} \: | \: \{ d_i \}, N) = \\
        & \frac{\int dR_\text{E} \: dR_{\text{QC}} \: R_{\text{QC}}^N \prod_i Z_{\text{QC}}(d_i)
        e^{-(R_\text{E} + R_\text{QC})} p(R_{\text{E}}, R_{\text{QC}})}{\int dR_\text{E}
        \: dR_{\text{QC}} \: \prod_j\left[ R_\text{QC} Z_{\text{QC}}(d_j) +R_\text{E}
        Z_\text{E} (d_j)\right] e^{-(R_\text{E} + R_\text{QC})} p(R_{\text{E}},
        R_{\text{QC}})}.
    \end{aligned}
\end{equation}
To compute the normalizing factor (denominator) exactly we need to integrate over
a product with many terms. We can instead compute the dominant terms of this integral to place
a lower bound on the normalizing factor. This allows us to place an upper bound on 
$p(\{ g_i = 0 \} \: | \: \{ d_i \}, N)$ thereby allowing us to place a lower bound
on $p_{\text{ecc, pop}}$. We discuss a method to compute the dominant terms in
this integral in Appendix \ref{sec:lower_bound_p_ecc_pop}.  

Finally, we can incorporate selection effects into this
framework. Selection effects occur because we do not observe every
BBH in our prior. Low mass or high distance events may not have a high enough
signal-to-noise ratio (SNR) to be seen by ground-based detectors. This has two effects
on the likelihood in Eq.~(\ref{eq:fgmc_likelihood}), which are derived in Ref.
\cite{Mandel:2018mve}. 

We define $\alpha_\text{E}$ and $\alpha_{\text{QC}}$ as the fraction of
events in the universe which would be detected under 
the eccentric and quasi-circular hypotheses respectively. Explicitly, $\alpha_{E}$ can 
be computed as:  
\begin{equation}
    \label{eq:alpha_selection}
    \alpha_{\text{E}} = \int d\bm{\vartheta} \: p_{\text{E, det}}(\bm{\vartheta}) p(\bm{\vartheta} | \mathcal{H}_{E}).
\end{equation}
where $p_{\text{E, det}}(\bm{\vartheta})$ is the probability of detecting a GW event
with parameters $\bm{\vartheta}$ under the eccentric hypothesis. The computation for
$\alpha_{\text{QC}}$ is analogous except using $p_{\mathrm{QC, det}}(\bm{\vartheta})$ (the probability to detect an event
with parameters $\bm{\vartheta}$ under the quasi-circular hypothesis). We can compute $p_{\mathrm{QC, det}}$ and
$p_{\mathrm{E, det}}$ using a threshold SNR and error functions \cite{Farr:2013yna}.

In general, $p_{\mathrm{E, det}}$ and $p_{\mathrm{QC, det}}$ are different. This is because eccentric waveforms are
not usually used for GW searches (see however, Refs. \cite{Wang:2021qsu,Nitz:2019spj,LIGOScientific:2019dag,LIGOScientific:2023lpe}). To compute $p_{\mathrm{E,det}}$,
to a high accuracy one would need to perform a suite of synthetic eccentric
injections into the LVK detection pipelines and count the number of observed triggers \cite{LIGOScientific:2021psn}.
Alternatively, one could compute the overlap between the eccentric signal and a
quasi-circular waveform maximizing over the parameters of the system. This was done in 
Ref.~\cite{Zevin:2021rtf} for an eccentric non-spinning fiducial injection with $d_L=$1Gpc and $\mathcal{M}=28 \mathrm{M}_\odot$ using 
the waveform model from Refs. \cite{Nagar:2021gss,Nagar:2018zoe,Albanesi:2022xge}. 
The authors of Ref.~\cite{Zevin:2021rtf} find that the deviation between $p_{\mathrm{E, det}}$ and $p_{\mathrm{QC, det}}$ is less 
than $0.05$ at $e_{\mathrm{10Hz}}$ for this injection.  
However, we must excercise caution as $p_{\mathrm{E, det}}$ and $p_{\mathrm{QC, det}}$ will be different 
for different injection configurations. One could imagine GWs whose SNR is near the boundary of the 
threshold SNR, for detection causing larger differences between $p_{\mathrm{E, det}}$ and $p_{\mathrm{QC, det}}$. However, 
as a detailed computation of $p_{\mathrm{E, det}}$ has not been done, we use the approximation $p_{\mathrm{E, det}} = p_{\mathrm{QC, det}}$ 
for this paper and leave a more robust computation for future work. 

According to Ref.~\cite{Mandel:2018mve} to incorporate selection effects we replace $R_{\text{E}}$ with $\alpha_{\text{E}} R_{\text{E}}$ and $R_{\text{QC}}$ with $\alpha_{\text{QC}} R_{\text{QC}}$. 
We need to also replace $Z_{\text{E}}$ with $Z_{\text{E}} / \alpha_{\text{E}}$ and $Z_{\text{QC}}$ with $Z_{\text{QC}} / \alpha_{\text{QC}}$. This amounts to replacing 
$\exp(R_{\text{E}} + R_{\text{QC}})$ with $\exp(\alpha_{\text{E}} R_{\text{E}} +
\alpha_{\text{QC}} R_{\text{QC}})$ in Eqs.~(\ref{eq:fgmc_likelihood})--(\ref{eq:fgmc_flag_marginalized}).
Thus our posterior over the rates including selection effects is:
\begin{equation}
    \begin{aligned}
        \label{eq:fgmc_posterior_selection_effects}
        p(\{ g_i \}, & R_{\text{E}}, R_{\text{QC}}, \: | \: \{ d_i \}, N) \\
        \propto & \left[  \prod_{\{ i | g_i = 1 \}}^{N_\text{E}} R_{\text{E}} Z_{\text{E}}(d_i) \right] \left[ \prod_{\{ i | g_i = 0 \} }^{N_\text{QC}} R_{\text{QC}} Z_{\text{QC}}(d_i) \right]   \\
        & \times e^{-(\alpha_{\text{E}} R_\text{E} + \alpha_{\text{QC}} R_\text{QC})}
        p(R_{\text{E}}, R_{\text{QC}}).
    \end{aligned}
\end{equation}
This can then be propagated into Eqs.~(\ref{eq:fgmc_flag_marginalized}) and (\ref{eq:p_ecc}).

\subsection{\label{sec:DINGO} \texttt{DINGO}}

We now summarize the main features of the machine-learning code \texttt{DINGO}
~\cite{dax2021group, Dax:2021tsq, Green:2020dnx}.
In particular, its use of 
normalizing flows\cite{Kobyzev:2019ydm, Papamakarios:2019fms} and 
importance sampling \cite{Dax:2022pxd,kong19}.

In many practical applications, obtaining $p(\bm{\vartheta} | d)$ in Eq.~(\ref{eq:Bayes_theorem}) 
is analytically impossible. But one can use techniques such as nested sampling
\cite{Speagle:2019ivv} or Markov-Chain Monte-Carlo (MCMC)
\cite{brooks2011handbook} to sample from $p(\bm{\vartheta} | d)$. The LVK Collaboration has developed
tools such as \texttt{LALInference}~\cite{Veitch:2014wba} and \texttt{Bilby} \cite{Ashton:2018jfp,Romero-Shaw:2020owr,Ashton:2021anp} 
for this task. However, these methods must evaluate
the likelihood $\mathcal{O} (10^{7-8})$ times per event. Thus, for time-domain waveform models 
produced by solving ordinary differential equations parameter estimation 
can be expensive. For example,
\texttt{SEOBNRv4EHM} takes $\mathcal{O} (100-700$ms) per
likelihood evaluation meaning inference can take $\mathcal{O}(\text{week})$ per event even parallelizing over 320 cores \cite{Ramos-Buades:2023yhy}.
It is then very computationally expensive to analyze the catalog of GW events with such waveform model.

We can instead use likelihood-free approaches to amortize the inference
\cite{Dax:2021tsq,Wong:2023lgb,Gabbard:2019rde}. Here we use \texttt{DINGO} \cite{Dax:2021tsq}, 
which has been shown to achieve results with the same 
accuracy as standard samplers \cite{Dax:2022pxd}. We pay the upfront cost of
training a neural network for $\mathcal{O}(\text{week})$, but then we can do
inference on any events within the trained priors in $\mathcal{O}(\text{hour})$.

\texttt{DINGO} learns a mapping $f_{d, S_n}: u \rightarrow \bm{\vartheta}$ 
from a simple-base distribution $p(u) = \mathcal{N}(0, 1)^D$ to the complex
target GW posterior $p(\bm{\vartheta} | d)$ (it is implied this is also conditioned on $\mathcal{H}_a$). This mapping is learned through a series
of composable functions parameterized by neural networks (a normalizing flow).
For each data sample $d$, the forward model is given by 
\begin{equation}
    \begin{aligned}
        \label{eq:normalizing_flow}
        q(\bm{\vartheta} | d, S_n)  &= [f_{d, S_n}]_{*} \mathcal{N}(0, 1)^D   
        \\[2ex]
        \quad \quad &= \mathcal{N}(0, 1)^D \left( f_{d, S_n}^{-1} (\bm{\vartheta}) \right) \; \left| \frac{\partial f_{d, S_n}^{-1}}{\partial \bm{\vartheta}} \right|,
    \end{aligned}
\end{equation}
where $D$ is the number of GW parameters and $(\cdot)_*$
denotes the push-forward operator. The $d$ and $S_n$ on the subscript of $f_{d, S_n}$ indicate that the mapping uses $d$ and $S_n$ as context to the 
neural network. The Jacobian in the last term of Eq.~ (\ref{eq:normalizing_flow}) is a
normalization factor to account for the fact that in general $f_{d, S_n}$ is not a volume preserving operation \cite{Papamakarios:2019fms}.

\texttt{DINGO} is unique from other flow approaches as the inference
is aided by conditioning the networks on the arrival times of GWs in the
interferometers. We train two separate networks: one to learn the distribution
$q(t_i | d)$ and one to learn the distribution $q(\bm{\vartheta} | d, S_n, \hat{t_i})$.
Here, $t_i$ are the arrival times of the GW in each detector and can be computed 
from $\bm{\vartheta}$. Then $\hat{t_i}$ are
``blurred'' versions of the arrival times. That is, $p(\hat{t}_i
| t_i) = \text{Unif(} t_i - 1 \text{ms}, t_i + 1 \text{ms})$. The process is to start 
with an initial guess for $t_i \sim q(t_i | d)$. Then using $t_i$, 
sample $\hat{t}_i \sim p(\hat{t}_i | t_i)$ and get an initial guess for $\bm{\vartheta}
\sim q(\bm{\vartheta} | d, S_n, \hat{t_i})$. After obtaining initial guesses, we can Gibbs sample
\cite{casella1992explaining} the distributions $p(\hat{t}_i | t_i)$ and
$q(\bm{\vartheta} | d, S_n, \hat{t_i})$ to obtain $q(\bm{\vartheta}, \hat{t}_i | d, S_n)$. Finally, we 
can drop $\hat{t}_i$ from $q(\bm{\vartheta}, \hat{t}_i | d, S_n, )$ to obtain $q(\bm{\vartheta}, | d, S_n)$. The
advantage in this process is that the network learning $q(\bm{\vartheta} | d, S_n,
\hat{t_i})$ never sees a GW with the merger greater than 1ms away from the start
of the data segment. To summarize, by conditioning on the blurred arrival times, we
standardize the data, allowing for smoother training.

\texttt{DINGO} also uses an embedding network to compress the high dimensional Fourier domain strain data into a feature
vector of length 128 for each detector. This embedding network is trained as
part of the mapping $f_{d, S_n}$.

After performing inference with the network, we need to have a guarantee that the distribution $q (\bm{\vartheta} | d, S_n)$
produced by \texttt{DINGO}, is a good approximation to the true posterior
distribution, $p(\bm{\vartheta} | d, S_n)$. To achieve this, we can importance sample the generated
proposal distribution to obtain samples from
the true distribution \cite{kong19}. Explicitly, we first sample $\bm{\vartheta}_i \sim q(\bm{\vartheta} |
d, S_n)$ and then weight each sample according to 

\begin{equation}
    \label{eq:importance_weights}
    w(\bm{\vartheta}_i) \propto \frac{p(d|\bm{\vartheta}_i, S_n) p(\bm{\vartheta}_i)}{  q(\bm{\vartheta} | d, S_n)}.
\end{equation}
Here we only need to evaluate the likelihood for as many points
as there are in the proposal posterior, which is orders of magnitude lower than a Nested
Sampling or MCMC run. Since we only need to compute the likelihood during importance sampling, 
this can be parallelized across an arbitrary number of cores. In
addition, since \texttt{DINGO} is trained on a forward Kullback-Leibler (KL) divergence
objective, it tends to cover regions of high density. This is preferred in
scientific domains, as we want to ensure a conservative 
as possible estimate before importance sampling. 

The \textit{effective sample size}, $n_{\text{eff}}$, can be computed from the importance weights using:
\begin{equation}
    \label{eq:ess}
    n_{\text{eff}} = \frac{(\sum_i w_i)^2}{\sum_i w_i^2}.
\end{equation}
This represents the number of samples drawn from the true posterior distribution
$p(\bm{\vartheta} | d)$ \cite{kong19}. Thus, any test statistic computed with the weighted samples 
is equivalent to using $n_{\text{eff}}$ samples from the true posterior. The importance weights can also be used to compute the
evidence, $p(d)$, (see Eqs.~ (2)---(8) in
Ref. \cite{Dax:2022pxd}). When reporting results with importance sampling, we label the result \texttt{DINGO-IS}.

Finally, in this paper we often compute the Jensen-Shannon divergence (JSD) between posterior distributions as a
comparison statistic. For example, this is used to compare \texttt{DINGO-IS} with \texttt{pBilby}.
The JSD is defined as a symmetrized version of the KL divergence that is between 0 and 1 bits. It is defined as a divergence between two
distributions $p(\bm{\vartheta})$ and $q(\bm{\vartheta})$:

\begin{equation}
    \label{eq:JSD}
    D_{\text{JS}} = \frac{D_{\text{KL}}(p | q)  + D_{\text{KL}}(q | p)}{2},
\end{equation}

where the KL divergence is

\begin{equation}
    \label{eq:KL}
    D_{\text{KL}}(p | q) = \int d \bm{\vartheta} \: p(\bm{\vartheta}) \log_2 \left(\frac{p(\bm{\vartheta})}{q(\bm{\vartheta})} \right).
\end{equation}
The KL divergence expresses the information loss we would accumulate if using distribution
$p$ to approximate a test statistic when $q$ was the true underlying distribution. 

\section{\label{sec:methods_waveform_models} Eccentric and quasi-circular waveform models}

In this paper we use mostly multipolar eccentric and quasi-circular aligned-spin waveform models, but 
for some applications also quasi-circular spin-precessing models. Several non-spinning and 
aligned-spin, eccentric inspiral-merger-ringdown models have been proposed in the 
literature~\cite{Huerta:2017kez,Hinder:2017sxy,Cao:2017ndf,Nagar:2021gss, Setyawati:2021gom,Nagar:2022fep,Islam:2021mha,Liu:2023dgl,Liu:2023ldr, Chiaramello:2020ehz}.
During the referee period of this paper even more more eccentric waveform models 
were developed \cite{Gamboa:2024hli, Planas:2025feq, Islam:2024rhm, Islam:2024bza, Islam:2024zqo, Islam:2025llx,
Paul:2024ujx, Morras:2025nlp}. Here, we employ the (time-domain) eccentric aligned-spin model \texttt{SEOBNRv4EHM}, developed in
Refs.~\cite{Khalil:2021txt,Ramos-Buades:2021adz}, the 
quasi-circular aligned-spin model \texttt{SEOBNRv4HM}
\cite{Bohe:2016gbl,Cotesta:2018fcv}, and the quasi-circular spin-precessing
model \texttt{SEOBNRv4PHM}~\cite{Buonanno:2005xu,Pan:2013rra,Babak:2016tgq,Cotesta:2018fcv,Ossokine:2020kjp}.  
For some studies, we also use the quasi-circular spin-precessing NR surrogate model \texttt{NRSur7dq4}
\cite{Varma:2019csw}.

The (EOB) formalism~\cite{Buonanno:1998gg,Buonanno:2000ef,Damour:2000we,Damour:2001tu,Buonanno:2005xu} 
maps the two-body dynamics onto the one of a test body in a deformed Kerr metric, the deformation 
being the symmetric mass ratio $\nu$. It relies on three key ingredients: (i) a Hamiltonian that describes 
the conservative dynamics, (ii) a radiation-reaction force, and (iii) the gravitational modes. These three 
components are built by resumming PN calculations. Furthermore, the EOB approach 
also provides the full inspiral-merger-ringdown waveforms using physically motivated 
ansatze for the merger, and results from BH perturbation theory for the ringdown. Finally, EOB 
waveforms are made highly accurate through calibration to NR simulations~\cite{Ossokine:2020kjp,Akcay:2020qrj,Gamba:2021ydi,Pompili:2023tna,Ramos-Buades:2023ehm}, and more recently 
from the gravitational self-force approach~\cite{vandeMeent:2023ols}.

To fully characterize GWs from eccentric BBHs, one needs to include two
additional parameters compared to the quasi-circular case. The 
\texttt{SEOBNRv4EHM}~\cite{Khalil:2021txt,Ramos-Buades:2021adz,Ramos-Buades:2023yhy} model 
adds two new parameters to the quasi-circular spin-aligned model,
\texttt{SEOBNRv4HM} \cite{Bohe:2016gbl,Cotesta:2018fcv}: 
the initial orbital eccentricity $e$ and relativistic anomaly
$\zeta$.  The model employs the quasi-Keplerian parametrization to express the
relationship between the radial separation $r$ and these parameters~\cite{Ramos-Buades:2021adz,Ramos-Buades:2023yhy}
\begin{equation}
    \label{eq:Keplerian_paramtrization}
    r = \frac{p}{1 + e \cos \zeta}, 
\end{equation}
where $p$ is the semi-latus rectum. In order to integrate the EOB Hamilton 
equations, one needs to specify initial conditions 
(see Eqs.~(12)--(14) in Ref.~\cite{Ramos-Buades:2023yhy}).  In 
\texttt{SEOBNRv4EHM} the eccentricity and relativistic anomaly are part of these
initial conditions, and they are specified at an orbit-averaged frequency. We note that 
one could specify the initial conditions at an instantaneous 
reference frequency. However, this strongly affects the time to merger and
creates a rapidly varying log-likelihood surface, which can cause difficulties for
parameter-estimation pipelines, as pointed out in Ref.~\cite{Ramos-Buades:2023yhy}.
In addition, unlike the instantaneous frequency, the orbit-averaged GW
frequency is related to twice the orbit-averaged orbital frequency
\cite{Ramos-Buades:2022lgf} for eccentric orbits. Thus we can use it to uniquely
specify the point along the orbit where we infer the eccentricity and
relativistic anomaly. We denote the quasi-Keplerian eccentricity and
relativistic anomaly measured at orbit-averaged GW frequency of 10Hz in the detector frame as $e_{\mathrm{10Hz}}$ and $\zeta_{\mathrm{10Hz}}$.  

The dynamics of the BBH is used to construct the waveform modes, $h_{\ell m}$.
The \texttt{SEOBNRv4EHM} model contains the $(\ell, |m|) = (2,2), (2,1), (3,3), (4,4), (5,5)$
multipoles. The inspiral waveform modes contain 2PN eccentric
corrections derived in Ref.~\cite{Khalil:2021txt}, and are enhanced during 
the late inspiral and plunge with non-quasi-circular terms that are fitted 
to quasi-circular NR simulations (see Eqs.~(7)--(10) in 
Ref.~\cite{Ramos-Buades:2021adz}). Note that in this model the radiation-reaction force
does \textit{not} have eccentric corrections, and the merger-ringdown modes 
are assumed to be quasi-circular~\cite{Cotesta:2018fcv}. This latter assumption is 
used by all the current eccentric waveform models, and it is supposed to hold 
for mild eccentricities \cite{hinder2008circularization,Huerta:2019oxn,Ramos-Buades:2019uvh}. 

The waveform modes can be used to construct the gravitational polarizations: 
\begin{equation}
    \label{eq:mode_decomposition}
    h_+ - i h_\times = \sum_{l=2}^{\infty}  \sum_{m=-l}^{m=l} {}_{-2} Y_{lm} (\varphi, \iota) h_{lm} (m_{1,2}, \chi_{1, 2}, e, \zeta; t),
\end{equation}
where for aligned-spins $(\varphi, \iota)$ are the azimuthal and polar angles to the observer 
in the source frame. Without loss of generality, $\varphi = \phi_{\mathrm{coal}}$ \cite{Schmidt:2017btt}. 

We remark that the \texttt{SEOBNRv4EHM} model has been tested against (mainly) non-spinning 
NR simulations up to $e_{20 \text{Hz}} = 0.3$~\cite{Ramos-Buades:2022lgf}, finding good agreement. 
Since we are generating waveforms with a starting frequency of 10Hz, we use \cite{peters1964gravitational} to 
map the maximum allowed eccentricity to $e_{\text{10Hz}} \leq 0.5$. Thus, we cannot
say anything about the accuracy of the waveform model for events that 
have eccentricities greater than $e_{\text{10Hz}} > 0.5$. We stress also that 
\texttt{SEOBNRv4EHM} does not simultaneously model spin-precession
and eccentricity\footnote{This is also true of other inspiral-merger-ringdown waveform
models in the literature with the exception of Ref.~\cite{Liu2023spinprececc}.} and
thus we cannot address the possibility that an event is both eccentric and
precessing.
  
Regarding the quasi-circular spin-precessing waveform model \texttt{SEOBNRv4PHM}
~\cite{Cotesta:2018fcv, Ossokine:2020kjp}, it integrates the time-domain EOB dynamics in
a co-precessing frame using the EOB spin-precession equations and a Hamiltonian
calibrated to aligned-spin NR simulations. In such a frame, it contains 
the $(\ell, |m|) = (2,2), (2,1), (3,3), (4,4), (5,5)$ multipoles. Then, the
waveform in the co-precessing frame is rotated to the inertial 
frame~\cite{Pan:2013rra,Babak:2016tgq}. In contrast, \texttt{NRSur7dq4} directly 
interpolates 1528 precessing NR waveforms 
with mass ratios $q \geq 1/4$ with $\chi_{1,2} \leq 0.8$ \cite{Varma:2018aht}. The model can also be used in the extrapolation 
region with $q \geq 1/6$ and $\chi_{1,2} \leq 1$. This model is typically used for heavier mass 
systems in parameter estimation studies as it is restricted by the length 
of the training waveforms. This model includes all $\ell \leq 4$ modes. 

Before we end the section, we briefly comment on the eccentricity parameter. 
Since eccentricity is not uniquely defined in GR there are multiple
ways to parametrize the orbit \cite{Blanchet:2013haa}. Waveform models
may use orbital parameters, compact object trajectories, energy and angular momentum 
or other parameterizations to define eccentricity \cite{Ramos-Buades:2018azo, Ciarfella:2022hfy,
Ramos-Buades:2021adz, Nagar:2021gss, Hinderer:2017jcs, Chiaramello:2020ehz,
Khalil:2021txt, Warburton:2011fk, Osburn:2015duj, VanDeMeent:2018cgn,
Chua:2020stf, Hughes:2021exa, katz2021fast, Lynch:2021ogr,
Buonanno:2010yk}.  

It is, however, possible to use a definition of eccentricity that 
can be extracted from the waveform. This was proposed in Refs.~\cite{Shaikh:2023ypz,
Ramos-Buades:2022lgf} (see also Ref.~\cite{Islam:2025oiv}), where this new eccentricity is denoted 
$e_{\text{gw}}$. This $e_{\text{gw}}$ is extracted by interpolating the
instantaneous GW frequency of the (2,2) mode, $\omega_{22}$, along
the pericenter and apocenter points. This is calculated as 
\begin{equation}
    \label{eq:psi_eccentricity}
    e_{\text{gw}} = \cos(\psi / 3) - \sqrt{3} \sin(\psi / 3),
\end{equation}
with 
\begin{equation}
    \label{eq:e_omega_22}
    \psi = \arctan \left( \frac{1 - e^2_{\omega_{22}}}{2 e_{\omega_{22}}} \right),
\end{equation}

and 
\begin{equation}
    \label{eq:e_omega_22}
    e_{\omega_{22}} = \frac{\omega_{22, p}^{1/2} - \omega_{22, a}^{1/2}}{\omega_{22, p}^{1/2} + \omega_{22, a}^{1/2}},
\end{equation}
where $\omega_{22, p}$ and $\omega_{22, a}$ are the GW frequency of the 22 mode at periastron 
and apocenter. 

We can similarly interpolate the GW mean anomaly
\begin{equation}
    \label{eq:mean_anomaly_times}
    l_{\text{gw}} = 2 \pi \frac{t - t_i^p}{t_{i+1}^p - t_{i}^p}.
\end{equation}
Here $t$ is the time at which we are measuring the mean anomaly and $t_i^p$ is the
time of the $i$th periastron passage measured using the (2,2) mode frequency.

We compute $e_{\text{gw}}$ using parallelized version of the \texttt{gw\_eccentricity}
package \cite{Ramos-Buades:2023yhy,Shaikh:2023ypz} \footnote{We use the ``Amplitude'' method 
in \texttt{gw\_eccentricity} and set \texttt{num\_orbits\_to\_exclude\_before\_merger = 1}} . We choose to infer the
eccentricity at a dimensionless frequency of $M_{\text{det}} f_{\text{ref}} $, where
$M_{\text{det}}$ is the detector frame total mass of the system and
$f_{\text{ref}} = \text{10 Hz}$. The reason to multiply by the total mass is
because we do not want our reported eccentricity to depend on the redshift of
the source. When reporting the GW eccentricity at 10Hz we denote this
as $e_{\mathrm{gw, 10Hz}}$. 

\section{\label{sec:event_results} Zero-noise synthetic-data  and real events}

\subsection{Trained networks and priors} 
\label{sec:trained_networks}

We train twenty-five \texttt{DINGO} networks on 80GB NVIDIA-A100 GPUs each for 11 days.
We train eleven networks for the eccentric aligned-spin case, eleven networks
for quasi-circular aligned-spin case and four networks for the quasi-circular precessing-spin case. 
The number of networks corresponds to the fact that one has
to train a separate network for each observing run (noise curve), detector
configuration, waveform model, choice of gravitational modes, mass prior and distance
prior. The mass and distance prior range is optional, but we find that training two separate
networks for the regimes $d_{\text{L}} < 3$ Gpc and $d_{\text{L}} < 6$ Gpc leads
to increased sample efficiency. Similarly, if we train a network with a smaller mass 
prior we achieve higher sample efficiency in the prior. But we also want to analyze heavy BBHs so we
train a separate network increasing the upper bound of the mass prior. Since
there are some GW events that only occur in one detector in the third-observing run, we also train
networks for LIGO-Hanford and LIGO-Livingston only. This is also useful for doing single-detector 
analyses of events with glitches. We train the networks in the range
$(f_{\text{min}}, f_{\text{max}}) = (20 \text{Hz}, 1024 \text{Hz})$. To
condition the data we use a Tukey window with a 0.4 second roll-off. We train the 
networks using an 8 second data stream, which implies a $\Delta f = 0.125$. We whiten
the signals by sampling over 4152 Welch PSDs estimated from 128 seconds of data obtained from
the GW Open Science Center (GWOSC) using the
\texttt{DCS-CALIB\_STRAIN\_CLEAN\_C01} channel. The networks have been trained assuming
Gaussian detector noise. The networks and their settings
have been made publicly available on Zenodo at \cite{zenodogwnetworks}. 

While we do train a large number of networks, it is important to realize
that the inference is amortized. This means the computational speed-up compared
to traditional samplers will scale as the number of events increases. For
example, in O1 where we only analyze only one GW event, there is no speed-up
compared to a traditional sampler. However, in O3 where we analyze 51 events,
the speed-up is on the order of 6 months. Another subtle point is that in GW
science, we rarely perform parameter estimation for each event once. We may want
to change the prior, $f_{\text{min}}$, $f_{\text{max}}$ or in the case of a
traditional sampler boost the sampler settings. We also may want to study
maximum likelihood injections for consistency. For example, in this study, we
obtained over 300 posteriors across different waveform models before generating
the final catalogue of 57 events. Thus, while we train a large number of
networks, the speed-up is still significant and scales well with a large number of
GW events.

We now discuss the subset of events we do not analyze. 
We do not analyze events with a chirp mass below 15 $M_\odot$. This includes neutron-star events
(although this is being addressed in an upcoming work \cite{dax2024bns}). This is
because low-mass systems have rapid Fourier-domain oscillations at low frequencies. 
As a consequence, if we want to represent the data, we need to use a much larger embedding network,
which leads to significant computational cost. As a low chirp mass corresponds to 
a longer signal, this constraint implies we do not analyze events longer than 8 seconds. 

We also do not report on events for which we were unable to obtain 5,000
effective samples. Note that we have purposely set this to a very high threshold
in order to be indistinguishable from other samplers like \texttt{pBilby}, and
have a smooth posterior.  We see that reducing this threshold leads to spikes in
the posterior density, which can lead to JSDs with \texttt{pBilby} larger than
0.002 bits (the expected stochastic deviation from identical nested sampling
runs \cite{Romero-Shaw:2020owr}).

We set an isotropic in component spin prior with $|\chi_{1,2}| < 0.9$. We also set
a uniform prior in $\zeta_{\text{10Hz}} \in [0, 2 \pi]$, a uniform prior in
$e_{\text{10Hz}} \in [0, 0.5]$ and a uniform in component mass prior on the mass
ratio $q \in [0.125, 1]$. The upper bound in the $e_{\text{10Hz}}$ is set by 
the restriction of the waveform model (see Sec.~\ref{sec:methods_waveform_models}).
Other than this, the priors are the same as in Ref.~\cite{Dax:2022pxd}.

We use a uniform prior on eccentricity as opposed to a log-uniform prior 
informed by astrophysics. We instead incorporate the astrophysics information by 
folding it into our computation of the prior odds (see Eq. (\ref{eq:posterior_odds})) or 
$p_{\text{ecc}}$ (see Eq. (\ref{eq:p_ecc})). 
Having a uniform prior is also useful because if we perform a population study in the future, we can importance sample to a
multitude of different priors \cite{LIGOScientific:2021psn}. In contrast, if we
use a light-tailed log-uniform prior, we could not easily convert to a different
prior. This is because there would be many high-weight points  at high
eccentricities leading to low sample efficiency. A separate reason to adopt a uniform prior
instead of an astrophysical log-uniform one is so that we do not have to
decide where to place the lower bound of a log-uniform prior. Changing
this lower bound can strongly affect Bayes factors as it gives
high weight to low eccentricities in the evidence integral. 

However ultimately, a feature of a good Bayesian analysis is using multiple reasonable
priors. For completeness, we also compute Bayes factors with a log-uniform prior on
$e_{10\text{Hz}} \in (10^{-4}, 0.5)$ for GW200129, GW190701 and GW200208\_22. We 
display the results in Table~\ref{tab:Bayes_table} and in
Fig.~\ref{fig:bf_lineplot}. We note that because these events are likelihood dominated, there is very little
shift in the posterior, but the Bayes factors for eccentricity do decrease. 

When performing the analysis with the quasi-circular spin-precessing \texttt{SEOBNRv4PHM} and \texttt{NRSur7dq4}
models, we utilize the same priors as in the previous paragraph except for the dimensionless 
spin magnitudes. As stated above, for the eccentric model we restrict the dimensionless spin magnitudes to 
$| \chi_{1,2} | < 0.9$. 
However, in the case of GW200129 with
\texttt{NRSur7dq4}, there is support for $a_{1,2} > 0.9$. Thus we perform the 
spin-precessing analysis with and without the spin-prior restriction and take 
the higher evidence. This is to avoid artificially lowering the evidence for precession \cite{Hannam:2021pit}.
For the spin angles we use the same priors as in Ref.~\cite{Dax:2022pxd}.

\subsection{Zero-noise synthetic injection}
\label{sec:zero_noise}

To validate the use of \texttt{DINGO} with eccentric waveform models, we first
perform  two synthetic-signal (injection) in zero-noise using two independent
samplers, \texttt{pBilby} and \texttt{DINGO}.  Due to the expensive nature of
doing \texttt{SEOBNRv4EHM} runs with \texttt{pBilby}, we use only the $(\ell,
|m|) = (2, 2)$ modes in the injection and recovery. 

We need to make some modifications to allow \texttt{DINGO} to work with zero noise. \texttt{DINGO}
is trained on Gaussian detector noise. Thus we cannot simply inject a zero-noise
signal into the data, as it will be flagged as out of distribution, and give a low
effective sample size. Instead, we first run \texttt{DINGO} on 100 Gaussian noise
realizations while fixing the injected parameters. We then pool the results. 
The idea is that when we average the \texttt{DINGO} posterior over infinite noise realizations, this should
be close (though generally broader) than the true zero-noise posterior. 

However, with this method we no longer have an estimate for the probability density of each sample as
there is no guarantee the set of pooled samples are drawn from a normalized
distribution. Thus, we train an unconditional density estimator on the set of
pooled samples in order to recover the density of $q_{\text{ZN}}(\bm{\vartheta} | h(f))$. We use 
one million pooled samples as a training dataset for the unconditional density estimator. 

Now that we have the density of each sample, we importance sample the set
of pooled samples, but using a zero noise likelihood, $p_{\text{ZN}}(h | \bm{\vartheta})$, when 
computing the importance weights. Thus, we have an exact reconstruction of the
zero-noise posterior distribution.

\begin{figure}
\centering

\captionsetup[subfigure]{labelformat=empty}
\subfloat[]{\includegraphics[width=0.95\linewidth]{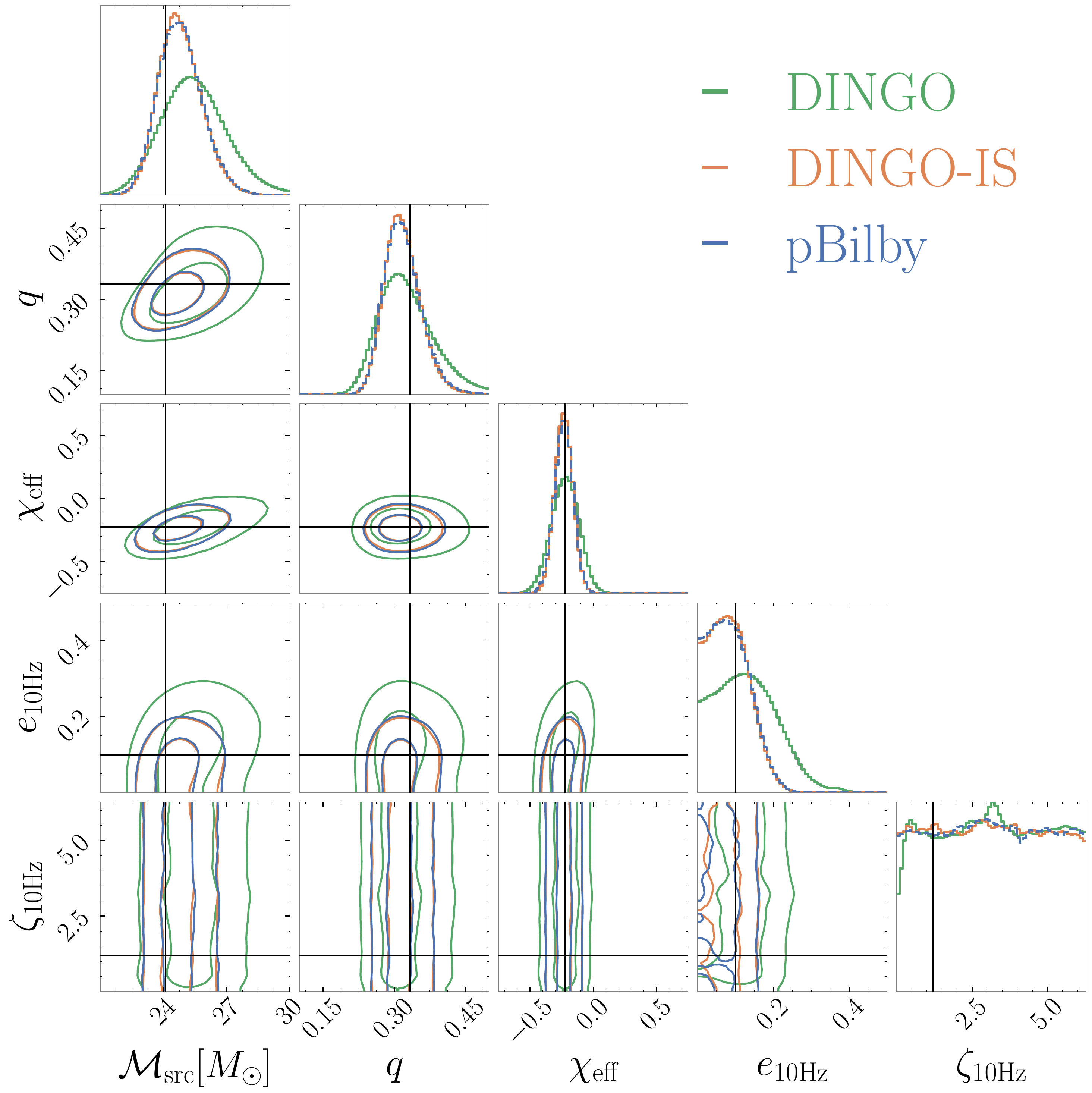}\label{fig:zero_noise_1}}

\subfloat[]{\includegraphics[width=0.95\linewidth]{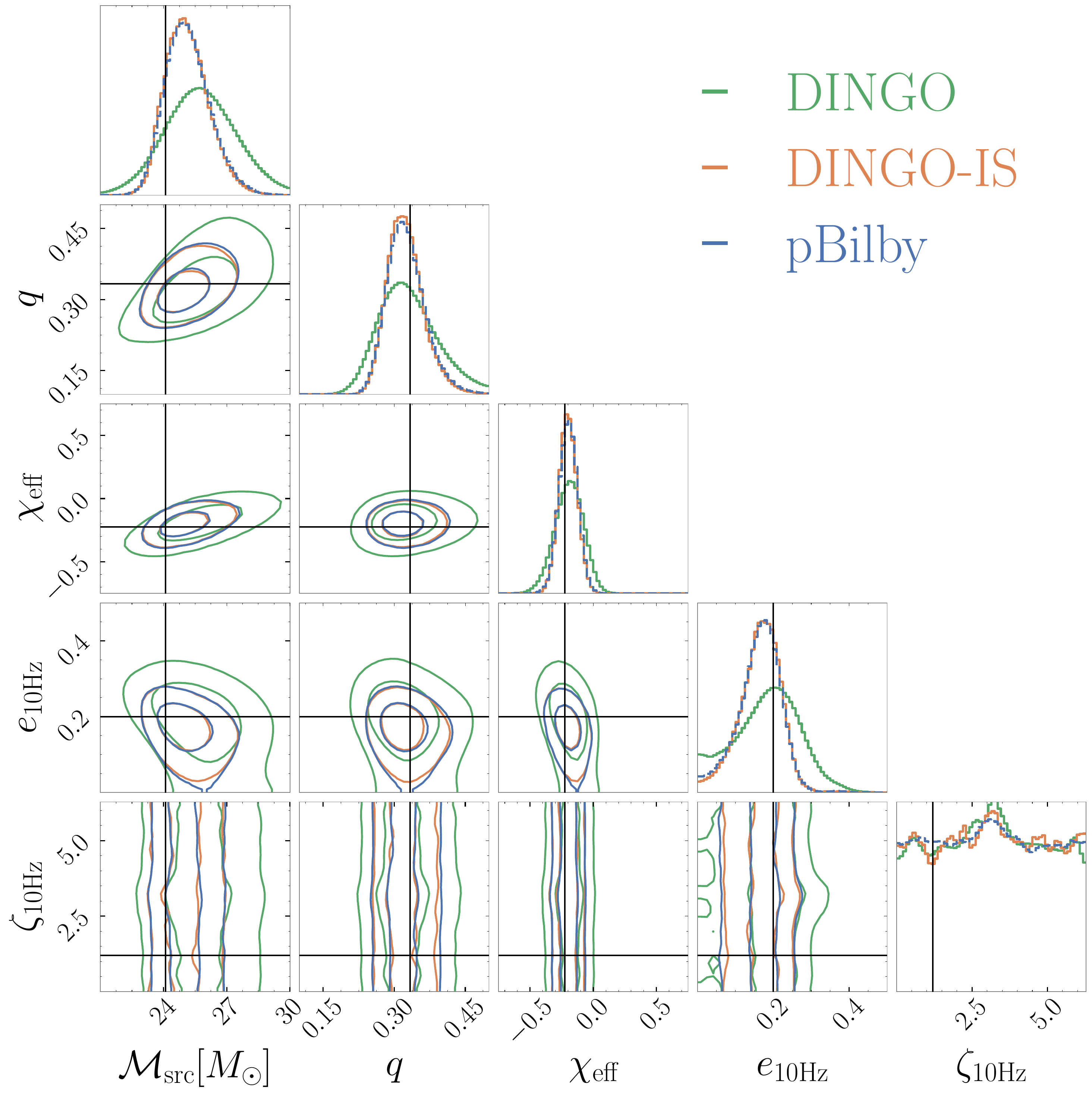}\label{fig:zero_noise_2}}

\caption{\label{fig:zero_noise} Zero-noise injections of \texttt{SEOBNRv4EHM} with parameters $\chi_{\text{eff}} = -0.23$,
$\mathcal{M}_{\text{det}} = 28 M_\odot$, $q=1/3$ and initial eccentricities $e_{10\text{Hz}}=0.1$ and $e_{10\text{Hz}}=0.2$ for the top and bottom figures, respectively. The injections are recovered with both \texttt{DINGO}, \texttt{DINGO-IS} (\texttt{DINGO} with importance sampling) and \texttt{pBilby}.}

\end{figure}

\begin{figure*}
\begin{center}
    \includegraphics[width=1\textwidth]{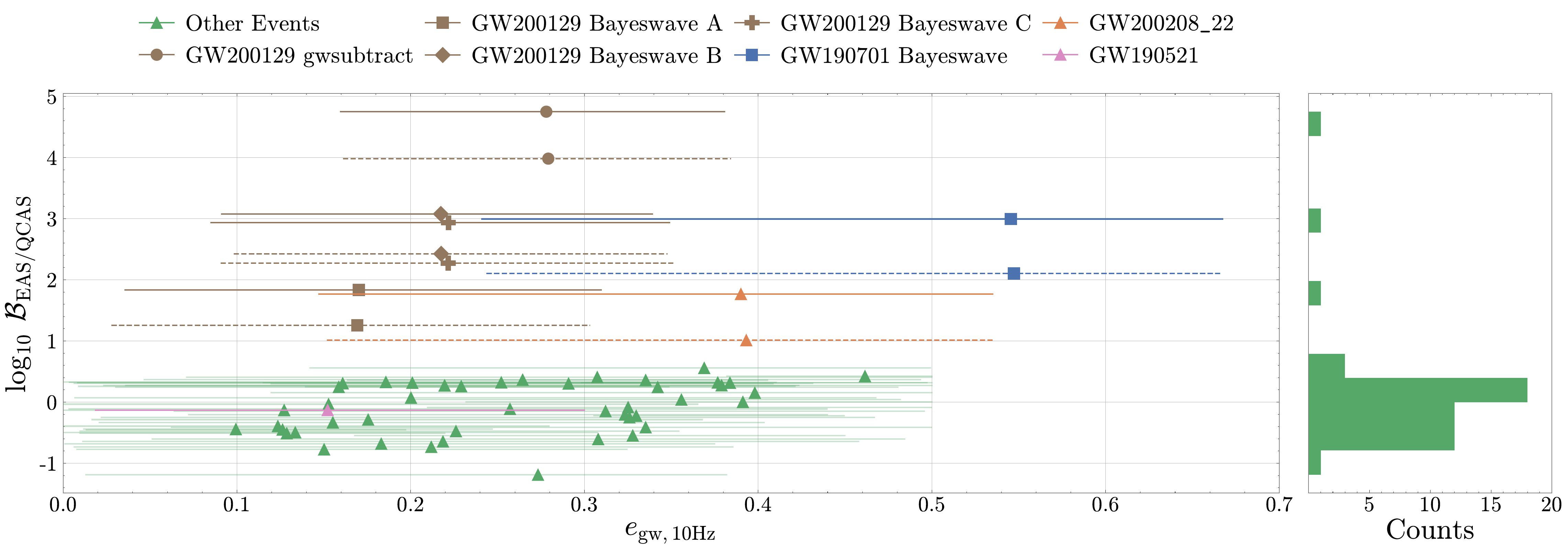}
\end{center}
\caption{\label{fig:bf_lineplot} Shown on the left are the $\log_{10} \mathcal{B}_{\text{EAS/QCAS}}$ and 90\%
highest-density intervals on $e_{\text{gw, 10Hz}}$ for the 57
events analyzed. Many events have $\log_{10}
\mathcal{B}_{\text{EAS/QCAS}} < 1$ and we color these green and reduce
their opacity. However, there are 3 events GW200129, GW190701 and GW200208\_22 which have $\log_{10}
\mathcal{B}_{\text{EAS/QCAS}} > 1$ which are labelled in the legend. 
We compute $e_{\text{gw, 10Hz}}$ only for these three events since for events with 
low support for eccentricity, $e_{\text{gw, 10Hz}} \approx
e_{\text{10Hz}}$ \cite{Shaikh:2023ypz}. 
We also color in pink GW190521 which has shown signs of eccentricity in previous
papers, but not in our analysis and the one in Ref.~\cite{Ramos-Buades:2023yhy}. 
For GW190521, we report the eccentricity at $5.5$Hz.
The different symbol shapes indicate variations on the
glitch mitigation algorithm employed in the analysis (see text). The solid lines
indicate a uniform prior on $e_{\text{10Hz}}$ from 
$[0, 0.5]$ while the dashed lines indicate a log uniform prior on
$e_{\text{10Hz}}$ from $[10^{-4}, 0.5]$. Shown on the right is a histogram of $\log_{10}$ Bayes
factors.} 
\end{figure*}

We present the analysis of two injections in Fig.~\ref{fig:zero_noise} (note
the result which should be used for comparison is the orange curve labeled
\texttt{DINGO-IS}). The injection parameters are $\chi_{\text{eff}} = -0.23$,
$\mathcal{M}_{\text{det}} = 28 M_\odot$, $q=1/3$ and $e=0.1$ or $e=0.2$. The JSD
between the 1D marginals of \texttt{pBilby} and \texttt{DINGO-IS} are all less
than 0.002 bits (the expected stochastic variation from GW sampling algorithms
\cite{Romero-Shaw:2020owr}). We note that while \texttt{DINGO} and
\texttt{pBilby} have a disagreement, after importance sampling this disagreement
disappears. We can also see that the proposal density generated by \texttt{DINGO} is
broader than the true posterior, a desirable property for importance
sampling. These two facts give us confidence that \texttt{DINGO-IS} and
\texttt{pBilby} have comparable results. We comment that for both samplers, there is some
bias in the recovered chirp mass and eccentricity, which is reminiscent of the behavior in
Ref.~\cite{Favata:2021vhw}. 

In order to further verify the accuracy of DINGO-IS with eccentric signals, we also perform a 
injection with simulated Gaussian noise and compare to \texttt{pBilby}. This 
test is shown in Appendix~\ref{sec:gaussian_noise}.

\subsection{GW200129}
\label{sec:GW200129}

\begin{figure}
\begin{center}
    \includegraphics[width=0.5\textwidth]{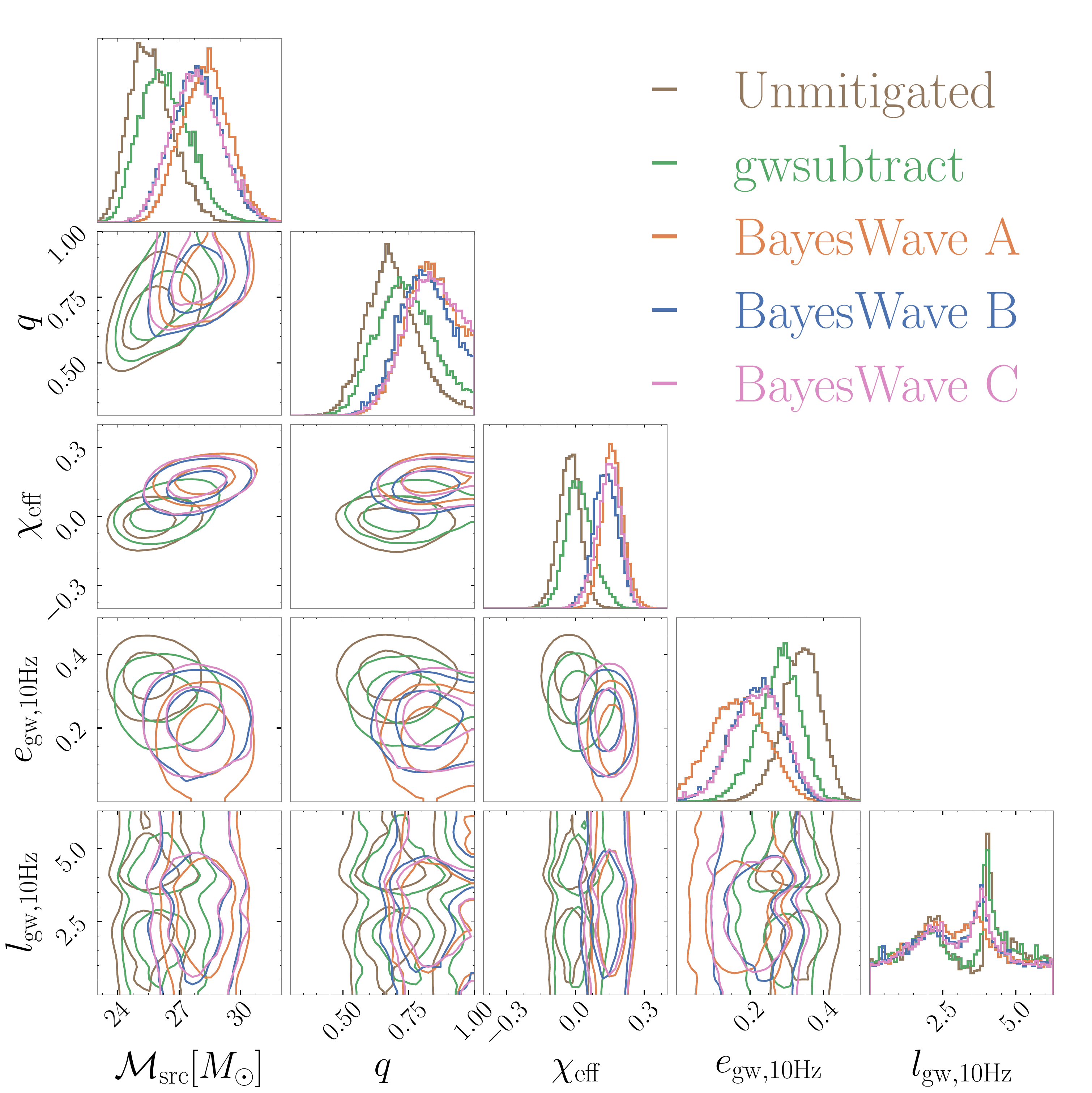}
\end{center}
\caption{\label{fig:glitch_subtraction} 
Posteriors of GW200129 using different glitch mitigation techniques. In green is the analysis
with \texttt{gwsubtract}, which uses auxillary detector channels and transfer functions to
subtract glitches. Shown in orange, blue and pink is the analysis carried
out using \texttt{BayesWave} de-glitching. This method involves generating a posterior of
glitches using and then
drawing individual glitches from this ``glitch posterior''. The A, B, and C
indicate different fair draws from the glitch distribution and
are publicly available at \cite{zenodogw200129}.
}
\end{figure}

There has been considerable interest around GW200129
due to the fact that it has shown signs of orbital precession \cite{Hannam:2021pit},
evidence for a measurable kick velocity \cite{Varma:2022pld}, false
violations of GR due to waveform systematics \cite{Maggio:2022hre}, 
and a glitch \cite{Davis:2022ird, Payne:2022spz}. Therefore, to
analyze this event we follow the treatment of Refs.~\cite{Davis:2022ird, Payne:2022spz,Hannam:2021pit}. Namely,
we incorporate multiple glitch variations
and analyze this event also for precession using both \texttt{SEOBNRv4PHM}
and \texttt{NRSur7dq4}.

When analyzed with \texttt{SEOBNRv4EHM}, GW200129 is an intermediate-mass BBH consistent
with zero effective spin.  In particular, We find $M_{\text{det}} =
78.3_{-3.1}^{+4.2}$, $M_{\text{src}} = 69.4_{-3.1}^{+4.2}$ and
$\chi_{\text{eff}} = 0.02_{-0.11}^{+0.12}$. We infer $e_{\text{gw, 10Hz}} = 0.27_{-0.12}^{+0.10}$, $e_{\text{gw, 20Hz}} = 0.16_{-0.05}^{+0.04}$ and $\log_{10} \mathcal{B}_{\text{EAS/QCAS}}$ in the
range $1.84$-$4.75$.

There is a striking variation in the Bayes factors when using different glitch
mitigation techniques. In particular, \texttt{gwsubtract}
\cite{Davis:2022ird,Davis:2018yrz}  favors eccentricity over
\texttt{BayesWave} \cite{Cornish:2014kda, Hourihane:2022doe} by 2+
orders of magnitude (see Table \ref{tab:Bayes_table} and Figs.~\ref{fig:bf_lineplot}, \ref{fig:glitch_subtraction} and \ref{fig:violinplot}). This makes interpreting the analysis challenging as one
has to decide which glitch mitigation technique to trust.

The \texttt{gwsubtract} mitigation technique uses a witness time series from an
auxillary channel in the interferometers to linearly subtract the noise. In
particular one uses the witness time series and the strain data to compute a
transfer function between the auxillary channel and true strain
\cite{Davis:2018yrz}. This transfer function is then used to estimate the
de-glitched strain data. This method was used by the LVK Collaboration to mitigate the glitch
in  LIGO-Livingston during the official analysis. This is because the witness
function was well estimated by the modulation control system at the time of
the detection (see Sec.~3 in Ref.~\cite{Davis:2022ird}).  We also note that there
have recently been additional studies of the impact of glitch mitigation on
GW200129 in Refs.~\cite{Macas:2023zdu, Macas:2023wiw} indicating that
\texttt{gwsubtract} may under-subtract the glitch.

\texttt{BayesWave} on the other hand, takes a data driven approach. It models the
astrophysical signal with a GW model\footnote{Notably, the quasi-circular aligned-spin \texttt{IMRPhenomD} \cite{Husa:2015iqa,Khan:2015jqa} is used, but more generally one should use an eccentric waveform model. However,
this is computationally expensive (see Ref.~\cite{Hourihane:2022doe}). } and
incoherent non-Gaussian noise with sine-Gaussian wavelets. It also models the PSD with a combination of cubic splines 
and Lorentzians. It then runs a trans-dimensional Reversible-Jump MCMC to infer a
posterior distribution over the signal, glitch and PSD. Finally, one takes a fair draw
from the inferred glitch distribution and subtracts it from the detector strain.
We can then run inference on the glitch 
subtracted strain data. Often in LVK analyses, only one glitch draw from the posterior is used (and we indeed 
follow this standard for most events analyzed). However, it is more accurate to draw several glitches from the glitch posterior and marginalize over them. This is in effect what is done in Ref.~\cite{Payne:2022spz} where they take three glitch draws, labelled ``A'', ``B'', and ``C'' and run inference on each mitigated frame. For this paper we utilize the same glitch draws as Ref.~\cite{Payne:2022spz}.

We run the analysis on both mitigation techniques for
completeness\footnote{We also experimented with truncating the 
glitch by setting the $f_{\text{min}} = 50$Hz. However, we note that the
majority of waveform difference from eccentricity with GW200129-like
parameters is contained in the $20-50$Hz regime so this type of test is
inconclusive. }. We also run the analysis using only LIGO-Hanford or only LIGO-Livingston. While there is 
a positive Bayes factor for eccentricity regardless of which detector combination is used, the 
evidence for eccentricity is dominated by LIGO Livingston (see Appendix \ref{sec:one_detector}).
We default to reporting results of GW200129 using
\texttt{gwsubtract} glitch subtracted data, as this is what is
used by official LVK analyses of this event. However, we explicitly indicate in the 
captions which glitch mitigation is being applied.  There is very strong evidence 
for eccentricity in the  \texttt{gwsubtract} case ($\log_{10} \mathcal{B}_{\text{EAS/QCAS}} > 3.5$), 
but we cannot rule out the possibility that \texttt{gwsubtract} under-subtracts the glitch or 
\texttt{BayesWave} over-subtracts the glitch. 
Since the conclusive evidence of eccentricity is contingent on the systematic uncertainty of glitch
mitigation, we do not claim this event to be a bona-fide eccentric event.
However, we note that greater than zero evidence for eccentricity is present irrespective
of the glitch mitigation (see Figs.~ 
 \ref{fig:glitch_subtraction}, \ref{fig:bf_lineplot} and Table \ref{tab:Bayes_table}). 

\begin{figure*}
    \begin{center}
        \includegraphics[width=1\textwidth]{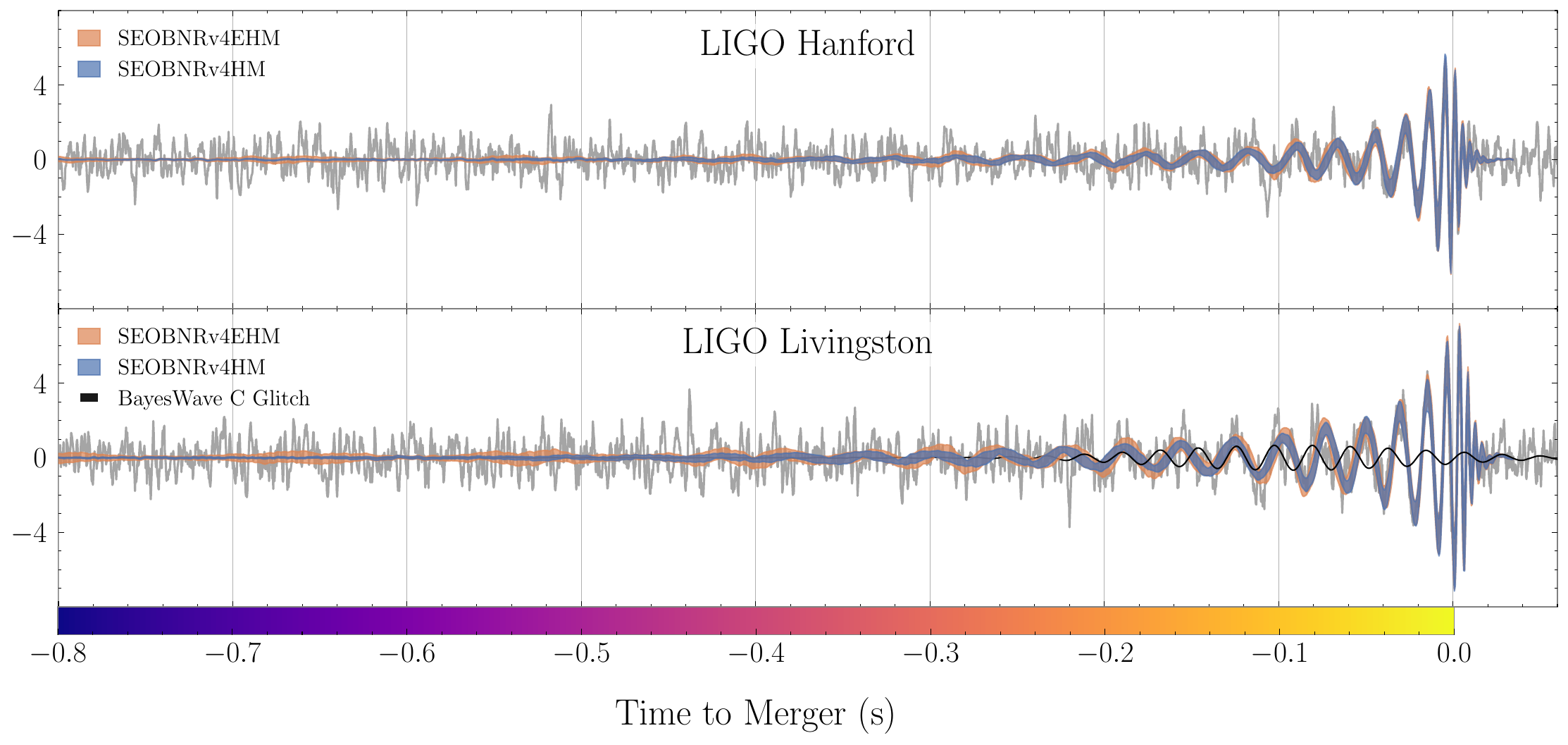}
    \end{center}
    \caption{\label{fig:detector_projection} The 90\% credible interval of the projected waveforms of GW200129 analyzed with 
    \texttt{SEOBNRv4EHM} (orange) and \texttt{SEOBNRv4HM} (blue) in LIGO-Hanford (top) and 
    LIGO-Livingston (bottom). The posterior was obtained using \texttt{BayesWave C} mitigated data. Shown in light gray is the whitened strain data without any glitch mitigation. We also over-plot
    the location of the \texttt{BayesWave C} glitch in the Livingston detector in black. On the bottom of the plot, we show a color bar 
    that corresponds to the trajectory represented in Fig. \ref{fig:dynamics}. The regions in the inspiral where the \texttt{SEOBNRv4EHM} 
    waveform deviates from \texttt{SEOBNRv4HM} waveform are due to periastron passages of the BBH. }
\end{figure*}

\begin{figure}
    \begin{center}
        \includegraphics[width=0.9\linewidth]{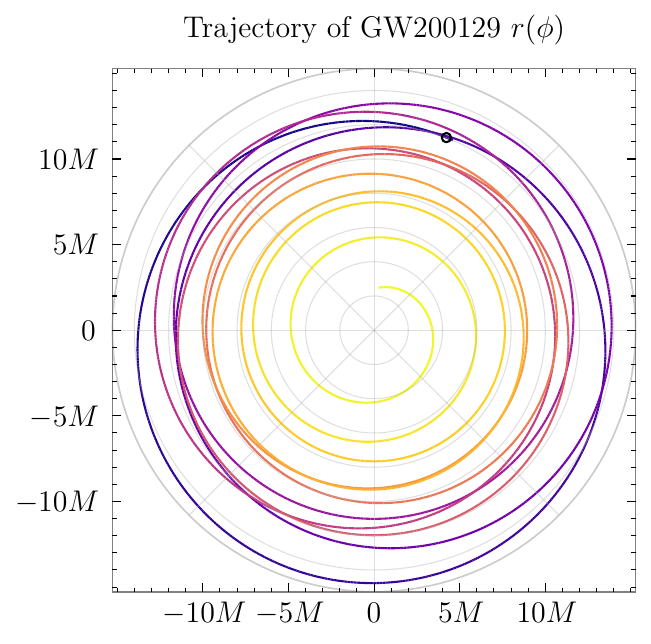}
    \end{center}
\caption{\label{fig:dynamics} Radial separation as a function of the coordinate angle of a binary with parameters corresponding to the maximum likelihood 
of the \texttt{SEOBNRv4EHM} analysis. The color of the orbit corresponds to the time to merger, and can be compared with the color bar in Fig. \ref{fig:detector_projection}. 
The separation is reported in units of the total mass of the system $M$. We mark the start of the orbit with a circle, which is chosen 
to be $t=-0.8$ seconds before merger. }

\end{figure}

\begin{figure*}
\begin{center}
    \includegraphics[width=1\textwidth]{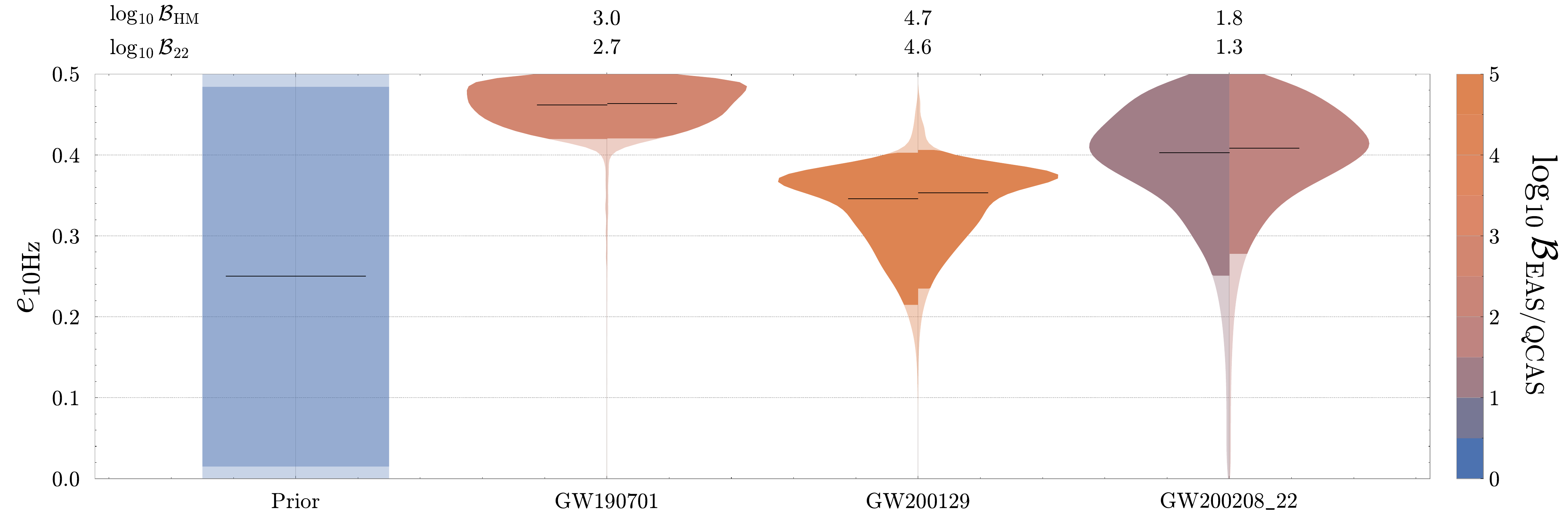}
\end{center}
\caption{\label{fig:violinplot} Eccentric violin plots for all events with $\log_{10} \mathcal{B}_{\text{EAS/QCAS}} > 1$ 
from the first through third observing runs of the LVK. The posterior distributions are obtained
with \texttt{DINGO} and then importance sampled. The eccentricity is
measured at 10 Hz in the detector frame using the quasi-Keplerian parameterization (see Eq.~ (\ref{eq:Keplerian_paramtrization})). For each event, the left violins are the posterior
distributions using only the $(\ell, |m|) = (2,2)$ mode, and the right violins are the posteriors using higher modes (see 
Eq.~ (\ref{eq:mode_decomposition})). Each violin is colored according to the $\log_{10}$ Bayes factor between the eccentric 
aligned-spin and quasi-circular aligned-spin hypothesis. Above the plot, 
we report the $\log_{10}$ Bayes factor when using higher modes ($\log_{10} \mathcal{B}_{\text{HM}}$) and 
when only using the $(\ell, m) = (2, 2)$ mode ($\log_{10} \mathcal{B}_{\text{22}}$). The dark shaded regions indicate the 90\% credible interval of the distribution and the black lines
indicate the median value of the eccentricity. 
}

\end{figure*}

For visualization we display the 90\% highest density interval of the projected waveforms of GW200129 in
LIGO-Hanford and LIGO Livingston in Fig. \ref{fig:detector_projection}. We  
use the posterior obtained with the \texttt{BayesWave C} glitch mitigation. We also 
show the location of the \texttt{BayesWave C} glitch. To generate this
figure, we draw samples from the 90\% highest density interval of the
\texttt{SEOBNRv4EHM} and \texttt{SEOBNRv4HM} posteriors, and project them onto
the detector. We then whiten the waveform, glitch and strain. By whitening we 
mean that we set the noise variance in each frequency bin to one using the PSD 
(see Ref. \cite{LIGOScientific:2019hgc} for further details).
We plot the strain data with a rolling window average with a window size of 1
millisecond for visualization purposes. In Fig. \ref{fig:dynamics} we show the radial separation 
as a function of the coordinate angle of a binary with parameters corresponding to the maximum likelihood 
of the \texttt{SEOBNRv4EHM} analysis of GW200129 with \texttt{BayesWave C} glitch mitigation. 

Due to the possibility that GW200129 is precessing we additionally compute the
Bayes factor against the quasi-circular precessing-spin case. This test is also
important as for short signals, it is
possible that spin-precession can mask the effect of eccentricity \cite{Romero-Shaw:2022fbf, CalderonBustillo:2020xms}. We first compute the Bayes factors against
\texttt{SEOBNRv4PHM} and find that the $\log_{10} \mathcal{B}_{\text{EAS/QCP}}$ lies in the range
$2.20-4.92$. We also compute the Bayes factor against \texttt{NRSur7dq4} and find
$\log_{10} \mathcal{B}_{\text{EAS/QCNRP}}$ in the range 
$1.43-4.0$. When using the \texttt{gwsubtract} glitch mitigation, we find the difference 
between the maximum $\log_{10}$ likelihoods of \texttt{SEOBNRv4EHM} and \texttt{NRSur7dq4} is $5.7$.

The range of Bayes factors is again due to using different glitch mitigation algorithms. One needs to interpret
the $\log_{10} \mathcal{B}_{\text{EAS/QCNRP}}$ with caution since eccentricity and precession are not the
only differences between \texttt{SEOBRv4EHM} and \texttt{NRSur7dq4}. In particular, the
underlying quasi-circular model of \texttt{SEOBNRv4EHM} is different from
\texttt{NRSur7dq4}. Nonetheless, there is evidence of eccentricity irrespective of the 
spin-precessing waveform model used. Despite the eccentric aligned-spin 
hypothesis being preferred over the quasi-circular spin-precessing hypothesis, we do not 
exclude the possibility that GW200129 is spin-precessing (it could be both eccentric and spin-precessing).

As another check if spin-precession can mimic the eccentricity in this event,
we also perform injections of the maximum likelihood waveform of
\texttt{NRSur7dq4} and recover with \texttt{SEOBNRv4EHM} (and vice versa).
We conclude for the maximum likelihood parameters of this event, injected precession cannot mimic the effect of eccentricity. Additionally,
injected eccentricity does not mimic the effect of precession. The result of this systematics study can 
be seen in Fig.~\ref{fig:cross_injections} in Appendix \ref{sec:cross_injections}. 

Since \texttt{SEOBNRv4EHM} is preferred over \texttt{NRSur7dq4} this
has important implications for the kick velocity of this event. In particular,
Ref.  \cite{Varma:2022pld} showed evidence for a measurable kick velocity in
GW200129. However, when using an eccentric aligned-spin model the evidence for a
measurable kick velocity substantially decreases (see
Fig.~\ref{fig:kick_velocity} and the discussion in Appendix
\ref{sec:kick_velocity}).

\subsection{GW190701}
\label{sec:GW190701}
We also see signs for eccentricity in GW190701. To our knowledge
GW190701 has not shown signs of eccentricity in previous studies. When analyzed 
with \texttt{SEOBNRv4EHM}, GW190701 is a heavy BBH
with $M_{\text{det}} = 183.5_{-39.1}^{+40.7}$, $M_{\text{src}} =
131.9_{-17.7}^{+18.0}$ and $\chi_{\text{eff}} = -0.04_{-0.26}^{+0.21}$. We find that $e_{\text{gw, 10Hz}} = 0.54_{-0.30}^{+0.12}$, $e_{\text{gw, 20Hz}} = 0.31_{-0.13}^{+-0.12}$
with Bayes factors of $\log_{10} \mathcal{B}_{\text{EAS/QCAS}} = 3.0$ and $\log_{10}
\mathcal{B}_{\text{EAS/QCP}} = 2.61$ (Table \ref{tab:Bayes_table}). We find the difference 
between the maximum $\log_{10}$ likelihoods of \texttt{SEOBNRv4EHM} and \texttt{SEOBNRv4PHM} is $5.2$.
Critically, the eccentricity rails against the upper bound 
of the prior (see Fig.~\ref{fig:violinplot}). Thus this Bayes
factor is a conservative lower bound since there is likelihood support above $e_{\text{10Hz}} >
0.5$. Like GW200129, this event also
contains a glitch, which is subtracted using one fair draw of a \texttt{BayesWave} glitch
model. However, we do not have access to the glitch distribution in this case so we 
perform the analysis using only one fair draw of the glitch posterior.

We tried to train a \texttt{DINGO} network in the regime $e_{\text{10Hz}} < 0.8$, to
investigate the high eccentricity of GW190701.  However, the gradients of the
network increase dramatically in the first 5 epochs. This indicates that the
additional hyper-parameters of the network, such as the learning rate need to be
further modified. Another possibility is that artifacts in the waveform due to high
eccentricity and spins \cite{Ramos-Buades:2021adz} affect substantially the
learning process. These issues could be potentially addressed in the future with
improved waveform models \cite{aldo2024} and network architectures
\cite{Dax:2023ozk}.

The maximum likelihood of GW190701 only has 3-4 inspiral cycles in band. This is primarily due to the high masses. The signal length is further shortened due to the high value of $e_{\text{10Hz}}$. This means one needs to interpret conclusions with caution (since the eccentric waveform model used assumes a quasi-circular merger-ringdown). We note that, independently of the number of cycles, confidently measuring eccentricity in such high mass events would require incorporating eccentric effects in the merger-ringdown, which we leave for future work. 

Furthermore, we need to back evolve the waveform in order to have enough cycles to compute
$e_{\text{gw, 10Hz}}$. However, this means evolving to values where
$e_{\text{10Hz}} > 0.5$. Here, the waveform model has not 
been tested against NR. Thus, one needs to interpret the $e_{\text{gw, 10Hz}}$ for 
this event with additional caution. With waveform models which probe and and tested against higher initial eccentricities, 
this issue can be mitigated \cite{aldo2024}.

\newcommand{\ColorOne}{\cellcolor[HTML]{FABCA4}} 
\newcommand{\ColorTwo}{\cellcolor[HTML]{FED7B0}} 
\newcommand{\ColorThree}{\cellcolor[HTML]{FFFAD6}} 
\newcommand{\ColorFour}{\cellcolor[HTML]{E6F4C1}} 
\newcommand{\ColorFive}{\cellcolor[HTML]{C8E7B3}} 
\newcommand{\ColorSix}{\cellcolor[HTML]{ABDAAF}} 
\newcommand{\ColorSeven}{\cellcolor[HTML]{92CEA9}} 
\newcommand{\ColorEight}{\cellcolor[HTML]{8DCCA8}} 
\newcommand{\ColorNine}{\cellcolor[HTML]{84BB9F}} 
\newcommand{\ColorTen}{\cellcolor[HTML]{79B597}} 
\newcommand{\ColorGray}{\cellcolor{lightgray}}

\begin{table*}
    \setlength{\tabcolsep}{6.6pt}
    \renewcommand{\arraystretch}{1.6}
    \begin{tabular}{cccccccc}

        \hline \hline 
        Glitch Subtraction & $e$ Prior & \multicolumn{1}{p{2cm}}{\centering \texttt{SEOBNRv4} \\ $\log_{10} \mathcal{B}$} & \multicolumn{1}{p{2cm}}{\centering \texttt{SEOBNRv4HM} \\ $\log_{10} \mathcal{B}$} & \multicolumn{1}{p{2cm}}{\centering \texttt{SEOBNRv4PHM} \\ $\log_{10} \mathcal{B}$} & \multicolumn{1}{p{2cm}}{\centering \texttt{NRSur7dq4} \\ $\log_{10} \mathcal{B}$} & $e_{\text{10Hz}}$ & $e_{\text{gw, 10Hz}}$ 
        \\   
        \hline  
        \addlinespace[2.5ex]
 
        \multicolumn{7}{c}{GW200129} \\ \hline \hline
         & & & & & & & \\ 
        \texttt{gwsubtract} & Uniform  & \multicolumn{1}{c}{\ColorTen$4.57$} & \multicolumn{1}{c}{\ColorTen$4.75$} & \multicolumn{1}{c}{\ColorTen$4.92$} & \multicolumn{1}{c}{\ColorNine$4.0$}  & $0.34^{+0.11}_{-0.06}$ & $0.27^{+0.10}_{-0.12}$ \\ 
        \texttt{BayesWave A} & Uniform & \multicolumn{1}{c}{\ColorFour$1.7$} & \multicolumn{1}{c}{\ColorFour$1.84$} & \multicolumn{1}{c}{\ColorFive$2.20$} & \multicolumn{1}{c}{\ColorFour$1.53$} & $0.24^{+0.10}_{-0.10}$ & $0.17^{+0.14}_{-0.13}$ \\ 
        \texttt{BayesWave B} & Uniform & \multicolumn{1}{c}{\ColorSix$2.92$} & \multicolumn{1}{c}{\ColorSeven$3.08$} & \multicolumn{1}{c}{\ColorSeven$3.43$} & \multicolumn{1}{c}{\ColorFive$2.35$} & $0.28^{+0.09}_{-0.11}$ & $0.22^{+0.12}_{-0.13}$ \\ 
        \texttt{BayesWave C} & Uniform & \multicolumn{1}{c}{\ColorSix$2.85$} & \multicolumn{1}{c}{\ColorSix$2.93$} & \multicolumn{1}{c}{\ColorSix$2.63$} & \multicolumn{1}{c}{\ColorThree$1.43$} & $0.27^{+0.09}_{-0.10}$ & $0.22^{+0.13}_{-0.14}$ \\ 
        \texttt{gwsubtract} & Log-Uniform & \multicolumn{1}{c}{\ColorNine$4.02$}  & \multicolumn{1}{c}{\ColorEight$3.98$} & \multicolumn{1}{c}{\ColorEight$3.99$} & \multicolumn{1}{c}{\ColorSeven$3.23$} & $0.33_{-0.11}^{+0.07}$ & $0.28_{-0.12}^{+0.11}$  \\ 
        \texttt{BayesWave A} & Log-Uniform & \multicolumn{1}{c}{\ColorFour$1.79$} & \multicolumn{1}{c}{\ColorThree$1.26$} & \multicolumn{1}{c}{\ColorFour$1.61$} & \multicolumn{1}{c}{\ColorTwo$0.94$} & $0.22_{-0.15}^{+0.13}$ & $0.17_{-0.13}^{+0.14}$ \\ 
        \texttt{BayesWave B} & Log-Uniform & \multicolumn{1}{c}{\ColorFive$2.28$} & \multicolumn{1}{c}{\ColorFive$2.43$} & \multicolumn{1}{c}{\ColorSix$2.78$} & \multicolumn{1}{c}{\ColorFour$1.70$} & $0.27_{-0.11}^{+0.09}$  & $0.22_{-0.12}^{+0.13}$ \\ 
        \texttt{BayesWave C} & Log-Uniform & \multicolumn{1}{c}{\ColorFive$2.22$} & \multicolumn{1}{c}{\ColorFive$2.27$} & \multicolumn{1}{c}{\ColorFour$1.97$} & \multicolumn{1}{c}{\ColorTwo$0.76$} & $0.26_{-0.11}^{+0.10}$ & $0.22_{-0.13}^{+0.13}$ \\ 
        \addlinespace[2.5ex]
 
        \multicolumn{7}{c}{GW190701} \\ \hline \hline
        & & & & & & & \\  
        \texttt{BayesWave} & Uniform & \multicolumn{1}{c}{\ColorSix$2.72$} & \multicolumn{1}{c}{\ColorSeven$3.0$} & \multicolumn{1}{c}{\ColorSix$2.61$} & -- & $0.46^{+0.04}_{-0.04}$ & $0.54^{+0.30}_{-0.12}$  \\ 
        \texttt{BayesWave} & Log-Uniform & \multicolumn{1}{c}{\ColorFour$1.86$} & \multicolumn{1}{c}{\ColorFive$2.11$} & \multicolumn{1}{c}{\ColorFour$1.71$} & -- & $0.45^{+0.05}_{-0.04 }$ & $0.53^{+0.31}_{-0.11}$ \\ 
        \addlinespace[2.5ex]
 
        \multicolumn{7}{c}{GW200208\_22} \\ \hline \hline
        & & & & & & & \\  
        Unmitigated Strain & Uniform & \multicolumn{1}{c}{\ColorThree$1.25$} & \multicolumn{1}{c}{\ColorFour$1.77$} & \multicolumn{1}{c}{\ColorThree$1.23$} & -- & $0.4^{+0.08}_{-0.15}$ &  $0.39_{-0.23}^{+0.23}$ \\ 
        Unmitigated Strain & Log-Uniform & \multicolumn{1}{c}{\ColorTwo$0.54$} & \multicolumn{1}{c}{\ColorThree$1.05$} & \multicolumn{1}{c}{\ColorOne$0.48$} & -- & $0.35^{+0.15}_{-0.33}$ & $0.37^{+0.19}_{-0.20}$ \\ 
        \addlinespace[2.5ex]
        \hline

    \end{tabular}
    \caption{ \label{tab:Bayes_table}Bayes factors of the three GW events with $\log_{10} \mathcal{B}_{\text{EAS/QCAS}} >
    1$. The first column indicates which glitch mitigation algorithm is used (see
    text). The second column indicates whether the prior on $e_{\text{10Hz}}$ is is uniform
    between [0.0, 0.5] or log-uniform between $[10^{-4}, 0.5]$. The third column indicates the Bayes factors between the eccentric
    aligned-spin $(\ell, m) = (2, 2)$ mode only model (\texttt{SEOBNRV4E}) and quasi-circular aligned-spin $(\ell, m) = (2, 2)$ mode only model
    (\texttt{SEOBNRv4}). The fourth through sixth columns indicate the Bayes factors between the eccentric
    aligned-spin model (\texttt{SEOBNRv4EHM}) against the quasi-circular aligned-spin (\texttt{SEOBNRv4HM}) or 
    quasi-circular precessing-spin (\texttt{SEOBNRv4PHM} or \texttt{NRSur7dq4}) models.
    The last two columns indicate the mean and 90\% highest density interval
    of the \texttt{SEOBNRv4EHM} posterior for $e_{\text{10Hz}}$ and $e_{\text{gw,
    10Hz}}$ respectively. The errors in each $\log_{10}$ Bayes factor are less than $0.04$. For entries with dashes, we do not compute Bayes factors 
    due to a lack of networks that cover the appropriate prior. 
    }
\end{table*}

\subsection{GW200208\_22}
\label{sec:GW200208_22}

We also see signs for eccentricity in GW200208\_22. The evidence 
for eccentricity in GW200208\_22 has been seen
previously in Refs. \cite{Romero-Shaw:2022xko,Iglesias:2022xfc}. 
When analyzed with \texttt{SEOBNRv4EHM}, we observe GW200208\_22 is
an intermediate-mass BBH with support for positive effective spin
($M_{\text{det}} = 94.7_{-12.5}^{+28.9}$, $M_{\text{src}} =
66.3_{-12.5}^{+17.5}$ and $\chi_{\text{eff}} =
0.12_{-0.29}^{+0.30}$). We find that $e_{\text{gw, 10Hz}} = 0.39_{-0.23}^{+0.23}$, 
$e_{\text{gw, 20Hz}} = 0.21_{-0.08}^{+0.08}$, $\log_{10} \mathcal{B}_{\text{EAS/QCAS}} = 1.77$ and $\log_{10}
\mathcal{B}_{\text{EAS/QCP}} = 1.23$. We find the difference 
between the maximum $\log_{10}$ likelihoods of \texttt{SEOBNRv4EHM} and \texttt{SEOBNRv4PHM} is $3.1$.

Importantly, this event has a relatively low value of $p_{\text{astro}}$ of $0.7$ and high 
false alarm rate (FAR) of 4.8 $\text{yr}^{-1}$ (recovered by \texttt{pycbc-bbh} \cite{alex_nitz_2023_7692098}).
Therefore, it sits just above the detection threshold set by the LVK. We note that these values are obtained using quasi-circular templates.
(though see Ref.~\cite{Wang:2025yac} which uses eccentric templates and was released during the referee period of this paper.)

We identify railing of the mass
ratio on the lower bound of $q = 0.125$ with \texttt{SEOBNRv4HM},
\texttt{SEOBNRv4PHM} and \texttt{SEOBNRv4EHM}. However, due to the lower 
bound mass restriction set by \texttt{DINGO}, we do not relax this lower bound. Instead,
we perform \texttt{pBilby} runs extending the lower bound on $q$ to $0.05$ to explore
the possibility of a secondary mode at low mass ratios. We do not see evidence
for such a mode, and the Bayes factors do not change significantly. We also see railing of eccentricity
against the eccentricity upper bound of $e_{\text{10Hz}} = 0.5$. But due to the
model restriction of \texttt{SEOBNRv4EHM} (i.e., $e < 0.5$), we do not increase this
upper bound. Like in the case of GW190701 $\log_{10}
\mathcal{B}_{\text{EAS/QCAS}}$ are then only 
conservative lower bounds of the Bayes factor. 

\subsection{GW190521}
\label{sec:GW190521}

We also train specialized $\texttt{SEOBNRv4EHM}$ and $\texttt{SEOBNRv4HM}$
networks to analyze GW190521. This event is interesting as it only has 4 cycles
in the detectors' bandwidth (assuming $f_{\text{min}}=11$Hz), and the SNR is dominated by the merger and ringdown. It
has been suggested that this event is a head-on collision with exotic compact
objects \cite{Sanchis-Gual:2018oui}, a non-spinning hyperbolic capture \cite{Gamba:2021gap}, a
merger within an active galactic nucleus (AGN)
\cite{Graham:2020gwr,Samsing:2020tda,Tagawa:2020qll} and eccentric
\cite{Romero_Shaw:2020thy, Gayathri:2020coq}. It has also been suggested that
this event has an electromagnetic counterpart detected by the Zwicky Transient
Facility (ZTF) \cite{Graham:2020gwr}.

In order to analyze GW190521, we train a specialized network with a larger upper bound on the
detector frame component masses ($m_{1,2} < 180 M_\odot$). The network also has a
starting frequency of $f_{\text{start}} = 5.5$ Hz. This is to make sure the higher modes, 
and in particular the (4,4) mode,  are in band at the minimum frequency, which we set to
$f_{\mathrm{min}} = 11$ Hz. Thus the eccentricity is sampled at 5.5
Hz $(e_{5.5 \text{Hz}})$. Accordingly, we adjust the prior in the eccentricity to be 
uniform between (0, 0.3). 

When using \texttt{SEOBNRv4EHM}, we recover $e_{\text{gw, 5.5Hz}} =0.12_{-0.12}^{+0.12}$ for GW190521 with a
posterior very close to the prior. We find a $\log_{10} 
\mathcal{B}_{\text{EAS/QCAS}}$ of 0.04. This is
likely due to the same reasons described in Ref.~\cite{Ramos-Buades:2023yhy}. Notably, GW190521 
is a merger-ringdown dominated signal, but the eccentric \texttt{SEOBNRv4EHM} waveform is the same as 
the quasi-circular \texttt{SEOBNRv4HM} waveform during the merger-ringdown. Thus it is 
difficult to measure eccentricity in the merger. So while 
we do not see signs of eccentricity in this analysis, the high eccentricity limit near 
merger is fairly unconstrained. We highlight that this is a common feature for all current eccentric 
waveform models in the literature. 

\begin{figure*}
\begin{center}
    \includegraphics[width=\textwidth]{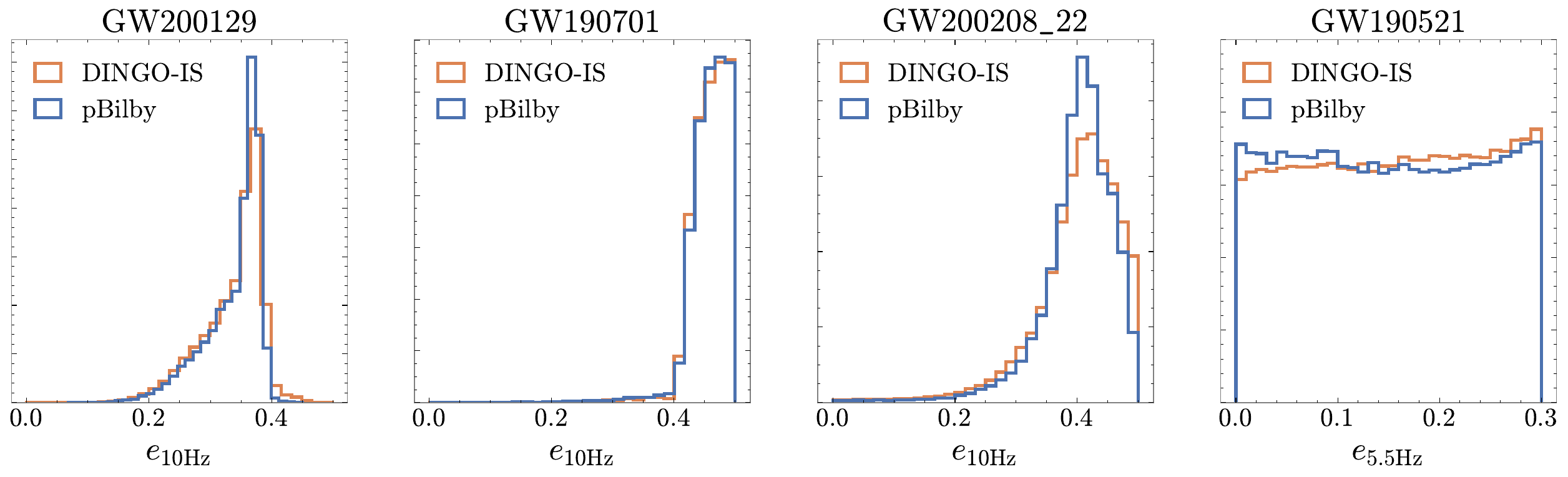}
\end{center}
\caption{\label{fig:eccentricity_comparisons} \texttt{DINGO-IS} versus \texttt{pBilby}
marginal eccentricity distributions for four of the events analyzed in the main text. 
We show that the presence of eccentricity is also found by nested samplers. 
Note that while, there are are slight differences between \texttt{DINGO-IS} and \texttt{pBilby} in
the GW200208\_22 posterior, this can be attributed to the railing of the eccentricity distribution 
at 0.5. When extending the prior of the \texttt{pBilby} run, we again find better agreement. }
\end{figure*}

Finally, to ensure the analysis of GW200129, GW190701,
GW200208\_22 and GW190521 do not depend on the sampler, we perform additional
\texttt{pBilby} analyses using \texttt{SEOBNRv4E} (\texttt{SEOBNRv4EHM} but turning 
off the higher modes) of these events in Fig.~\ref{fig:eccentricity_comparisons}.


\subsection{Events with mild eccentricity support}
\label{sec:mild_support}

We find marginal evidence for eccentricity in GW190620\_030421. When analyzed
with \texttt{SEOBNRv4EHM}, we observe this is a heavy BBH with support for a
positive effective spin with $M_{\text{src}} = 168.3_{-34.9}^{+36.4}$,
$M_{\text{det}} = 126.6_{-17.9}^{+18.7}$ and $\chi_{\text{eff}} =
0.20_{-0.27}^{+0.26}$. Unlike the previously discussed events in this paper,
GW190620\_030421 was only detected in LIGO Livingston. We find that $\log_{10}
\mathcal{B}_{\text{EAS/QCAS}} = 0.56$ with $e_{\text{gw, 10Hz}} =
0.30_{-0.19}^{+0.19}$. Due to the low Bayes factor, we do not consider it
as having significant evidence for eccentricity. This event has shown similar Bayes factors
for eccentricity in Refs.~\cite{Iglesias:2022xfc,Romero-Shaw:2022xko}. 

We also analyze GW191109\_010717, which has shown signs of eccentricity in Ref.
\cite{Romero-Shaw:2022xko}. When analyzed with \texttt{SEOBNRv4EHM}, we find this is a
heavy BBH with $M_{\text{src}} = 175.6_{-30.0}^{+34.0}$, $M_{\text{det}} =
137_{-14.4}^{+14.0}$ and $\chi_{\text{eff}} = -0.25_{-0.27}^{+0.24}$. 
We find $\log_{10} \mathcal{B}_{\text{EAS/QCAS}} = 0.07$ with $e_{\text{gw,
10Hz}} = 0.26_{-0.22}^{+0.24}$ for this event. Interestingly, the posterior does
rail against the upper bound of the prior, but due to the eccentricity limit
enforced by \texttt{SEOBNRv4EHM}, we do not increase the upper bound. This event 
also has two peaks in the eccentricity posterior with one at $e_{\text{gw,
10Hz}} = 0.27$ and the other and the edge of the prior $e_{\text{gw, 10Hz}} =
0.50$. This was also noted in Ref.~\cite{Romero-Shaw:2022xko}. 

While these two events individually do not have a high $\mathcal{B}_{\text{EAS/QCAS}}$, 
they do contribute to the population level analysis (see Sec.~\ref{sec:p_ecc}).

\begin{figure*}
\begin{center}
    \includegraphics[width=\textwidth]{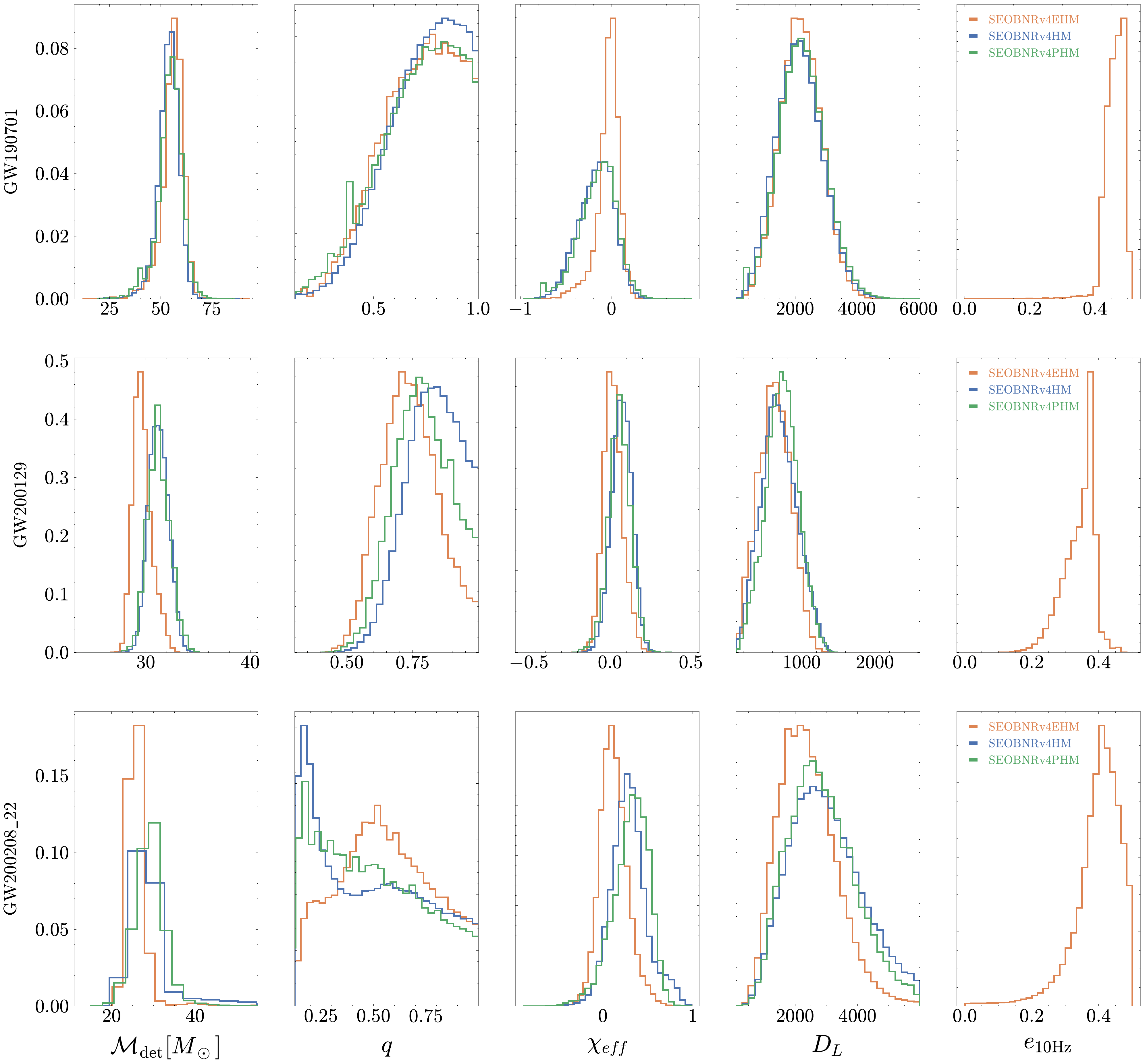}
\end{center}
\caption{\label{fig:systematics}  
Posterior distributions for the 3 candidate eccentric events analyzed with \texttt{SEOBNRv4EHM}, \texttt{SEOBNRv4PHM},
and \texttt{SEOBNRv4HM}. The posteriors were obtained by DINGO and then importance 
sampled (see \ref{sec:DINGO}). We note the \texttt{SEOBNRv4PHM} posteriors are in qualitative agreement with the 
LVK results \cite{KAGRA:2021vkt}. When ignoring the effects of eccentricity, we see large differences in the $\mathcal{M}_{\text{det}}$ and $\chi_{\text{eff}}$. 
Noting that eccentricity decreases the time to merger with respect to
quasi-circular templates, the biases can be understood as compensations
by $\mathcal{M}_{\text{det}}$ and $\chi_{\text{eff}}$ to keep the time to merger fixed. }
\end{figure*}

\section{Waveform systematics}
\label{sec:systematics}

\subsection{Biases from neglecting eccentricity}
\label{sec:neglecting_eccentricity}

When inferring source parameters with a waveform model that does not include the
effect of eccentricity, one can bias the inferred source parameters. This effect
is particularly pronounced in the three events that show signs for eccentricity
where biases in the chirp mass and effective spin are observed. The results of
this analysis are shown in Fig.~\ref{fig:systematics}. One of the main effects
is that the eccentricity reduces the time to merger. To compensate for this,
other parameters must change keeping the true time to merger fixed. In the cases
of GW200129 and GW200208\_22, the chirp mass decreases to keep a fixed time to
merger. We observe mean chirp-mass differences between \texttt{SEOBNRv4EHM} and
\texttt{SEOBNRv4HM} of 5.2\% ($-1.54 M_\odot$) for GW200129, 3.6\% ($2.00
M_\odot$) for GW190701 and 11.8\% ($-3.08 M_\odot$) GW200208\_22. This
corresponds to a JSD differences of $3.78 \times 10^{-1}$ bits, $4.01 \times
10^{-2}$ bits and $1.99 \times 10^{-1}$ bits for GW200129, GW190701 and
GW20028\_22 respectively. We observe chirp-mass differences between
\texttt{SEOBNRv4EHM} and \texttt{SEOBNRv4PHM} of 5.2\% ($-1.56 M_\odot$) for
GW200129, 2.4\% ($1.32 M_\odot$) for GW190701, and 11.6\% ($-3.03 M_\odot$) for
GW200208\_22. This corresponds to JSD differences of $3.55 \times 10^{-1}$ bits,
$2.21 \times 10^{-2}$ bits, and $3.03 \times 10^{-1}$ bits for GW200129,
GW190701 and GW200208\_22 respectively.

There is also a bias in $\chi_{\text{eff}}$ across the events. Comparing
\texttt{SEOBNRv4EHM} and \texttt{SEOBNRv4HM} we observe differences of $-0.04$,
$0.12$ and $-0.19$ in the mean $\chi_{\text{eff}}$ for GW200129, GW190701 and
GW200208\_22 respectively. This corresponds to JSD differences of $1.27 \times
10^{-1}$ bits, $1.50 \times 10^{-1}$ bits and $1.58 \times 10^{-1}$ bits for
GW200129, GW190701 and GW200208\_22 respectively. In conclusion, when including
the effect of eccentricity in these events, the $\chi_{\text{eff}}$ distribution
tends towards 0.

\subsection{Biases from neglecting higher modes in eccentric waveforms}
\label{sec:hms}

We investigate the impact of neglecting higher modes beyond the $(\ell, |m|) =
(2, 2)$ mode in the estimation of source parameters. It has been shown that neglecting them in quasi-circular 
models can lead to biases in parameter estimation~\cite{LIGOScientific:2020ibl,Capano:2013raa,Varma:2014jxa,VanDenBroeck:2006ar,VanDenBroeck:2006qu,Littenberg:2012uj}. Here we extend the study to eccentric waveforms.  
We quantify this by
calculating the JSD between the analysis performed with \texttt{SEOBNRv4E} and
\texttt{SEOBNRv4EHM}. For comparison we also compute the JSDs for
\texttt{SEOBNRv4} and \texttt{SEOBNRv4HM}. We follow Ref.~\cite{LIGOScientific:2020ibl}
and consider deviations beyond 0.007 bits as significant biases. For a Gaussian, 
this corresponds to a 20\% shift in the mean measured in standard deviations of the Gaussian. 
(We remark that 0.002 bits is the expected stochastic variation from GW sampling
algorithms \cite{Romero-Shaw:2020owr}).

\begin{figure*}
\begin{center}
    \includegraphics[width=\textwidth]{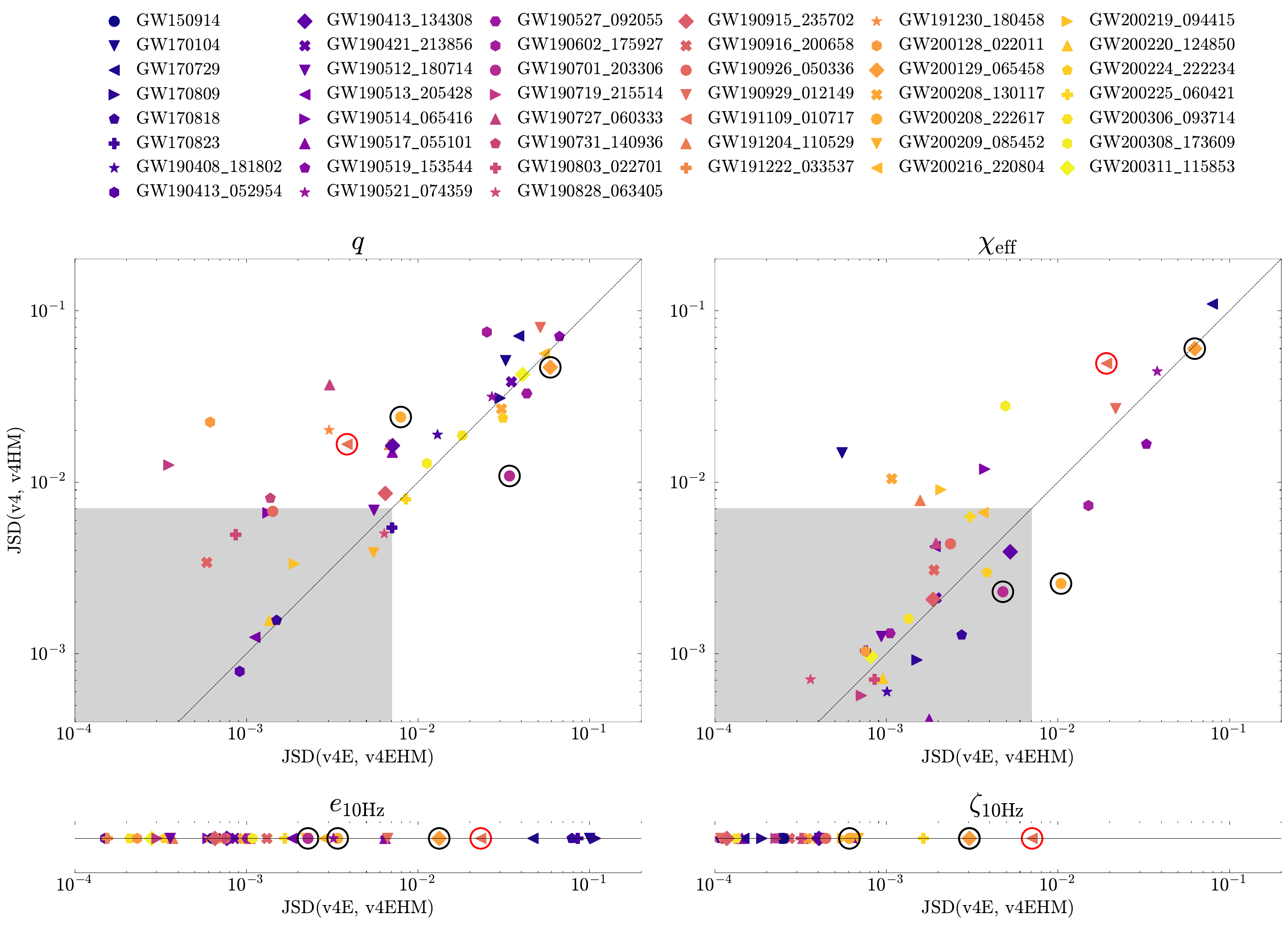}
\end{center}
\caption{\label{fig:higher_mode_jsds} JSDs of 1D
marginals when including or not including higher modes in GWs. Each
point represents a GW event and contains four (top) or two (bottom) posterior distributions. In the top 
plot, the $y$-axis shows the JSD between \texttt{SEOBNRv4} and \texttt{SEOBNRv4HM} in bits. On the top and bottom plots 
the $x$-axis shows the JSD between \texttt{SEOBNRv4E} and \texttt{SEOBNRv4EHM} in bits. The title of the plots indicate  which 1D marginal is being compared (in order mass ratio, effective spin, eccentricity
and relativistic anomaly). The gray section indicates the region in which the JSD is less than 0.007 bits. For a Gaussian, this corresponds to a 20\% shift in the mean measured in standard deviations of the Gaussian. Events to the right of the diagonal line represent
events for which the eccentric higher-modes are important. For events like GW190701, GW200208\_22 and GW200129 (circled in black), which show signs of eccentricity, there are larger JSDs between
\texttt{SEOBNRv4E} and \texttt{SEOBNRv4EHM} than \texttt{SEOBNRv4} and \texttt{SEOBNRv4HM}. In addition, higher modes are important for the analysis of GW191109\_010717 (circled in red)
as they cause large deviations in the eccentricity and relativistic anomaly posterior. }
\end{figure*}

First, we find that there are biases in the eccentricity distribution when not
including higher modes. This is especially true for GW191109\_010717, which can
be seen in Fig.~\ref{fig:higher_mode_jsds}. In particular, we see that
$e_{\text{10Hz}}$ is boosted to higher values and leads to a JSD of 0.025 bits when using only
the $(\ell, |m|) = (2, 2)$ mode. To rephrase, neglecting eccentricity results in increasing the
mean eccentricity by $0.034$. For this event neglecting higher
modes also leads to larger Bayes factors.  When only using the $(\ell, |m|) =
(2, 2)$ mode, $\log_{10} \mathcal{B}_{\text{EAS/QCAS}} = 0.32$ but when using
higher modes $\log_{10} \mathcal{B}_{\text{EAS/QCAS}} = 0.07$. 

Furthermore, neglecting higher modes in eccentric waveforms can lead to biases
in source parameters other than the eccentricity. For example, the inclusion of higher modes in eccentric waveforms for GW190701 leads to a bias (0.03 bits) in the
mass ratio. In contrast, when neglecting higher modes in the quasi-circular waveforms (that is comparing \texttt{SEOBNRv4} to \texttt{SEOBNRv4HM}), 
we see lower difference (0.01 bits). This is also true of the 
$\chi_{\text{eff}}$ distribution of GW200208\_22 where neglecting higher modes in eccentric waveforms 
leads to a 0.01 bit difference whereas neglecting higher modes in quasi-circular waveforms 
only leads to a 0.002 bit difference.

\section{Odds ratios and $p_{\text{ecc}}$}
\label{sec:odds_ratio}

\subsection{Prior odds and odds ratio}
\label{sec:prior_odds}

We now turn our attention to computing the odds ratio in Eq.~(\ref{eq:posterior_odds}) 
by taking into account the effect of astrophysical 
prior odds on our observations.  This is important, as the current prior (uniform in eccentricity) and prior odds (equal odds) assume that each event is equally likely to be eccentric or not. To do this we first put a prior on the rate of
eccentric BBHs, $R_{\text{E}}$, and quasi-circular BBHs, $R_{\text{QC}}$, that we expect. 
We then take the maximum probability of this prior to estimate the prior odds. 

We define $R_{\text{E}}$ as the number of events with $e_{10\text{Hz}} > 0.05$
per year within a luminosity distance of 6000 Mpc ($z < 0.9$) \cite{Lower:2018seu}. The reason we
select $z< 0.9$ is because this is the maximum distance we consider in the
analysis. Similarly, we define $R_{\text{QC}}$ as the number of events 
with $e_{10\text{Hz}} < 0.05$, again within $z<0.9$.

The first step is to place a prior on the merger rate of eccentric and quasi-circular
events based on theoretical predictions\footnote{An important 
caveat is that the $e_{10\text{Hz}}$ reported in astrophysical studies 
is \textit{not} the same as the $e_{10\text{Hz}}$ in the quasi-Keplerian parameterization (Eq.~\ref{eq:Keplerian_paramtrization}) \cite{Vijaykumar:2024piy}. 
Since the point which we use to define the rate is $e_{10\text{Hz}} \approx 0.05$ and the discrepancy between the two
definitions becomes relevant at $e_{10\text{Hz}} \geq 0.2$, we do not expect this to be a significant 
issue for defining the heuristic prior.}. We consider four binary formation environments 
that may lead to BBHs. The principal difference between these formation environments is the 
cluster escape speed, which increases the number of dynamical encounters
\cite{heggie1975binary}. We consider nuclear-star clusters (NSCs),
globular clusters (GCs), young-star clusters (YSCs) and isolated binaries (IBs). We assume that
each channel is independent and contributes some rate of eccentric events,
$R_{\text{E}}^{\text{c}}$, and some rate of quasi-circular events,
$R_{\text{QC}}^{\text{c}}$. The superscript $c$ denotes the name of the channel.
Each of these channels contribute to the total rate of eccentric or
quasi-circular events:

\begin{equation}
    \label{eq:eccentric_rate_sum}
    \begin{aligned}
        R_{\text{E}} &= R_{\text{E}}^{\text{GC}} + R_{\text{E}}^{\text{NSC}} + R_{\text{E}}^{\text{YSC}} + R_{\text{E}}^{\text{IB}}  \\
        R_{\text{QC}} &= R_{\text{QC}}^{\text{GC}} + R_{\text{QC}}^{\text{NSC}} + R_{\text{QC}}^{\text{YSC}} + R_{\text{QC}}^{\text{IB}}. 
    \end{aligned}
\end{equation}

We can then write the prior over the total rates as a convolution over the channel rates. 

\begin{equation}
    \label{eq:convolution}
    \begin{aligned}
        p(R_{\text{E}}, R_{\text{QC}}) &= 
        p(R_{\text{E}}^{\text{GC}}, R_{\text{QC}}^{\text{GC}}) * p(R_{\text{E}}^{\text{NSC}}, R_{\text{QC}}^{\text{NSC}})  \\
        &*p(R_{\text{E}}^{\text{ysc}}, R_{\text{QC}}^{\text{ysc}}) *
        p(R_{\text{E}}^{\text{IB}}, R_{\text{QC}}^{\text{IB}}).
    \end{aligned}
\end{equation}
In the literature, the total rates from each channel
(denoted $R^c$) and the fraction of the total rate that have $e_{\text{10Hz}} >
0.05$ (denoted $\gamma^c$) are reported
\cite{DallAmico:2023neb,Samsing:2017xmd,Rodriguez:2017pec,Gondan:2020svr}.
Thus, we place a prior on $R^c$ and $\gamma^c$ and do a change of variables to
obtain $p(R^c_{\text{E}}, R^c_{\text{QC}})$. Explicitly, this change of variables
is: 

\begin{equation}
    \label{eq:change_of_variables}
    p(R_E^c, R_{QC}^c) = p(R^c, \gamma^c) \left[ \frac{1}{\gamma^c R^c} - \frac{1}{(1 - \gamma^c) R^c} \right],
\end{equation}
where the term in the brackets is the Jacobian transformation between
$(R^c_{\text{E}}, R^c_{\text{QC}})$ and $(R^c, \gamma^c)$.

We first discuss the priors on the fractional rates, $\gamma^c$.  
NSCs can host BBH mergers due to the high density ($2
\times 10^4 \: \mathrm{pc}^{-1}$) of BHs \cite{Tagawa:2020jnc,
Gondan:2017wzd, bahcall1976star, bahcall1977star, Miralda-Escude:2000kqv,
hailey2018density, Sigurdsson:1993tui, Sigurdsson:1994ju, Kulkarni:1993fr, PortegiesZwart:1999nm}. Mergers can also be induced by dynamical friction in
the galactic disc \cite{Bartos:2016dgn, Stone:2016wzz,
McKernan:2020lgr, Rowan:2022ehz, Rozner:2022ykb, Li:2022jbj, Ford:2021kcw}. The high BH density in NSCs leads to GW captures, which lead to high eccentricities ($e_{\mathrm{10Hz}} > 0.1$)
\cite{Rasskazov:2019gjw, Takatsy:2018euo, OLeary:2008myb,Gondan:2017wzd,Samsing:2017xmd,Gondan:2018khr,Chattopadhyay:2023pil}.
If binary-single BH interactions are constrained to the nuclear disc,
10-70\% of mergers have $e_{\mathrm{10Hz}} > 0.1$, whereas if interactions are 
isotropic, then 8-30\% of mergers have $e_{\mathrm{10Hz}} > 0.03$ \cite{Tagawa:2020jnc}. Thus we place 
a uniform prior on $\gamma^{\mathrm{NSC}}$ with a minimum of 8\% and a maximum of 70\%.

GC environments can also host BBH mergers due to the high density of BHs \cite{
Sigurdsson:1993tui, Sigurdsson:1994ju, Kulkarni:1993fr,PortegiesZwart:1999nm}
Ref.~\cite{Rodriguez:2018pss} finds $\gamma_{\text{GC}} \sim 6\%$ (Ref.
\cite{Samsing:2017xmd} finds similar results with a PN calculation).  This
method includes PN corrections to encounters between BHs and BBHs using
techniques developed in
Ref.~\cite{Rodriguez:2017pec,Amaro-Seoane:2015umi,Antognini:2013lpa}, which lead
to non-negligible eccentricity. Thus, we place a truncated Gaussian prior around
$\gamma^{\text{GC}} = 6\%$ with a standard deviation of $3\%$, which comes from 
the variation in the cluster escape velocity (see Fig.~3 of
Ref.~\cite{Samsing:2017xmd}).

Studies on the rate of eccentric mergers in YSCs are less common than NSCs or
GCs (see however, Refs.~\cite{Britt:2021dtg} and \cite{DallAmico:2023neb}).
Ref.~\cite{DallAmico:2023neb} finds less than 0.08\% of mergers in YSCs have
$e_{\mathrm{10Hz}} > 0.1$ from binary-single interactions. Thus we neglect this
effect and set $p(\gamma^{\mathrm{YSC}}) = \delta(0)$. Similarly, in IBs, we
expect $p(\gamma^{\text{IB}})= \delta(0)$ as the binaries circularize before
merger \cite{peters1964gravitational,hinder2008circularization}. 

We now turn to placing a prior over the total rates from each channel, $p(R^c)$. There have been a 
number of studies estimating the rate density of NSCs \cite{Miller:2008yw,Petrovich:2017otm},
GCs \cite{Rodriguez:2018rmd, Rodriguez:2018pss, Rodriguez:2021nwd, Kremer:2019iul}, 
YSCs \cite{DiCarlo:2020lfa, Banerjee:2021xzp,Santoliquido:2020bry}, and IBs
\cite{belczynski2016first, Santoliquido:2020bry,
Mapelli:2017hqk, vanSon:2021zpk}. In the following, we use
the rate estimates from Ref.~\cite{Mapelli:2021gyv} as they model the IB and
NSC, GC and YSC pathways with the same input physics/assumptions. See Tab. 1 in Ref.~\cite{Mapelli:2021gyv} 
for the initial conditions/priors of the simulations.
Since, we are interested in the merger rate of BBHs within $z < 0.9$ and not the
merger rate density as a function of redshift, which is computed in \cite{Mapelli:2021gyv}, we may write $R^c$ as  

\begin{equation}
    \label{eq:merger_rate}
    R^c = \int_0^{0.9} dz \: \mathcal{R}(z) \frac{dV}{dz} \frac{1}{1 + z}. 
\end{equation}
Here, $\mathcal{R}(z)$ is the volumetric rate density in units of
mergers $\mathrm{Gpc}^{-3} \mathrm{yr}^{-1}$ as a function of redshift and 
${dV}/{dz}$ is the differential co-moving volume. The $\frac{1}{1+z}$ accounts for the 
fact that $\mathcal{R}(z)$ is in the source frame whereas we want $R^c$ in the observer frame. 
We can perform the integration for the set of simulations listed in Ref.~\cite{Mapelli:2021gyv} and 
set lower and upper bounds on the rates. Then, we assume that $R^c$ is drawn from uniform distribution 
between these bounds. The final priors on $\gamma^c, R^c$ are summarized in Table \ref{tab:prior_table}.
The prior can be seen in 1D marginals on the left plot in Fig.~\ref{fig:rates}.

\setlength{\tabcolsep}{8pt}
\begin{table}[h]
    \centering
    \renewcommand*{\arraystretch}{2.0}
    \begin{tabular}{c|c|c}
        formation environment & $p(\gamma^c)$ & \multicolumn{1}{p{2cm}}{\centering $p(R^c)$ \\ mergers/yr} \\  \hline \hline  
        nuclear-star cluster & $\text{Unif}[0.08, 0.7]$ & $\text{Unif}[119, 292]$ \\ [1.0ex]
        globular cluster & $\tilde{\mathcal{N}}(0.06, 0.03)$ & $\text{Unif}[375, 1492]$ \\  [1.0ex]
        young-star cluster & $\delta(0)$ & $\text{Unif}[502, 2325]$ \\ [1.0ex]
        isolated binary & $\delta(0)$  &  $\text{Unif}[639, 31939]$ 
    \end{tabular}
    \caption{ \label{tab:prior_table} Priors on the fractional rate of eccentric
    BBH mergers, $\gamma^c$ and total rate of BBH mergers $R^c$ in each
    formation channel within $z<0.9$. Here $\text{Unif}[a, b]$ is a uniform
    distribution with upper and lower bound of $a$ and $b$, respectively.
    $\mathcal{\tilde{N}}[c, d]$ is a truncated normal distribution with mean and
    standard deviation $c$ and $d$, respectively, with truncation bounds of 0 and
    1.} 
\end{table}

The maximum probability of the prior occurs when $R_{\text{E}} / R_{\text{QC}} = 0.023$, and thus 
we quote this value for our prior odds. Now using Eq.~ (\ref{eq:posterior_odds}) we obtain  

\begin{align}
    \nonumber
     \log_{10} \mathcal{O}_{\text{EAS/QCAS}} &= 0.2-3.11 \quad  \text{for GW200129}, \\
    \nonumber
        &= \text{1.36} \quad \qquad \,\,\text{for GW190701}, \\
    \nonumber
&= \text{0.13} \quad \qquad \,\,\text{for GW200208\_22}. \\
    \nonumber
\end{align}

\subsection{Computing $p_{\text{ecc}}$ for three events}
\label{seca:p_ecc}

We now compute the rate of eccentric and quasi-circular events using GW
observations. We then use these rates to estimate the probability an event is
eccentric for the three GW events considered in Sec.~\ref{sec:event_results}.
This is done by using the formalism in Eqs.~(\ref{eq:posterior_over_flags})--
(\ref{eq:fgmc_posterior_selection_effects}). The rates obtained with this analysis using an astrophysical prior and a uniform prior over
the rates is shown in Fig.~\ref{fig:rates}. 

\begin{figure*}
    \label{fig:rates}
    \includegraphics[width=\textwidth]{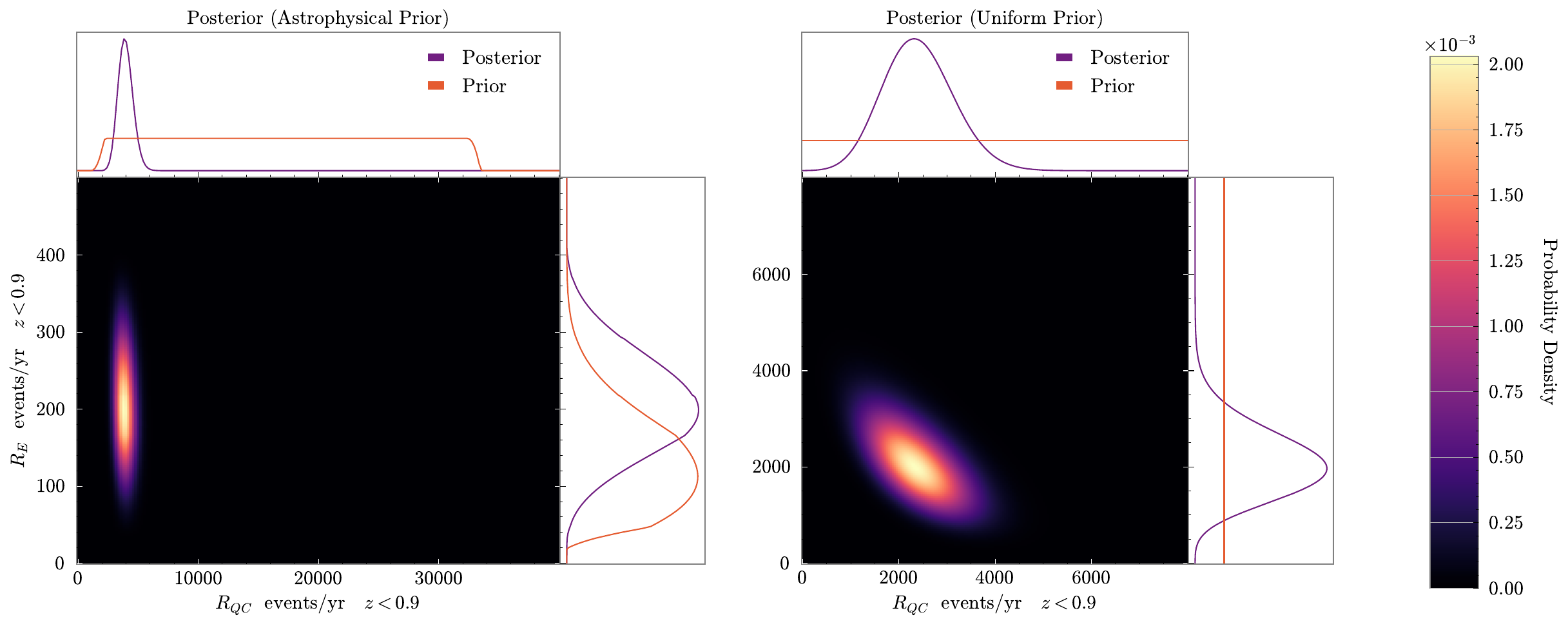}
    \caption{\label{fig:rates} Posterior distribution for the rate of eccentric $R_{\text{E}}$ and quasi-circular $R_{\text{QC}}$ events 
    per year within $z < 0.9$. This is computed using the formalism of Ref.~\cite{Farr:2013yna} 
(see Sec. \ref{sec:p_ecc}). Shown on the left is the rate estimate combining the astrophysical prior 
    with the GW data. One can see that the posterior on $R_{\text{E}}$ shifts towards higher values of 
    compared to the prior on $R_{\text{E}}$. Shown on the right is the rate 
    distribution estimated from GWs when using a uniform prior for $R_{\text{E}}$ and $R_{\text{QC}}$. In this case, the quasi-circular 
    and eccentric rates are similar since for the majority of events we cannot distinguish if $e_{\text{10Hz}} < 0.05$ or $e_{\text{10Hz}} > 0.05$
    from the data alone. Note the scales on the plots are different. }
\end{figure*}

We also include selection effects in
this estimate. To do so we compute $\alpha_{\text{QC}}$ and $\alpha_{\text{E}}$ with 
Eq.~(\ref{eq:alpha_selection}). For the prior, we use a power-law--peak distribution with
hyperparameters of the maximum-likelihood point of the LVK analysis
\cite{LIGOScientific:2021psn}. In addition, for any parameters that are outside
the prior, ($m_1 < 10 M_\odot$, $m_2 < 10 M_\odot$, $q < 0.125$, or $d_L >
6,000$ Mpc), we set $p_{\mathrm{QC, det}}(\bm{\vartheta}) = p_{\mathrm{E, det}}(\bm{\vartheta}) =
0$. Due to the fact that the selection function also depends on the PSD, we
compute $\alpha_{\text{QC}}$ and $\alpha_{\text{E}}$ for each observing run. We then
take a weighted average of the $\alpha$'s where the weights are determined by
the joint duty cycle of LIGO Hanford and LIGO Livingston for each observing run (49 days,
118 days, 107 days and 96 days for O1, O2, O3a and O3b, respectively). We also 
computed $\alpha_{\text{QC}}$ and $\alpha_{\text{E}}$ by marginalizing over the population uncertainty. That 
is, we also sampled over the population posterior when computing Eq.~\ref{eq:alpha_selection}.
However, given that there is only a 2\% error compared to the maximum-likelihood estimate we use 
the maximum-likelihood estimate of $\alpha_{\text{QC}}$ and $\alpha_{\text{E}}$ in the rest of the calculations.

There are a few interesting observations to make about Fig.~\ref{fig:rates}.
First, there is a large difference in $p(R_{\text{E}} | \{d_i\}, N)$ when using
an astrophysical prior versus a uniform one. Specifically, the astrophysical
prior has a strong influence over the eccentric rate. 
The reason for this 
that the uncertainty on the fraction of eccentric BBHs quoted in
Sec.~\ref{sec:prior_odds} is small compared to the uncertainty on the fraction of 
eccentric BBHs from GWs. The latter uncertainty can be 
seen by noticing that in Fig.~\ref{fig:bf_lineplot} the majority of events have $-1 < \log_{10} \mathcal{B}_{\text{EAS/QCAS}} < 1$
as opposed to $\log_{10} \mathcal{B}_{\text{EAS/QCAS}} \ll -1$. In other words, if just considering 
the data, it is hard to say the majority of events are definitely \textit{not} eccentric. With 
better detector sensitivities, perhaps we can better separate the populations. This 
is also the reason that when using a uniform prior on the rate of eccentric and quasi-circular
BBHs, the mean of $p(R_{\text{E}} | \{d_i\}, N)$ is only larger than $p(R_{\text{QC}} | \{d_i\}, N)$ by 14\% (361 events/yr).

In contrast, the astrophysical prior does not have the same effect on $p(R_{\text{QC}} | \{d_i\}, N)$. Here,
$p(R_{\text{QC}} | \{d_i\}, N)$ is dominated by the information from the data. This
effect can be attributed to the large uncertainty in the rate of quasi-circular events 
from astrophysical simulations \cite{Mapelli:2021gyv}. 

We can compute the probability that particular GWs are eccentric,
$p_{\text{ecc}}$, using Eq.~(\ref{eq:p_ecc}). Again due to the variety of glitch
mitigation techniques in GW200129, we have a range of $p_{\text{ecc}}$'s. This
uncertainty now also propagates into events other than GW200129 as
$p_{\text{ecc}}$ for each event is also conditioned on the GW data from
GW200129. When using an astrophysical prior on the rates, we obtain  

\begin{align}
    \nonumber
    p_{\text{ecc}} &= 0.7490-0.9995 \quad\text{for GW200129}, \\ 
    \nonumber
      &= 0.9763-0.9774 \quad \text{for GW190701}, \\ 
    \nonumber
      &= 0.7197-0.7286 \quad \text{for GW200208\_22}.\\
\end{align}

When using a uniform prior on the rates, we find that 

\begin{align}
    \nonumber
    p_{\text{ecc}} &= 0.9800-0.99998 \quad \text{for GW200129}, \\ 
    \nonumber
      &= 0.9986-0.9986 \quad \,\,\, \text{for GW190701}, \\ 
    \nonumber
      &= 0.9769-0.9771 \quad \,\,\, \text{for GW200208\_22}. \\
\end{align}
The value of $p_{\text{ecc}}$ from the uniform prior provides
a useful comparison to quantify the effect of the prior, while the value of
$p_{\text{ecc}}$ from the astrophysical prior, as explained above, is more
realistic. 

We can now compute the probability that there exists at least one event which is eccentric in the
population. First, we give a back-of-the-envelope computation of $p_{\text{ecc, pop}}$ 
using only $\mathcal{O}_{\text{EAS/QCAS}}$. We stress that this is \textit{not} a robust estimate as it 
assumes that the inferred rate of eccentric and quasi-circular BBHs is not influenced by GW observations. In addition, it does not marginalize 
over the astrophysical uncertainty on the rates. This calculation is meant to give an intuition and sanity check on the 
machinery in Sec.~\ref{sec:p_ecc}. We can use Eq.~(\ref{eq:p_ecc_intuition}) to map the odds ratios of the analyzed events 
to the interval $[0, 1]$. We can then combine the events by computing 
\begin{equation}
    \label{eq:back_of_envelope_p_ecc_pop}
    \tilde{p}_{\text{ecc, pop}} = 1 - \prod_{i=1}^{N} \left[ 1 - \tilde{p}_{\text{ecc}}(i) \right].
\end{equation}
If we use \texttt{BayesWave A} glitch mitigation on GW200129, we recover
$\tilde{p}_{\text{ecc, pop}} = 99.8\%$. We can also perform the same
calculation, but only use the three events that show signs of eccentricity. In
this case, we find $\tilde{p}_{\text{ecc, pop}} = 99.3\%$.

Finally, we can use the machinery in Eq.~(\ref{eq:p_ecc_population}), and Eqs.~(\ref{eq:flag_normalization_1})--
(\ref{eq:flag_normalization_5}) to robustly compute the probability that there is at least one eccentric event in the population. When using an
astrophysical prior on the rates we find

\begin{align}
    \nonumber
    p_{\text{ecc, pop}} &> 1 - 8.4 \times 10^{-8} \: \text{using \texttt{gwsubtract}},  \\
    \nonumber
     &> 1 - 4.7 \times 10^{-3} \: \text{using \texttt{BayesWave A}}. \\
\end{align}

Just for comparison we also compute $p_{\text{ecc, pop}}$ when 
using a uniform prior on the rates,

\begin{align}
    \nonumber
    p_{\text{ecc, pop}} &> 1 - 7.6 \times 10^{-14} \: \text{using \texttt{gwsubtract}},  \\
    \nonumber
     & > 1 - 6.2 \times 10^{-11} \: \text{using \texttt{BayesWave A}}.\\ 
\end{align}

We can also perform leave one out studies for to see how much impact 
each individual event has on the $p_{\text{ecc, pop}}$. In the following we only 
report the numbers when using an astrophysical prior on the rates. In addition,
we report the results using the \texttt{BayesWave A} glitch mitigation which is the most 
conservative case. We find 

\begin{align}
    \nonumber
    p_{\text{ecc, pop}} &> 1 - 1.7 \times 10^{-2} \: \text{leaving out GW200129},  \\
    \nonumber
    p_{\text{ecc, pop}} &> 1 - 1.6 \times 10^{-1} \: \text{leaving out GW190701},  \\
    \nonumber
    p_{\text{ecc, pop}} &> 1 - 1.6 \times 10^{-2} \: \text{leaving out GW200208\_22}.  \\
    \nonumber
\end{align}

\section{Conclusion}
\label{sec:conclusion}

In this paper we performed an analysis of 57 GWs using the multipolar
eccentric non-precessing waveform model \texttt{SEOBNRv4EHM}
\cite{Ramos-Buades:2022lgf}. This waveform model includes eccentricity
corrections to the GW multipoles up to the 2PN order, and allows for sampling in
both the eccentricity and relativistic anomaly.  This large scale analysis was
made possible with the normalizing-flow approach available with
\texttt{DINGO}~\cite{Dax:2021tsq}. 

We compare \texttt{DINGO} with \texttt{pBilby} \cite{Smith:2019ucc} using two
zero-noise injections, and find consistency between the samplers. The posteriors
are completely consistent with JSDs in the 1D marginals less than 0.002 bits.
This gives us confidence that \texttt{DINGO} is able to achieve the same
accuracy as standard samplers for eccentric waveforms, while requiring orders
of magnitude lower compute.

There are signs of eccentricity in three candidate events: GW200129 GW190701 and
GW200208\_22. We find $\log_{10}$ eccentric aligned-spin against
quasi-circular aligned-spin Bayes factors of 1.84-4.75 for GW200129, 3.0 for
GW190701 and 1.77 for GW200208\_22. We cannot say conclusively that these
three binaries are eccentric, however, the eccentric hypothesis fits the data better than
the quasi-circular aligned-spin and quasi-circular spin-precessing hypotheses. These
signs indicate that eccentricity is an integral consideration in future and
current parameter-estimation studies. 

To our knowledge, this is the first time GW200129 has been fully sampled with an
eccentric waveform model. In the Appendix of Ref.~\cite{Romero-Shaw:2022xko} the authors
tried to sample this event using the re-weighting technique introduced in Ref.~\cite{Payne:2019wmy} with the waveform from Ref.~\cite{Cao:2017ndf}. The
re-weighting in Ref.~\cite{Payne:2019wmy} involves taking a quasi-circular
proposal distribution and importance sampling to an eccentric target
distribution. However, for this event the re-weighting led to an effective sample size
of one. This may be the case for events where the quasi-circular 
proposal is significantly biased by the presence of eccentricity. By directly performing inference
on the eccentric parameter, we are able to obtain over 5,000 effective
samples across all priors, glitch mitigation techniques and waveform models for this event.

We do see very strong evidence for eccentricity in GW200129 when using
\texttt{gwsubtract} glitch mitigation. But when taking into account the
astrophysical prior odds and other methods of glitch mitigation, we cannot
conclude that GW200129 is eccentric. In particular the $\log_{10}$
odds ratio between the eccentric aligned-spin and quasi-circular aligned-spin 
waveform is between $0.2-3.11$. This
implies the importance of robust glitch mitigation techniques
and thorough comparisons between said techniques is crucial in current and future observing runs \cite{Davis:2018yrz,Davis:2022ird, Payne:2022spz,
Macas:2023zdu, Macas:2023wiw}.

We also see signs of eccentricity in GW190701. However, because
$\log_{\text{10}}$ eccentric aligned-spin against quasi-circular spin-precessing
odds ratio for this event is $<3.5$, we cannot conclude this event is eccentric.
This event also has a glitch in it, however, we only have one glitch draw from
the \texttt{BayesWave} glitch posterior \cite{Cornish:2014kda}. In future
studies, it will be critical to explicitly marginalize over the glitch posterior
for this event, as was done in \cite{Payne:2022spz}. Given the nature of
\texttt{DINGO}, marginalizing over different glitch realizations is
straightforward.

We also find signs of eccentricity in GW200208\_22. This has been
reported previously in Ref.~\cite{Romero-Shaw:2022xko}. We see mild support for
eccentricity in GW190620, which has been flagged as potentially eccentric first
in Ref.~\cite{Romero-Shaw:2022xko}, and then in Ref.~\cite{Iglesias:2022xfc}.

We do not observe evidence of eccentricity in GW190521 in the region
$e_{\text{10Hz}} < 0.5$. This is in contention with Refs.
\cite{Romero_Shaw:2020thy}. However, \cite{Iglesias:2022xfc} which uses the \texttt{TEOBResumS-DALI} waveform model
\cite{Nagar:2021gss,Nagar:2018zoe,Albanesi:2022xge}, finds similar results. It is at present 
unclear what the cause of this discrepancy is, and further work in this direction is needed. 
Note, our result is not immediately in contention with
Ref.~\cite{Gamba:2021gap} which probes the parameter space of non-spinning
hyperbolic captures which is separate from eccentric inspiral-merger-ringdown waveforms. 
It is also not immediately in contention with Ref.~\cite{Gayathri:2020coq} 
which probes the eccentricity at merger with numerical relativity simulations and a maximum likelihood test.

We also see marginal support for eccentricity in GW190620\_030421 and GW191109\_010717 which is broadly
in agreement with previous eccentric parameter estimation studies \cite{Romero-Shaw:2022xko, Iglesias:2022xfc}. 

We also perform a study of the effect of higher modes beyond the $(\ell, m) = (2,2)$
mode in eccentric waveform models. We find that there are biases in the intrinsic parameters when
one neglects higher modes in parameter-estimation studies. In the quasi-circular
case, this has been found in Refs.~\cite{LIGOScientific:2020ibl,Capano:2013raa,Varma:2014jxa,VanDenBroeck:2006ar,VanDenBroeck:2006qu,Littenberg:2012uj}.
Our contribution is to study the biases which occur in the presence of eccentricity. We
have quantified this effect by comparing the loss of information accumulated when
neglecting higher modes in eccentric waveforms with the loss of
information accumulated when neglecting higher modes in quasi-circular waveforms. We find with GW191109\_010717 that ignoring higher modes leads to biases in the eccentricity distribution of 0.025 bits. For context, with a Gaussian, an 0.007 bit difference
corresponds to a 20\% shift in the mean measured in standard deviations of the Gaussian. 
In addition, if we neglect higher modes in GW191109\_010717, the 
$\log_{10}$ Bayes factor changes from 0.07 to 0.32. We do not find in the events
analyzed that neglecting higher modes leads to incorrectly finding evidence for
eccentricity.

We also investigated the systematic effects of ignoring eccentricity completely
in parameter-estimation studies. Similar studies have been performed in Refs.~\cite{Cho:2022cdy, Divyajyoti:2023rht, Bonino:2022hkj,
Wu:2020zwr,Lower:2018seu, Ramos-Buades:2023yhy, OShea:2021faf,
Favata:2021vhw, Favata:2013rwa, Iglesias:2022xfc,
Sun:2015bva, Wagner:2024ecj, LIGOScientific:2016ebw}. Our new contribution is to report the bias in the masses and
spins for real events with support for $e_{\text{10Hz}} > 0.1$. We find that
when ignoring eccentricity, the chirp mass is biased by 5.2\%  for GW200129, 3.6\%  for GW190701 and 
11.8\%  GW200208\_22 with respect to the aligned-spin quasi-circular case. Similarly, 
the $\chi_{\text{eff}}$ is biased by $-0.04$ for GW200129, $0.12$ for GW190701 and $-0.19$ for GW200208\_22 (again for the 
aligned-spin quasi-circular case).

Finally, we have performed a computation to measure the
probability that a GW event is eccentric. This method is more robust than
simply considering odds ratios, as it combines information from 
the ensemble of GW events with astrophysical predictions. It is 
analogous to $p_{\text{astro}}$ in LVK searches \cite{Farr:2013yna}. As more GWs are observed, 
this method will give tighter constraints on the rate of eccentric and quasi-circular events than 
simply using an astrophysical prior. This will lead to smaller uncertainty on
the probability that individual events are eccentric. Currently, this method is limited to 
differentiating eccentric aligned-spin and quasi-circular aligned-spin binaries. This is because there are no population synthesis codes that coherently model eccentricity and spin-precession, so we cannot put reasonable priors on the relative rate between eccentric aligned-spin versus quasi-circular spin-precessing binaries. Thus we caveat population statements with the fact that we neglect spin-precession when reporting $p_{\text{ecc}}$ and  $p_{\text{ecc, pop}}$. This issue will be addressed in future work.  

We find that
$p_{\text{ecc}} = 0.7490-0.9995$ for GW200129, $p_{\text{ecc}} = 0.9763-0.9774$
for GW190701, and $p_{\text{ecc}} = 0.7197-0.7286$ for GW200208\_22 when using
astrophysical priors. We extend this formalism to evaluate the 
probability that there exists an eccentric event in the population. We find this to 
be $p_{\text{ecc, pop}} > 1 - 4.7 \times 10^{-3}$ when using \texttt{BayesWave A} glitch mitigation 
on GW200129 and $p_{\text{ecc, pop}} > 1 - 8.4 \times 10^{-8} $ when using \texttt{gwsubtract} 
glitch mitigation on GW200129.

\begin{acknowledgments}

It is our pleasure to thank H\'ector Estell\'es for useful discussions during 
the project especially related to the computation of the kick velocity of
GW200129.  We would like to thank Alexandre Toubiana for providing the code to
generate samples from the power-law-peak LVK distribution. We thank Vivien
Raymond for chairing the LVK Collaboration internal review of \texttt{DINGO}. 
We thank Alan Knee and Jess McIver for their useful interactions in the
early stages of this project. We would like to thank Arif Shaikh for the LSC
Publication \& Presentation Committee review of this manuscript. 

The computational work for this manuscript was carried out
on the compute clusters Saraswati, Lakshmi and Hypatia at the Max
Planck Institute for Gravitational Physics in Potsdam. Training of networks 
was also carried out on the Atlas cluster at the Max Planck Institute for Intelligent Systems in T{\"u}bingen.
\texttt{DINGO} is publicly available through the \texttt{Python} package
dingo-gw at \texttt{https://github.com/dingo-gw/dingo}. Stable
versions of \texttt{DINGO} are published through conda, and can be installed via 
conda install -c conda-forge dingo-gw

A.R.B. was supported in the last stages of this work by the Dutch Research Council (NWO) through the Veni project  VI.Veni.222.396. This research has made use of data or software obtained from the Gravitational
Wave Open Science Center (gwosc.org), a service of the LIGO Scientific
Collaboration, the Virgo Collaboration, and KAGRA. This material is based upon
work supported by NSF's LIGO Laboratory which is a major facility fully funded
by the National Science Foundation, as well as the Science and Technology
Facilities Council (STFC) of the United Kingdom, the Max-Planck-Society (MPS),
and the State of Niedersachsen/Germany for support of the construction of
Advanced LIGO and construction and operation of the GEO600 detector. Additional
support for Advanced LIGO was provided by the Australian Research Council. Virgo
is funded, through the European Gravitational Observatory (EGO), by the French
Centre National de Recherche Scientifique (CNRS), the Italian Istituto Nazionale
di Fisica Nucleare (INFN) and the Dutch Nikhef, with contributions by
institutions from Belgium, Germany, Greece, Hungary, Ireland, Japan, Monaco,
Poland, Portugal, Spain. KAGRA is supported by Ministry of Education, Culture,
Sports, Science and Technology (MEXT), Japan Society for the Promotion of
Science (JSPS) in Japan; National Research Foundation (NRF) and Ministry of
Science and ICT (MSIT) in Korea; Academia Sinica (AS) and National Science and
Technology Council (NSTC) in Taiwan.

Virgo
is funded, through the European Gravitational Observatory
(EGO), by the French Centre National de Recherche Scien-
tifique (CNRS), the Italian Istituto Nazionale di Fisica Nu-
cleare (INFN) and the Dutch Nikhef, with contributions by
institutions from Belgium, Germany, Greece, Hungary, Ireland, Japan, Monaco, Poland, Portugal, Spain. KAGRA is
supported by Ministry of Education, Culture, Sports, Science
and Technology (MEXT), Japan Society for the Promotion
of Science (JSPS) in Japan; National Research Foundation
(NRF) and Ministry of Science and ICT (MSIT) in Korea;
Academia Sinica (AS) and National Science and Technology
Council (NSTC) in Taiwan.

M.P. was supported by NSF Grants AST-2407453 and PHY-2512902.
\end{acknowledgments}

\appendix
\label{sec:appendix}

\section{Eccentricity estimates from single detector networks}
\label{sec:one_detector}

\begin{figure}
    \begin{center}
        \includegraphics[width=0.5\textwidth]{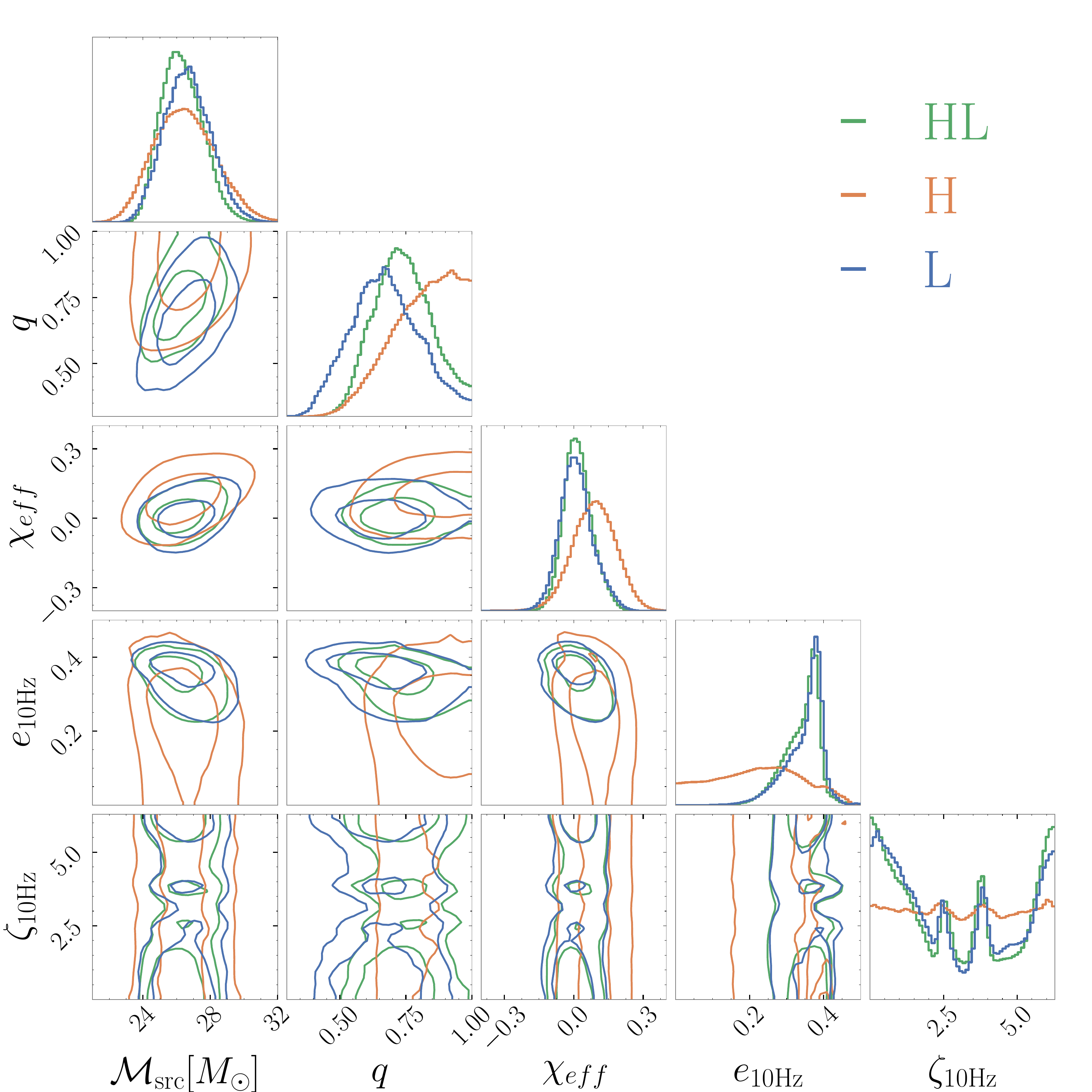}
    \end{center}
    \caption{\label{fig:gw200129_detectors} Analysis of
    GW200129 using LIGO-Hanford and LIGO-Livingston (HL), LIGO Hanford (H) and LIGO
    Livingston (L). We observe that the evidence for eccentricity comes largely
    from the LIGO-Livingston detector. The \texttt{gwsubtract} glitch subtraction is being 
    used when using the Livingston detector. There is no glitch in the Hanford detector. }
\end{figure}

GW200129 was detected in LIGO Hanford, LIGO Livingston and VIRGO. In the main
text, we perform inference using the combined data from LIGO Hanford and LIGO
Livingston. However, due to the glitch in LIGO Livingston, it is useful to
seperate the analysis into single detector analyses. This was also performed in
Ref.~\cite{Payne:2022spz}. As shown in Fig.~\ref{fig:gw200129_detectors}, 
the majority of evidence for eccentricity comes from LIGO Livingston. There is 
still marginal support for eccentricity in the Hanford only analysis with
$\log_{10} \mathcal{B}_{\text{EAS/QCAS}} = 0.86$. Interestingly, the maximum
likelihood of the Hanford only analysis peaks at $e_{\text{10Hz}}=0.40$.

The difference in the Bayes factors between Hanford and Livingston can be 
explained by the lack of SNR in Hanford. The SNR in Livingston is 22 whereas the SNR in
Hanford is only 14. The difference in SNR is due to the 
particular antenna patterns for this event. We can take the max-likelihood
waveform of GW200129 analyzed in both detectors and compute $F_+^2 + F_\times^2$
\cite{Schutz:2011tw} to get an order of magnitude estimate of the impact of
the detector projection. We find that for Hanford $F_+^2 + F_\times^2 = 0.53$ and for Livingston $F_+^2 + F_\times^2 = 0.92$. 

We can then ask if the lack of evidence of eccentricity in Hanford is due to low
SNR. To do this we perform an injection recovery using the max likelihood parameters of GW200129 analyzed
with Hanford. The maximum likelihood point of the Hanford only analysis peaks at $e_{\text{10Hz}}=0.40$. If 
we decrease $d_L$ of the injection to artificially set the SNR to 22, we see a peak in the eccentricity posterior. 
However, performing the same injection with lower SNR yields a eccentricity posterior much closer to the prior. This 
implies the reason there is only a small bump in the eccentricity in Hanford is due to the lack of SNR in the 
detector rather than the lack of evidence for eccentricity.

\section{Lower Bound on $p_{\text{ecc, pop}}$}
\label{sec:lower_bound_p_ecc_pop}

In order to compute $p_{\text{ecc, pop}}$ we must compute $p(\{ g_i = 0 \} \: |
\: \{ d_i \}, N)$. However, to do this we need to normalize the distribution 
$p(\{ g_i \} \: | \: \{ d_i \}, N )$. Note that 

\begin{equation}
    \label{eq:flag_normalization_1}
    \begin{aligned}
       1 = & \sum_{\{ g_i \} } p(\{ g_i \} | \{ d_i \}, N) \\
        &= \sum_{\{ g_i \} } \int dR_{\text{E}} \: dR_{\text{QC}} \: p(\{ g_i \}, R_{\text{E}}, R_{\text{QC}} | \{ d_i \}, N).
    \end{aligned}
\end{equation}

We don't have the normalization over $p(\{ g_i \}, R_{\text{E}}, R_{\text{QC}} | \{ d_i \}, N)$, 
but we can re-write Eq.~ (\ref{eq:flag_normalization_1}) as 
\begin{equation}
    \label{eq:flag_normalization_2}
    \begin{aligned}
        \beta = \sum_{\{ g_i \} } \left[ \prod_{\{i | g_i = 1 \}}^{N_{\text{E}}} Z_{\text{E}}(d_i) \right]
        \left[\prod_{\{i | g_i = 0 \}}^{N_{\text{QC}}} Z_{\text{QC}}(d_i) \right]
        \: I(\{ g_i \})
    \end{aligned}
\end{equation}
where $\beta$ is the normalization of $p(\{ g_i \}, R_{\text{E}}, R_{\text{QC}} | \{ d_i \}, N)$ 
and $I(\{ g_i \})$ is an evidence independent integral over the rates 

\begin{equation}
    \label{eq:integral_for_flags}
    I(\{ g_i \}) = \int  dR_{\text{E}} \: dR_{\text{QC}} \: R_{\text{E}}^{\sum g_i} R_{\text{QC}}^{\sum (1-g_i)}  e^{-(\alpha_{\text{E}} R_{\text{E}} + \alpha_{\text{QC}} R_{\text{QC}})} p(R_{\text{E}}, R_{\text{QC}}),
\end{equation}

which depends only on the number of eccentric and quasi-circular flags and the prior over the rates. 

In principle we can compute the sum in Eq.~ (\ref{eq:flag_normalization_2}) exactly. However, the sum has $2^N$
elements so if all the summands are important this will become infeasible.
Instead, if we can identify dominant terms in the sum we can place a lower bound on $\beta$.
Then we can we can put an upper bound on $p(\{ g_i = 0 \} \: | \: \{ d_i \} )$ which
translates to a lower bound on $p_{\text{ecc, pop}}$. So the larger the lower
bound we can place on $\beta$ the larger the lower bound we can place on
$p_{\text{ecc, pop}}$. Thus our goal is to now eliminate small terms from the
sum to get as large a lower bound as possible.

There are two scenarios to consider: using  a uniform prior or using an
astrophysical prior. If we use a uniform prior, each $I(\{ g_i \})$ is of the
same order in Eq.~ (\ref{eq:flag_normalization_2}). Thus the $\{ g_i \}$ which
contribute the most in the sum are those which select the events with
$Z_{\text{E}} \gg Z_{\text{QC}}$ or $Z_{\text{E}} \ll Z_{\text{QC}}$. Given that
we have no events with $Z_{\text{E}} \ll Z_{\text{QC}}$ (see Fig.~\ref{fig:bf_lineplot}) we focus on cases where $Z_{\text{E}} \gg Z_{\text{QC}}$.
Let $\{j_1, \dots j_{N_\text{E}} \}$ be the indexes of the $N_\text{E}$ events
with $Z_{\text{E}} \gg Z_{\text{QC}}$. We can remove any terms in Eq.~
(\ref{eq:flag_normalization_2}) which do not have at least one event with
$Z_{\text{E}} \gg Z_{\text{QC}}$. That is we remove $\{ g_i \}$ with

\begin{equation}
    \label{eq:flag_set_removal_condition}
    \sum_{i \in \{j_1, \dots j_{N_\text{E}}\}} g_i = 0,
\end{equation}

from the sum. The normalization $\beta$ becomes 
\begin{equation}
    \label{eq:flag_normalization_3}
    \begin{aligned}
        \beta > & \sum_{k=1}^{N_{\text{E}}}  \sum_{\{ j_k \} }^{N_{\text{E}} \choose k}  \quad \left[ \prod_{j_l \in \{ j_k \} } Z_{\text{E}}(d_{j_l}) \right] \\
        & \sum_{\{g_i | i \notin \{j_k \} \}} \left[\prod_{\{i | g_i = 1 \wedge i \not\in \{ j_k \} \}} Z_{\text{E}}(d_i) \right]
        \left[ \prod_{\{i | g_i = 0 \wedge i \notin \{ j_k \} \}} Z_{\text{QC}}(d_i) \right] \: \\
        & \quad \quad \times I(\{ g_i \}). 
    \end{aligned}
\end{equation}

Where it is implied that $I(\{ g_i \})$ here explicitly sets the $j_l$ flags to
1. In the first sum we sum over the number of flags, $k$, which are set to 1 and
have $Z_{\text{E}} \gg Z_{\text{QC}}$. The second sum represents the sum over
the $N_{\text{E}}$ choose $k$ combinations of
flags with $Z_{\text{E}} \gg Z_{\text{QC}}$. Each, $\{j_k \}$ is 
shorthand for creating a set choosing $k$ elements from $\{j_1, \dots j_{N_{\text{E}}} \}$. 
These first two sums serve to separate the events with 
$Z_{\text{E}} \gg Z_{\text{QC}}$ from the events with $Z_{\text{E}} \sim Z_{\text{QC}}$.
Finally, the last
sum is over the $\{g_i \}$
with $Z_{\text{E}} \sim Z_{\text{QC}}$. 

This last sum is still intractable, but we can note that each term in the last
sum of Eq.~ (\ref{eq:flag_normalization_3}) is at least greater than $I_s
\prod_i Z_s(d_i)$ where $Z_s(d_i) = \min(Z_{\text{E}}(d_i), Z_{\text{QC}}(d_i))$
and $I_s = \min_{\{ g_i \}} {I(\{ g_i \})}$. $I_s$ can be computed directly from
Eq.~\ref{eq:integral_for_flags}. We find that $I(\{ g_i \})$ is minimized when $\sum g_i = N$
(in the uniform prior case this is also equivalent to the case where all the $\sum g_i = 0$
via symmetry). Thus we can place a bound on $\beta$ of 

\begin{equation}
    \label{eq:flag_normalization_4}
    \begin{aligned}
        \beta > \sum_{k=1}^{N_{\text{E}}} & \sum_{\{ j_k \}}^{N_{\text{E}} \choose k} \quad \left[ \prod_{j_l \in \{ j_k \}} Z_{\text{E}}(d_{j_l}) \right] \\
        \times \:  & 2^{N - N_\text{E}} \left[ \prod_{i}^{N-k} Z_s(d_i) \right] I_s. 
    \end{aligned}
\end{equation}

Here, the $2^{N - N_\text{E}}$ comes from counting the number of terms in the last 
sum of Eq.~\ref{eq:flag_normalization_3}. In our case with $N_{\text{E}} = 3$, this is a sum over only $3 + 1 + 3 = 7$ elements. 

However, if we use a astrophysical prior 
note $R_\text{QC}$ has support for values much larger than $R_\text{E}$ in the prior (see top left of
Fig.~\ref{fig:rates}). Thus since $I(\{g_i \})$ has terms like
$R_{\text{QC}}^{\sum 1 - g_i}$ in the integrand, the $\{ g_i \}$ with the
largest $\sum 1 - g_i$ will dominate Eq.~ (\ref{eq:flag_normalization_2}). Thus 
we can group terms which have all $g_i = 0$, terms which have all but one $g_i = 0$ 
terms which have all but two $g_i = 0$ and so on. In each grouping, the terms which contribute the most are those with 
at least one $g_{j_l} = 1$ with $j_l \in \{j_1 \dots j_k \}$.
The lower bound then becomes 

\begin{equation}
    \label{eq:flag_normalization_5}
    \begin{aligned}
        \beta > & \sum_{k=0}^{N_{\text{E}}}  \sum_{\{ j_k \} }^{N_{\text{E}} \choose k}  \quad \left[ \prod_{j_l \in \{ j_k \} } Z_{\text{E}}(d_{j_l}) \right]
        \left[\prod_{i \not \in \{ j_k \}} Z_{\text{QC}}(d_i) \right] I(\{ g_i \} )
    \end{aligned}
\end{equation}

\section{Background distribution of Bayes factors}
\label{sec:background_distribution}

\begin{figure}
\begin{center}
    \hspace*{-0.03\textwidth}\includegraphics[width=0.47\textwidth]{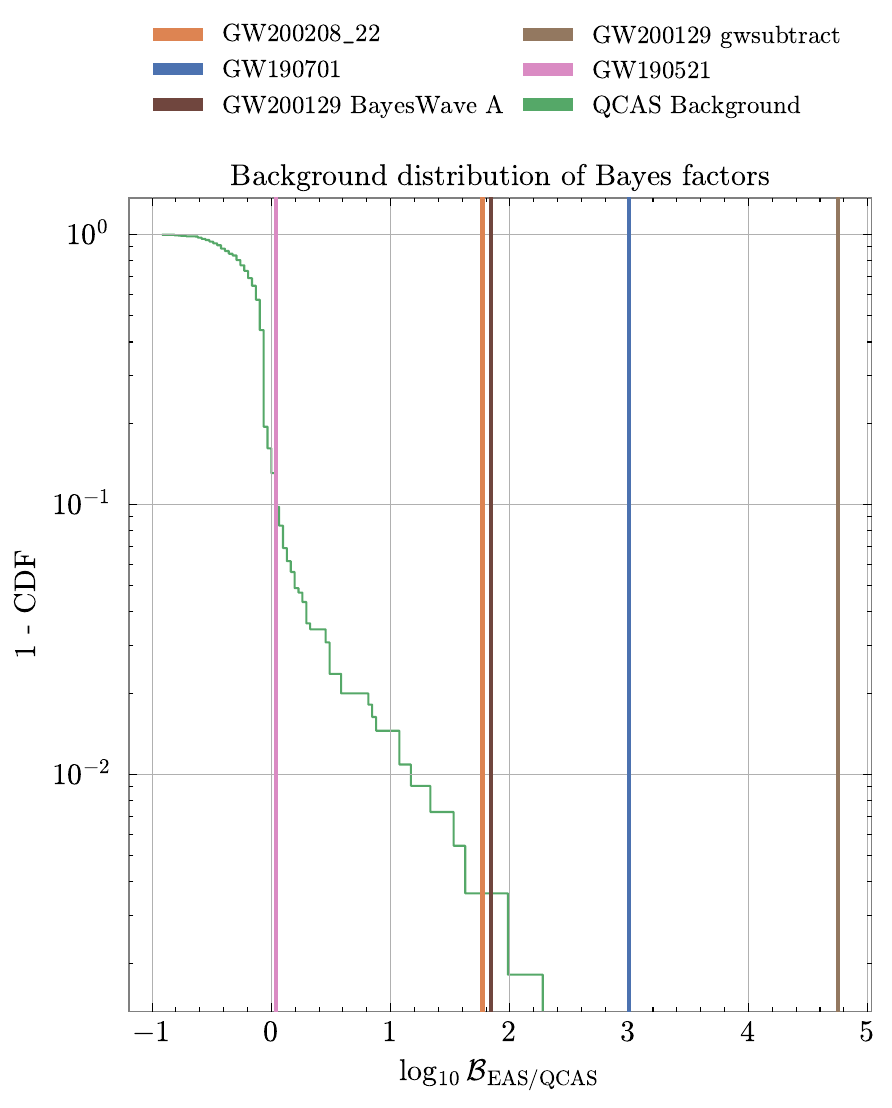}
\end{center}
\caption{\label{fig:bf_significance_level} Significance levels of GW200129, GW190701 and GW200208\_22 based 
on a quasicircular aligned-spin background distribution. In green, we show 1 minus the CDF of 550 draws from the population
distribution in Ref.~\cite{LIGOScientific:2021psn}. By finding where the Bayes factors of the individual events intersect 
with the green line, we can compute the significance of each event. }
\end{figure}

We further characterize the significance level of our observations by comparing 
our measured Bayes factors to an astrophysical background distribution.  This is
useful for mitigating the effect of the choice of eccentricity prior in the 
analysis as the background distribution is subject to the same prior effects. 
We do this test by first drawing samples from the maximum likelihood of the
power-law--peak mass distribution with the DEFAULT spin model of the LVK
\cite{LIGOScientific:2021psn}. Then we apply the selection function described
in Eq.~(91) of Ref.~\cite{Thrane:2018qnx} and Sec.~\ref{sec:p_ecc} to this injection set. Here we are
using a non-eccentric selection function since we want to 
simulate the effect of a non-eccentric background distribution on non-eccentric
LVK searches. We draw 550 injections (after selection) as this allows us to put significance levels on GW200208\_22 and GW200129.
We then inject \texttt{SEOBNRv4HM} waveforms and recover
the posterior with \texttt{SEOBNRv4EHM}. Finally, we use the Savage-Dickey
ratio to compute the Bayes factor between \texttt{SEOBNRv4HM} and
\texttt{SEOBNRv4EHM}\footnote{We could also obtain the Bayes factors by
analyzing the injections with \texttt{SEOBNRv4HM}, however, since we are only
interested in the Bayes factor and not the posteriors for this test, the
Savage-Dickey ratio gives us a cheap alternative.  This is possible because
\texttt{SEOBNRv4HM} is equivalent to \texttt{SEOBNRv4EHM} in the $e \rightarrow
0$ limit.}.  We show the cumulative distribution function (CDF) of 
the recovered Bayes factors in Fig.~\ref{fig:bf_significance_level}. By assessing where the
measured Bayes factors intersect with the CDF of the population, we can compute 
how much of an outlier each event is with respect to a quasi-circular aligned-spin population. We find
GW200129 with \texttt{BayesWave A} glitch mitigation has a significance of $3.6
\times 10^{-3}$, GW200208\_22 has a significance of $3.6 \times 10^{-3}$ and
GW190521 has a significance of $1.3 \times 10^{-1}$. Due to the limited number
of injections we can only claim that GW200129 with \texttt{gwsubtract} glitch
mitigation and GW190701 lie beyond the $1.8 \times 10^{-3}$ level. 

We also attempted to inject a quasicircular precessing-spin population and
recover the distribution of $\mathcal{B}_{\text{EAS/QCAS}}$.
This test would measure how much of an outlier the $\mathcal{B}_{\text{EAS/QCAS}}$ of
individual events are under the assumption that all events are actually
spin-precessing. However, because the \texttt{SEOBNRv4EHM} DINGO network is not
trained on \texttt{SEOBNRv4PHM} waveforms, we fail to get $n_{\text{eff}}>5,000$
for a subset (40\%) of injections.  As this may cause an incorrect estimate of
the CDF of the background distribution we leave this for future work. 

\section{Gaussian noise synthetic injection}
\label{sec:gaussian_noise}
\begin{figure}[h]
    \begin{center}
        \includegraphics[width=0.5\textwidth]{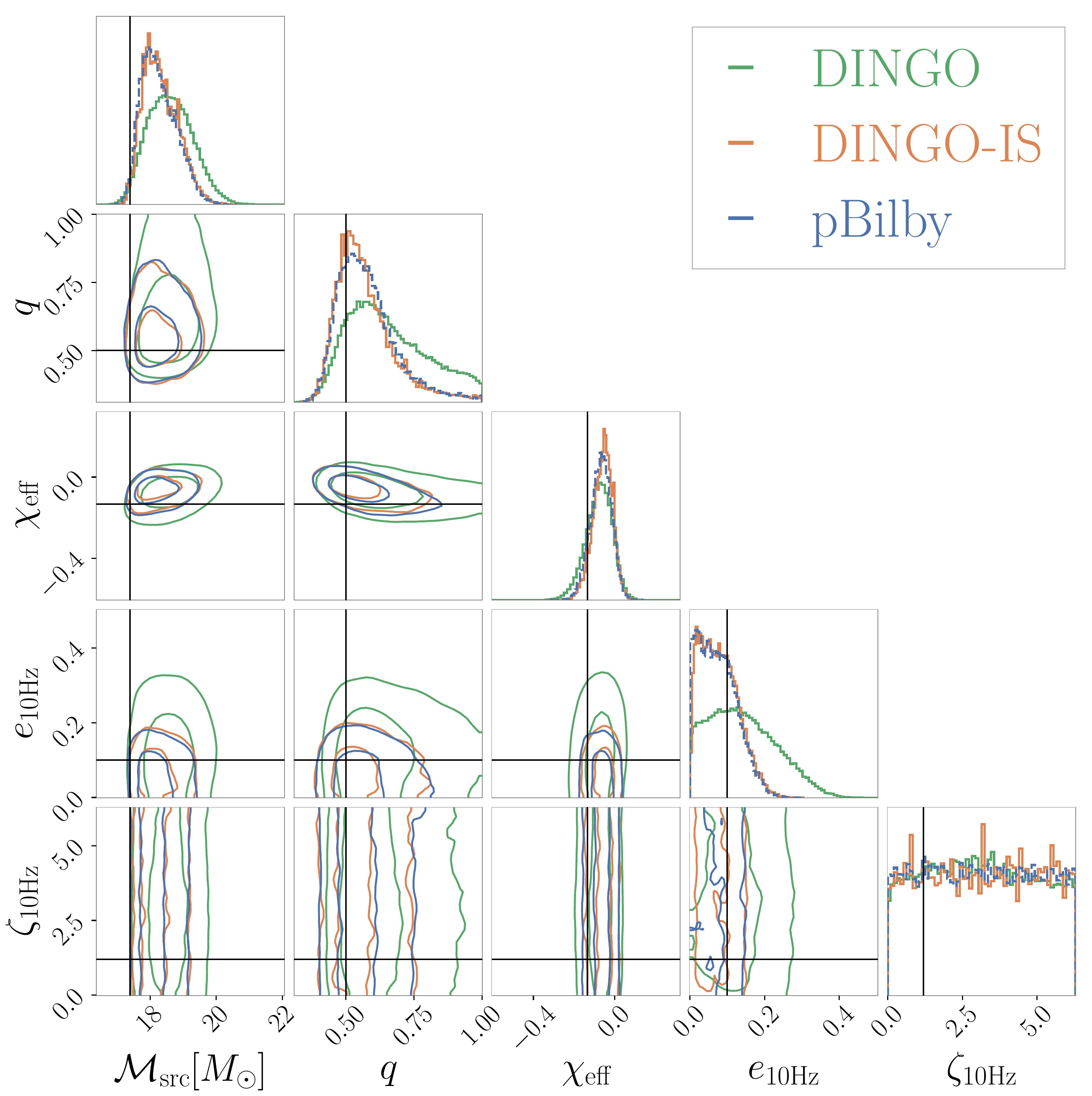}
    \end{center}
    \caption{\label{fig:gaussian_noise} Posterior distributions for a Gaussian noise injection 
    with \texttt{SEOBNRv4EHM} with parameters $\chi_{\text{eff}} = -0.13$, $\mathcal{M}_{\text{src}} = 17.2 M_\odot$, $q=0.5$, 
    $e_{10\text{Hz}}=0.1$ and $\zeta_{10\text{Hz}} = 1.2$. The sample efficiency importance sampling on this event is 7\%.}
\end{figure}

To verify the accuracy of DINGO in the presence of Gaussian noise, we perform a
Gaussian noise injection using \texttt{SEOBNRv4EHM}. The cornerplot comparing 
DINGO-IS to pBilby is shown in Fig.~\ref{fig:gaussian_noise}. When comparing the pBilby posterior to the DINGO-IS posterior 
we see all JSDs
are less than $2 \times 10^{-3}$ bits. It may seem strange that the 
injected value of the chirp mass is at the 90\% credible level for both pBilby and DINGO-IS. However, note that this is not unexpected as 10\% of injections in gaussian noise should fall outside the 90\% interval of the marginal distribution. 
 
\section{\label{sec:kick_velocity} Kick velocity of GW200129}

\begin{figure*}
\begin{center}
    \includegraphics[width=\textwidth]{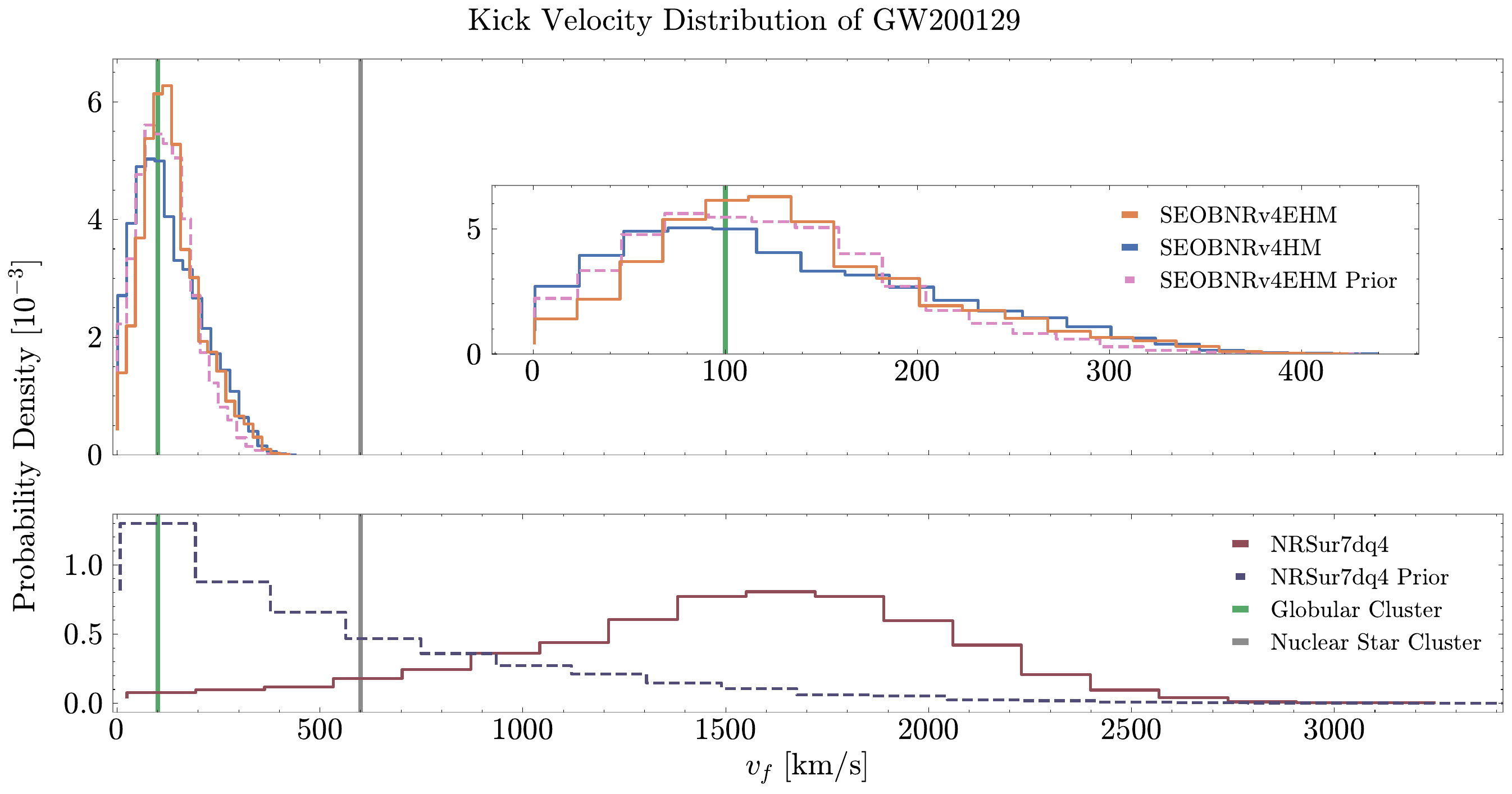}
\end{center}
\caption{\label{fig:kick_velocity} Kick velocity of GW200129 in km/s. Shown on the top (bottom) figure 
is the posterior distribution of the kick velocity using samples from the
\texttt{SEOBNRv4EHM} and \texttt{SEOBNRv4HM} (\texttt{NRSur7dq4}) analyses using \texttt{gwsubtract}
glitch mitigation. The green (grey) vertical lines show the typical escape
velocity of a globular (nuclear) cluster as shown in Ref.
\cite{Varma:2022pld}. }
\end{figure*}

As discussed in the main text, when analyzing GW200129 with \texttt{SEOBNRv4EHM}
we find a low kick velocity.  We compute the kick velocity for
\texttt{SEOBNRv4EHM} by integrating the momentum radiated throughout the orbit,
$\boldmath{P}^{\text{rad}}$, using the waveform modes $h_{lm}$
\cite{Ruiz:2007yx}. We then determine the remnant mass, $m_f$, using a fit to
numerical relativity as described in Ref.  \cite{Varma:2019csw}. Then the
remnant kick velocity is simply $v_f = \boldmath{P}^{\text{rad}} / m_f$. For
\texttt{NRSur7dq4} we can simply use the \texttt{NRSur7dq4Remnant} model from
Ref.~\cite{Varma:2019csw} with the \texttt{surfinBH} package to find $v_f$. 

However, obtaining a low kick velocity when analyzing GW200129 with
\texttt{SEOBNRv4EHM} does \textit{not} prove that this event has a negligible
kick. The large kicks in BBHs come from asymmetric momentum emission. The
highest momentum asymmetry occurs with misaligned spins
\cite{Campanelli:2007cga,Bruegmann:2007bri}. Thus with an aligned-spin model
like \texttt{SEOBNRv4EHM} or \texttt{SEOBNRv4HM} the recoil velocity has to be
small \textit{by construction} \footnote{It has also been that in low eccentricity NR simulations
kicks are only $\sim 25\%$ larger than the quasi-circular case
\cite{Radia:2021hjs}.}. But if one were to include both eccentricity and
spin-precession, we may find that GW200129 still has a measurable kick velocity.

\section{The effect of the occam penalty}
\label{sec:occam_penalty}

An eccentric aligned-spin model contains less degrees of freedom 
than a quasi-circular spin-precessing model. 
Therefore, the Bayes factors will slightly favor
the eccentric aligned-spin hypothesis, even if the true parameters of the event 
are quasi-circular and aligned-spin \footnote{We note the Occam penalty should 
also cause $\log_{10} \mathcal{B}_{\text{EAS/QCAS}}$ to favor the quasi-circular aligned-spin model
by the same argument.}. This is known as the Occam penalty 
and it is a feature of Bayesian model comparison. We stress that one should 
not try to cancel out the Occam penalty, it quantifies the fact that 
if the data is uninformative, the model with less degrees of freedom should 
be preferred (this is also sometimes called Occam's razor).

Nonetheless, we can compute the expected $\mathcal{B}_{\text{EAS/QCP}}$ had the true event parameters 
been quasi-circular aligned-spin. If these are similar to the $\mathcal{B}_{\text{EAS/QCP}}$
reported in Tab.~\ref{tab:Bayes_table}, we know that the $\mathcal{B}_{\text{EAS/QCP}}$ are 
driven by the fact that the eccentric aligned-spin model has less degrees of freedom and 
rather than information from the data. 

We perform three zero-noise injections using the maximum likelihood parameters of GW200129, GW190701 and GW200208\_22
analyzed with \texttt{SEOBNRv4EHM} but setting the eccentricity to zero. We then compute 
$\log_{10} \mathcal{B}_{\text{EAS/QCP}}$ for these injections. We find that $\log_{10} \mathcal{B}_{\text{EAS/QCP}} = 
0.1, 0.06, 0.07$ for GW200129, GW190701 and GW200208\_22 respectively. In all cases, we
we see that there is a minor preference for the eccentric aligned-spin hypothesis. We note that 
this mild preference for the eccentric aligned-spin hypothesis is orders of magnitude smaller than the
Bayes factors reported in Tab.~\ref{tab:Bayes_table} and Fig.~\ref{fig:bf_significance_level}. This indicates that the prior volume difference alone cannot account for the observed Bayes factors.

\section{Precessing injections with eccentric recovery and vice versa}
\label{sec:cross_injections}

\begin{figure*}
\captionsetup[subfigure]{labelformat=empty}

\subfloat[]{\includegraphics[width=0.5\linewidth]{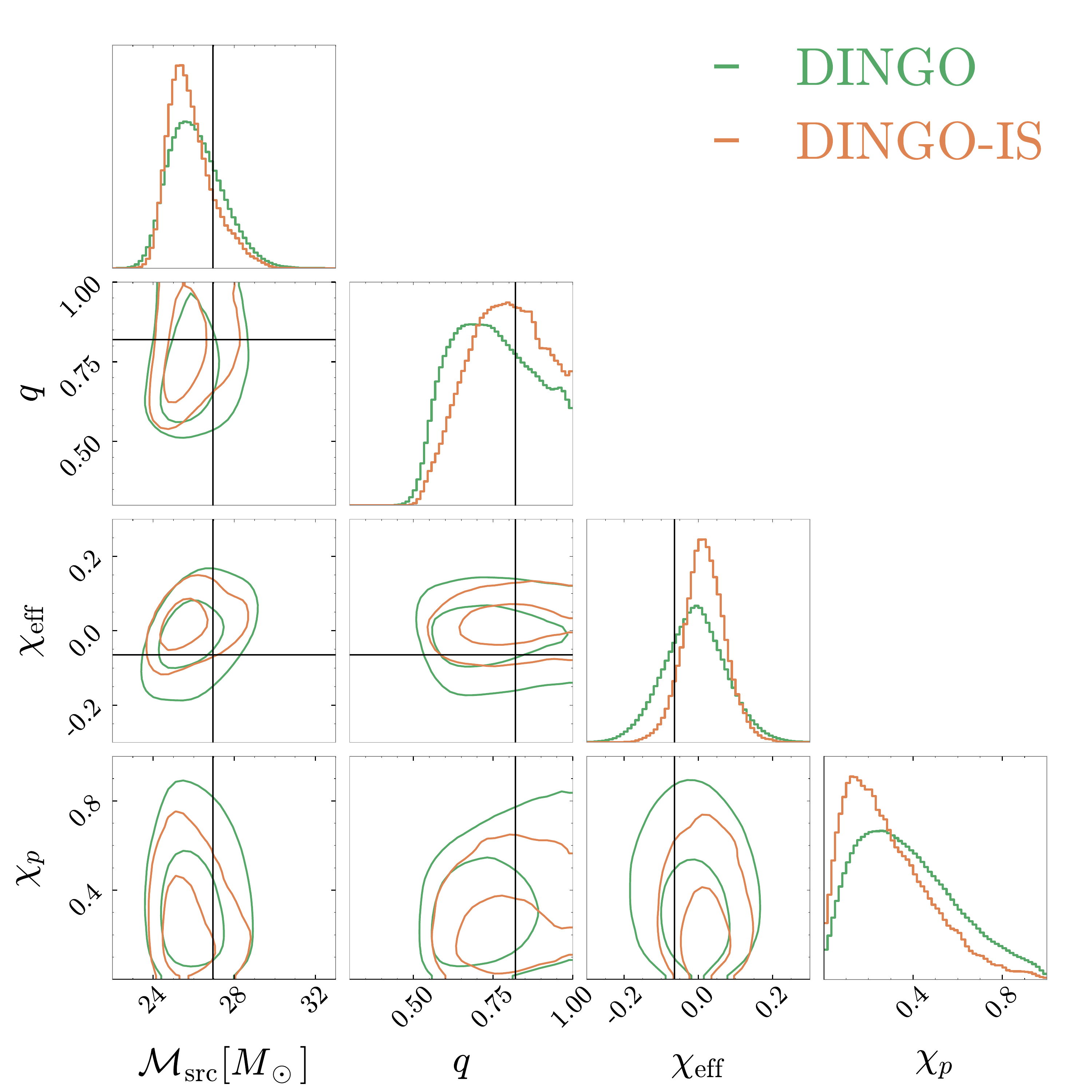}\label{fig:ehm_into_nr_injection}} 
\subfloat[]{\includegraphics[width=0.5\linewidth]{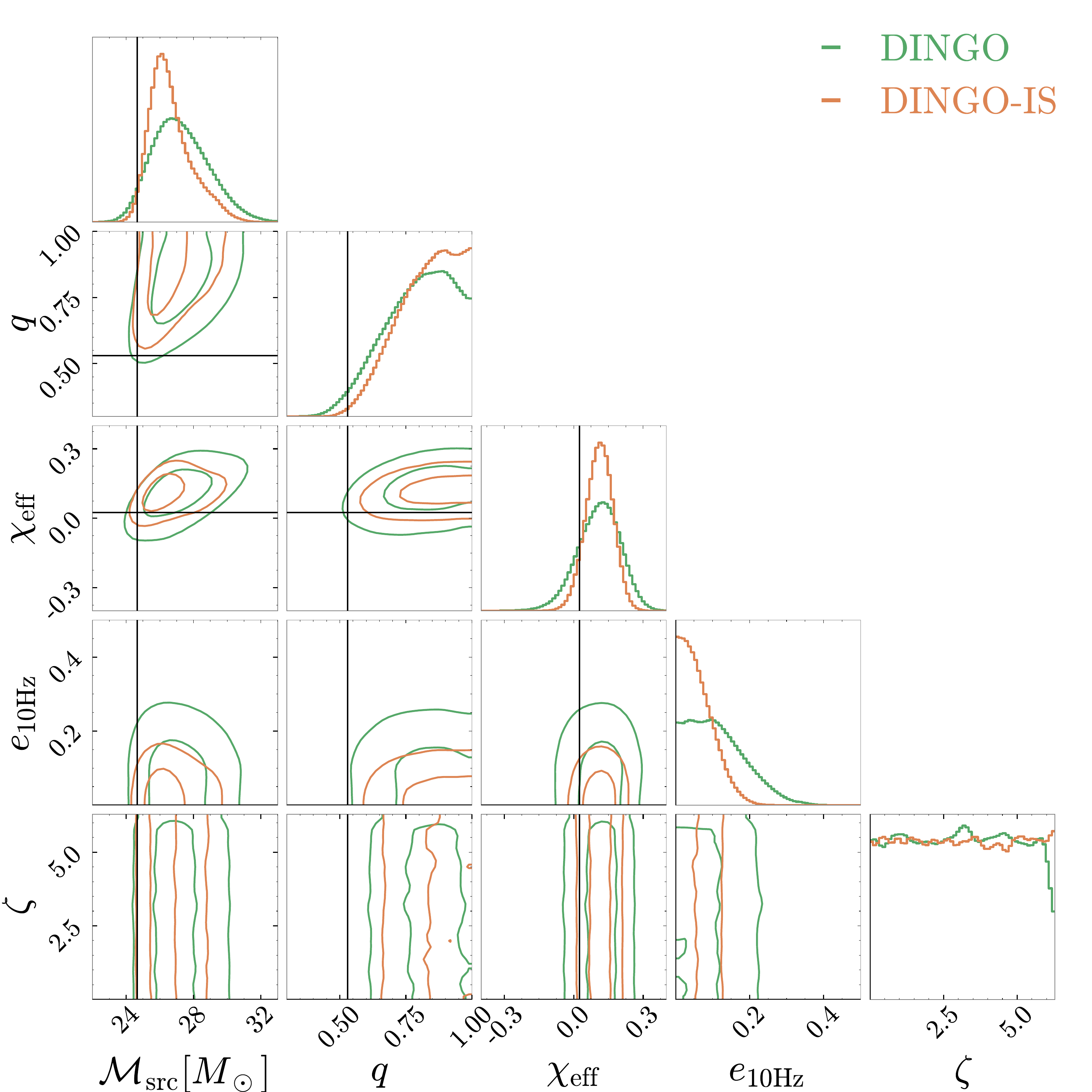}\label{fig:nr_into_ehm_injection}}

\caption{\label{fig:cross_injections} 
Posterior distributions obtained with \texttt{NRSur7dq4} (left) and \texttt{SEOBNRv4EHM} (right)
when injecting \texttt{SEOBNRv4EHM} (left) or \texttt{NRSur7dq4} (right). The 
injection parameters were chosen to be the maximum likelihood point of GW200129 with \texttt{gwsubtract}
glitch mitigation analyzed with \texttt{SEOBNRv4EHM} (left) and \texttt{NRSur7dq4} (right). 
}
\end{figure*}

Due to the possibility that eccentricity could be confused for spin-precession in short 
signals \cite{Romero-Shaw:2022fbf, CalderonBustillo:2020xms}, we perform an injection recovery campaign using 
\texttt{SEOBNRv4EHM} and \texttt{NRSur7dq4}. In this case we are most interested in 
GW200129. Therefore, we first analyze GW200129 with \texttt{gwsubtract} using both 
\texttt{SEOBNRv4EHM} and \texttt{NRSur7dq4}. We then find the maximum likelihood 
point from each analysis. Finally, we inject a precessing \texttt{NRSur7dq4} waveform 
with the maximum likelihood parameters and recover it with \texttt{SEOBNRv4EHM} (and vice versa). 
The results of this analysis are shown in Fig.~\ref{fig:cross_injections}. With the 
parameters considered here, we do not find evidence that spin-precession can be mistaken 
for eccentricity (or vice-versa).

\clearpage

\rowcolors{2}{gray!25}{white}
\setlength{\tabcolsep}{6.6pt}
\renewcommand{\arraystretch}{1.6} 
\begin{longtable*}{lcccccccc}

    \centering
    Event Name & $\text{FAR}_{\text{min}}$ ($\text{yr}^{-1}$) & $p_{\text{ecc}}$ & $\log_{10} \mathcal{B}_{\text{EAS/QCAS}}^{22}$ & $\log_{10} \mathcal{B}_{\text{EAS/QCAS}}^{\text{HM}}$ & $e_{\text{10Hz}}$ & $n_{\text{eff}}$ & $n_{\text{eff}} / N \times 100$ \\
    \hline

    GW150914 & $<1.00\times 10^{-5}$ & $0.02$ & -- & -0.44 & $0.07_{-0.07}^{+0.11}$ & 21668 & 4.3\% \\
    GW170104 & $<1.00\times 10^{-5}$ & $<0.01$ & -0.31 & -0.78 & $0.17_{-0.17}^{+0.22}$ & 8868 & 1.8\% \\
    GW170729 & $1.80\times 10^{-1}$ & $0.01$ & -0.18 & -0.65 & $0.19_{-0.19}^{+0.23}$ & 22520 & 4.6\% \\
    GW170809 & $<1.00\times 10^{-5}$ & $0.01$ & -0.30 & -0.73 & $0.17_{-0.17}^{+0.22}$ & 17561 & 3.6\% \\
    GW170814 & $<1.00\times 10^{-5}$ & $0.10$ & -0.27 & 0.66 & $0.09_{-0.09}^{+0.13}$ & 5280 & 0.6\% \\
    GW170818 & $<1.00\times 10^{-5}$ & $0.01$ & -0.30 & -0.68 & $0.17_{-0.17}^{+0.23}$ & 39222 & 8.1\% \\
    GW170823 & $<1.00\times 10^{-5}$ & $0.01$ & -0.18 & -0.61 & $0.19_{-0.19}^{+0.23}$ & 27691 & 5.7\% \\
    GW190408\_181802 & $<1.00\times 10^{-5}$ & $0.02$ & -0.45 & -0.48 & $0.16_{-0.16}^{+0.15}$ & 14498 & 3.3\% \\
    GW190413\_052954 & $8.20\times 10^{-1}$ & $0.09$ & -0.10 & 0.25 & $0.20_{-0.20}^{+0.19}$ & 31394 & 1.0\% \\
    GW190413\_134308 & $1.81\times 10^{-1}$ & $0.10$ & -0.08 & 0.32 & $0.23_{-0.23}^{+0.22}$ & 138986 & 2.8\% \\
    GW190421\_213856 & $2.83\times 10^{-3}$ & $0.03$ & -0.27 & -0.25 & $0.15_{-0.15}^{+0.22}$ & 76219 & 15.6\% \\
    GW190503\_185404 & $<1.00\times 10^{-5}$ & $0.06$ & -- & 0.04 & $0.23_{-0.23}^{+0.17}$ & 6037 & 2.1\% \\
    GW190512\_180714 & $<1.00\times 10^{-5}$ & $0.02$ & -0.37 & -0.40 & $0.11_{-0.11}^{+0.17}$ & 69513 & 14.2\% \\
    GW190513\_205428 & $<1.00\times 10^{-5}$ & $0.11$ & 0.30 & 0.36 & $0.26_{-0.24}^{+0.14}$ & 6546 & 0.3\% \\
    GW190514\_065416 & $2.80\times 10^{0}$ & $0.10$ & -0.01 & 0.32 & $0.23_{-0.23}^{+0.21}$ & 295948 & 6.3\% \\
    GW190517\_055101 & $3.47\times 10^{-4}$ & $0.02$ & -0.29 & -0.21 & $0.19_{-0.19}^{+0.20}$ & 13103 & 2.7\% \\
    GW190519\_153544 & $<1.00\times 10^{-5}$ & $0.02$ & -0.43 & -0.41 & $0.15_{-0.15}^{+0.25}$ & 44089 & 9.4\% \\
    GW190521 & $<1.00\times 10^{-5}$ & $0.04$ & -- & -0.04 & $0.15_{-0.13}^{+0.15}$ & 261513 & 26.7\% \\
    GW190521\_074359 & $1.00\times 10^{-2}$ & $0.04$ & -0.00 & -0.13 & $0.15_{-0.14}^{+0.08}$ & 148692 & 30.4\% \\
    GW190527\_092055 & $2.28\times 10^{-1}$ & $0.05$ & 0.08 & 0.01 & $0.33_{-0.29}^{+0.17}$ & 29131 & 6.2\% \\
    GW190602\_175927 & $<1.00\times 10^{-5}$ & $0.04$ & -0.19 & -0.15 & $0.19_{-0.19}^{+0.21}$ & 75130 & 15.7\% \\
    GW190620\_030421 & $1.12\times 10^{-2}$ & $0.15$ & -- & 0.56 & $0.32_{-0.21}^{+0.18}$ & 32211 & 6.6\% \\
    GW190630\_185205 & $<1.00\times 10^{-5}$ & $0.02$ & -- & -0.45 & $0.09_{-0.09}^{+0.12}$ & 5580 & 1.1\% \\
    GW190701\_203306 & $5.71\times 10^{-3}$ & $0.98$ & 2.72 & 3.00 & $0.46_{-0.04}^{+0.04}$ & 19337 & 1.0\% \\
    GW190706\_222641 & $<1.00\times 10^{-5}$ & -- & 0.37 & -- & $0.46_{-0.41}^{+0.04}$ & 14690 & 1.5\% \\
    GW190719\_215514 & $6.30\times 10^{-1}$ & $0.11$ & -0.18 & 0.37 & $0.18_{-0.18}^{+0.23}$ & 15033 & 0.8\% \\
    GW190727\_060333 & $<1.00\times 10^{-5}$ & $0.10$ & -0.33 & 0.31 & $0.12_{-0.12}^{+0.19}$ & 5833 & 1.2\% \\
    GW190731\_140936 & $3.30\times 10^{-1}$ & $0.09$ & -0.20 & 0.26 & $0.17_{-0.17}^{+0.23}$ & 10976 & 2.2\% \\
    GW190803\_022701 & $7.32\times 10^{-2}$ & $0.09$ & -0.21 & 0.27 & $0.17_{-0.17}^{+0.22}$ & 5154 & 1.1\% \\
    GW190828\_063405 & $<1.00\times 10^{-5}$ & $0.10$ & 0.28 & 0.31 & $0.25_{-0.22}^{+0.13}$ & 68326 & 14.0\% \\
    GW190828\_065509 & $<1.00\times 10^{-5}$ & -- & 0.31 & -- & $0.31_{-0.24}^{+0.18}$ & 8007 & 0.2\% \\
    GW190910\_112807 & $2.87\times 10^{-3}$ & $0.0$ & -- & 0.08 & $0.16_{-0.16}^{+0.14}$ & 14053 & 4.7\% \\
    GW190915\_235702 & $<1.00\times 10^{-5}$ & $0.02$ & -0.45 & -0.55 & $0.23_{-0.22}^{+0.14}$ & 40476 & 9.0\% \\
    GW190916\_200658 & $4.70\times 10^{0}$ & $0.10$ & -0.18 & 0.33 & $0.17_{-0.17}^{+0.22}$ & 5965 & 1.2\% \\
    GW190926\_050336 & $1.10\times 10^{0}$ & $0.09$ & -0.03 & 0.25 & $0.23_{-0.23}^{+0.21}$ & 5257 & 1.1\% \\
    GW190929\_012149 & $<1.00\times 10^{-5}$ & $0.04$ & -0.34 & -0.09 & $0.20_{-0.20}^{+0.22}$ & 157953 & 3.6\% \\
    GW191109\_010717 & $1.80\times 10^{-4}$ & $0.06$ & 0.33 & 0.15 & $0.26_{-0.22}^{+0.24}$ & 15211 & 7.9\% \\
    GW191127\_050227 & $2.49\times 10^{-1}$ & -- & 0.06 & -- & $0.30_{-0.27}^{+0.20}$ & 7193 & 0.7\% \\
    GW191204\_110529 & $<1.00\times 10^{-5}$ & $0.03$ & -0.29 & -0.34 & $0.12_{-0.12}^{+0.21}$ & 21785 & 4.5\% \\
    GW191215\_223052 & $<1.00\times 10^{-5}$ & -- & 0.12 & -- & $0.28_{-0.27}^{+0.17}$ & 9083 & 0.9\% \\
    GW191222\_033537 & $<1.00\times 10^{-5}$ & -- & -0.30 & -- & $0.13_{-0.13}^{+0.20}$ & 1720155 & 35.1\% \\
    GW191230\_180458 & $5.02\times 10^{-2}$ & $0.10$ & -0.21 & 0.32 & $0.16_{-0.16}^{+0.23}$ & 8923 & 0.3\% \\
    GW200112\_155838 & $2.04\times 10^{-1}$ & $0.02$ & -- & -0.51 & $0.08_{-0.08}^{+0.13}$ & 38358 & 7.7\% \\
    GW200128\_022011 & $4.29\times 10^{-3}$ & $0.10$ & -0.21 & 0.32 & $0.16_{-0.16}^{+0.18}$ & 7198 & 1.5\% \\
    GW200129\_065458 & $<1.00\times 10^{-5}$ & $1 - 5 \times 10^{-3}$ & 4.57 & 4.75 & $0.34_{-0.06}^{+0.11}$ & 6204 & 0.1\% \\
    GW200208\_130117 & $3.11\times 10^{-4}$ & $0.03$ & -0.25 & -0.22 & $0.15_{-0.15}^{+0.21}$ & 7732 & 1.6\% \\
    GW200208\_222617 & $4.80\times 10^{0}$ & $0.73$ & 1.25 & 1.77 & $0.4_{-0.15}^{+0.08}$ & 31246 & 0.5\% \\
    GW200209\_085452 & $4.60\times 10^{-2}$ & $0.13$ & 0.07 & 0.42 & $0.25_{-0.23}^{+0.23}$ & 45766 & 0.9\% \\
    GW200216\_220804 & $3.50\times 10^{-1}$ & $0.12$ & 0.00 & 0.41 & $0.26_{-0.26}^{+0.19}$ & 1538557 & 15.4\% \\
    GW200219\_094415 & $9.94\times 10^{-4}$ & $0.05$ & -0.36 & -0.03 & $0.14_{-0.14}^{+0.22}$ & 6297 & 0.6\% \\
    GW200220\_124850 & $3.00\times 10^{1}$ & $0.09$ & -0.07 & 0.27 & $0.21_{-0.21}^{+0.24}$ & 7529 & 0.8\% \\
    GW200224\_222234 & $<1.00\times 10^{-5}$ & $0.03$ & -0.27 & -0.29 & $0.13_{-0.13}^{+0.14}$ & 135039 & 27.6\% \\
    GW200225\_060421 & $<1.00\times 10^{-5}$ & $<0.00$ & -1.19 & -1.18 & $0.18_{-0.18}^{+0.20}$ & 5797 & 0.2\% \\
    GW200302\_015811 & $1.12\times 10^{-1}$ & $0.07$ & -- & 0.24 & $0.15_{-0.15}^{+0.21}$ & 20635 & 4.1\% \\
    GW200306\_093714 & $2.40\times 10^{1}$ & $0.04$ & -0.12 & -0.12 & $0.17_{-0.17}^{+0.22}$ & 37297 & 7.8\% \\
    GW200308\_173609 & $2.40\times 10^{0}$ & $0.03$ & -0.27 & 0.07 & $0.14_{-0.14}^{+0.25}$ & 126454 & 3.3\% \\
    GW200311\_115853 & $<1.00\times 10^{-5}$ & $0.02$ & -0.50 & -0.50 & $0.09_{-0.09}^{+0.15}$ & 160548 & 32.9\% \\

   \midrule

    \caption{\label{tab:sample_data} Table showing the set of all BBHs analyzed. 
    When available we report the results using higher modes. However, for a few of the events, we do 
    not achieve greater than 5,000 effective samples using higher modes and therefore we report these results using 
    only the 22 mode. We report the minimum FAR for each event following \cite{LIGOScientific:2021psn}. }
\end{longtable*}

\clearpage
\bibliography{references}

\end{document}